\documentclass[times,authoryear]{elsarticle}

\usepackage{framed,multirow}

\usepackage{amssymb}
\usepackage{latexsym}

\usepackage{url}
\usepackage[x11names]{xcolor}
\definecolor{newcolor}{rgb}{.8,.349,.1}

\journal{Journal of Computational Physics}
\usepackage[
 pdfstartview=FitH,%
  bookmarks=true,%
  bookmarksopen=true,%
  breaklinks=true,%
  colorlinks=true,%
  linkcolor=blue,anchorcolor=blue,%
  citecolor=blue,filecolor=blue,%
  menucolor=blue,%
  urlcolor=black,
  pdfauthor=Alexandre Fournier
]{hyperref}

\usepackage{amsmath}
\usepackage{amssymb}
\usepackage{bm}
\usepackage{fleqn}
\usepackage{multirow}
\usepackage{booktabs}
\usepackage{manfnt}

\usepackage{geometry}
\newcommand{\nonlinop}{\mathcal{N}}
\newcommand{\tlnonlinop}{\mathcal{N}'}
\newcommand{\linop}{\mathcal{L}}
\newcommand{\constrainop}{\mathcal{G}}
\newcommand{\totalop}{\mathcal{Q}}
\newcommand{\taunl}{\tau_{\mbox{\scriptsize N}}}
\newcommand{\taul}{\tau_{\mbox{\scriptsize L}}}
\newcommand{\statev}{\mathbf{x}}
\newcommand{\algebraicv}{\mathbf{z}}
\newcommand{\prandtl}{\mathrm{Pr}}
\newcommand{\rayleigh}{\mathrm{Ra}}
\newcommand{\nusselt}{\mathrm{Nu}}
\newcommand{\reynolds}{\mathrm{Re}}
\newcommand{\power}{\mathrm{P}}
\newcommand{\cost}{\mbox{cost}}
\newcommand{\eff}{\mbox{eff}}
\newcommand{\epstau}{\epsilon_\tau}
\newcommand{\epsdt}{\epsilon_{\Delta t}}
\newcommand{\epslambda}{\epsilon_\lambda}
\newcommand{\real}[1]{\Re\left[ #1\right]}

\makeatletter
\newcommand*\dashline{\rotatebox[origin=c]{90}{\bfseries ${\dabar@\dabar@\dabar@\dabar@}$}}
\makeatother
\usepackage{lineno}
\usepackage{ulem}
\newcommand{\addRone}[1]{\textcolor{black}{#1}}
\newcommand{\rmRone}[1]{{\textcolor{white}{}}}
\newcommand{\addRtwo}[1]{\textcolor{black}{#1}}
\newcommand{\rmRtwo}[1]{{\textcolor{white}{}}}

\begin{document}



\begin{frontmatter}

\title{
An assessment of implicit-explicit time integrators for the
       pseudo-spectral approximation of Boussinesq thermal convection in an annulus
}%

\author[1]{Venkatesh Gopinath\fnref{fn1}}
\cortext[cor1]{Corresponding author: e-mail: 
  \htmladdnormallink{fournier@ipgp.fr}{mailto:fournier@ipgp.fr}
  }
\author[1]{Alexandre Fournier\corref{cor1}}
\fntext[fn1]{Present address: Bosch Engineering and Business Solutions, India. e-mail: 
\htmladdnormallink{gopinath.venkatesh2@in.bosch.com}
{mailto:gopinath.venkatesh2@in.bosch.com}}
\author[1]{Thomas  Gastine \fnref{fn3}}
\fntext[fn3]{e-mail: \htmladdnormallink{gastine@ipgp.fr}{mailto:gastine@ipgp.fr}}

\address[1]{Universit\'{e} de Paris, Institut de physique du globe de Paris, CNRS, F-75005 Paris, France}


\begin{abstract}
We analyze the behaviour 
of an ensemble of time integrators applied
to the semi-discrete problem resulting
from the spectral discretization
of the equations describing Boussinesq
thermal convection in a cylindrical annulus.
The equations are
cast in their vorticity-streamfunction formulation
that 
 yields a differential algebraic equation (DAE).
The ensemble comprises 28 members:   
4~implicit-explicit multistep schemes, 22 implicit-explicit Runge-Kutta (IMEX-RK)
 schemes, and 2 fully explicit schemes \rmRone{punctually} 
used for reference. 
The schemes whose theoretical order varies from 2 to 5 are assessed
for 11~different physical setups that cover
 laminar and turbulent regimes. Multistep
  and order 2 IMEX-RK
 methods
 exhibit their expected order of convergence
 under all circumstances. 
 IMEX-RK methods of higher-order show occasional order
 reduction that impacts both algebraic
 and differential field variables.
  We ascribe
 the order reduction to the stiffness
 of the problem at hand and, to a larger extent,
 the presence of the DAE.
 Using the popular
 Crank-Nicolson Adams-Bashforth of order 2 (CNAB2)
 integrator as reference,
 performance is 
 defined by the ratio
 of maximum admissible time step to the cost of performing one iteration;
 the maximum admissible time step is determined by inspection of the time
 series of viscous dissipation within the system, which 
 guarantees a physically acceptable solution.
 Relative performance is bounded between 0.5 and 1.5 across
 all studied configurations. 
  Considering accuracy jointly with performance, we find that 6
 schemes consistently outperform CNAB2, meaning that in addition
 to allowing for a more efficient calculation, the accuracy that
 they achieve at their operational\addRone{, dissipation-based} 
  limit of stability yields a lower error.
  In our most turbulent setup, where
  the behaviour of the methods is almost entirely dictated
 by their explicit component,
  13 IMEX-RK integrators outperform CNAB2 in terms of
 accuracy and efficiency.
\end{abstract}


\end{frontmatter}



\section{Introduction}
This study is concerned with the evaluation of implicit-explicit (IMEX) 
time marching methods applied to the numerical simulation of thermal convection
for geophysical or astrophysical bodies. 
Thermal convection is an ubiquitous process in natural systems of large size; 
it drives the internal and external evolution 
of planets and stars as they receive and shed heat from and to outer space. In the case of Earth's internal
dynamics, two envelopes undergo convection: the rocky mantle and 
the  liquid outer core underneath it, which is essentially composed of an 
Iron-Nickel alloy.  
Both systems host vigorous time-dependent convective currents, over vastly different
time scales, since the mantle turnover time, $\mathcal{O}(10^8)$~years, is 
a million times larger than that of the core, $\mathcal{O}(10^2)$~years. 

Numerical models of 
mantle convection appeared in the nineteen-seventies
\citep[e.g.][]{mckenzie1974convection,sato1976finite,kopitzke1979finite,jarvis1984time} 
and have grown steadily in size and complexity since
\citep[see e.g.][for a review]{zhong2015numerical}. Efforts have been carried
out in view 
of handling both complex geometries and rheologies, in order for instance to simulate
 plate tectonics in an inertialess framework that necessitates the solve
 of a modified Stokes problem for the flow. This led 
to the development of multilevel elliptic solvers and the implementation of
adaptive mesh refinement techniques. The design of high-order integration 
schemes has logically not been the focus of attention, since
the priority was to design methods able in particular to cope 
with viscosity contrasts spanning several decades. 
Two freely available
codes whose development continues today and used to simulate mantle convection in two or three dimensions are 
fluidity \citep{davies2011fluidity} and ASPECT \citep{kronbichler2012high}. 
With regard to the advection-diffusion of temperature anomalies, 
fluidity resorts to an implicit $\theta$-method \citep[][\S D.2.3]{chqz2006} for the diffusive and advective terms, 
and has $\theta=1/2$, which corresponds  to the second-order Crank-Nicolson (CN henceforth) method. 
The  ASPECT code 
 opted for a compromise between stability and accuracy 
in choosing a backward-difference formula of second order \citep[BDF2, e.g.,][\S D.2.4]{chqz2006}.
 BDF2 was also praised by these authors on the account of its ``efficiency of implementation (higher-order schemes often become unwieldy as they require
complicated initialization during the first few time steps, and require
the storage of many solution vectors from previous time steps)"
\citep{kronbichler2012high}. This statement does 
not consider high-order self-restart time integration methods of the implicit-explicit Runge
-Kutta (IMEX-RK henceforth) type, to be discussed below. 

In contrast, self-consistent
models of core dynamics, which comprise in their full form 
an electromagnetic component responsible for the generation
of the geomagnetic field by dynamo action, involve a Newtonian fluid 
of constant viscosity whose dynamics is strongly affected by 
planetary rotation. Their complexity lies in their inherent three-dimensional, global 
character and in the nonlinear coupling between field variables (velocity, pressure, temperature, magnetic field).  
These models came to the fore in the nineteen-nineties
 \citep{glatzmaier1995three,kageyama1995computer}, in the wake of the 
pioneering work of \cite{glatzmaier1984numerical} on the solar dynamo. Most codes to date
are part of  Glatzmaier's legacy; they 
rely in the horizontal directions on a spectral representation of field variables using
spherical harmonics. Nonlinear terms are computed in physical space, which requires 
forward and inverse spherical harmonic transforms to be performed at each time step. This
step represents the most expensive computation of three-dimensional spherical simulations. 
Efforts to improve code performances
focused on increasing the efficiency of spectral transforms by parallelization,
using strategies based either on distributed-memory \citep{clune1999computational}
or shared-memory  \citep{schaeffer2013efficient}. In comparison, little work was devoted
to reducing the time-to-solution using efficient time integration schemes. The performance
benchmark by \cite{matsui2016performance} 
offers an interesting perspective on this state of affairs, as
it reports the performance of $13$ spectral codes on various laminar test problems 
 (the study also includes two finite element codes). 
All spectral codes rely on a 
 implicit-explicit scheme that treats linear terms, at the exception of the Coriolis force,
 implicitly,  and nonlinear terms explicitly. 
 The  majority of codes resort to a $\theta$-method for the linear term and
 have the Crank-Nicolson value 
 $\theta=1/2$, 
 \cite[although some may use the stabler first-order $\theta=0.6$, see e.g.][around Eq.~(21)]{hollerbach2000spectral}. 
 Nonlinear and Coriolis terms are evaluated in $9$ instances out of $13$ with
 a second-order Adams-Bashforth (AB2 henceforth) method. 
 Most codes
 combine CN for the implicit part with AB2 for the explicit part, 
 forming what will we refer to in the following 
  the traditional CNAB2 method.   
  This popularity appears at odds with the flaws of the CNAB2 method, 
  reported for instance by \cite{tilgner1999spectral} in his analysis
  of time integrators for fluid flow problems in spherical shell 
  geometry: there may exist circumstances under which 
  ``the unreasonably popular CNAB2 '' \citep{ascher1997implicit} may not damp 
  oscillatory modes at the expected physical rate (meaning they are not
  damped enough), which can then lead
  to a global instability if non-linearities are present \citep[see also the discussions in 
  ][, \S D.2.2.]{ascher1995implicit,chqz2006}. 
   An alternative  consists of 
  a second-order predictor-corrector approach,
 with the predictor and corrector stages
 based on AB2 and CN, respectively \citep[as used in the Leeds code, see~][]{willis2007thermal}. 
  The focus of the study by \cite{matsui2016performance}
 is the scalability of codes 
 towards using petascale computers, based on the assessment of the 
 efficiency of the various spatial parallelization 
 strategies followed by the various contributors to the performance benchmark. 
 It is intriguing to note that the paper does not even mention
 the benefit (in terms of efficiency and also accuracy, in view of performing turbulent
 simulations using spectral methods in space) that one could potentially gain from using more accurate and 
 stabler time-schemes, on the condition that the gain compensates the extra cost. 

A survey of literature reveals that 
alternatives to CNAB2 
 were considered \rmRone{punctually} \addRone{occasionally} in the past, for three-dimensional Boussinesq
 thermal convection in axisymmetric (cylindrical or spherical)
 domains \citep[][where a BDF2/AB3 IMEX scheme was used]{fournier2005fourier},
  convection-driven dynamo action in Cartesian domains 
 \citep[][where the SBDF2, sometimes called extended Gear of
 order 2, combination was employed]{stellmach2008efficient}, or rotating 
convection under the anelastic approximation in 2-D and 3-D Cartesian domains
\citep[][where SBDF3 was used]{verhoeven2014compressional}. The influence of 
the chosen time integrator on the accuracy of the solution was also recently 
put forward by \cite{lecoanet2019tensor}. Comparing CNAB2 and SBDF4, they 
showed that using the latter fourth order scheme allowed to assess a refined 
convergence of the reference values for benchmarks of convection and dynamo 
action in full spheres \citep{marti2014full} up to ten decimal places. 
 Following up on the investigation of \cite{tilgner1999spectral}, 
  \cite{livermore2007implementation} analyzed the performance
 of second-order exponential time differencing (ETD) applied to dynamo
 action in spherical geometry, finding that for order 2, ETD methods were
 not to be preferred over the traditional CNAB2, given that their performance was similar, and
 the implementation of the latter much easier. \cite{livermore2007implementation}
 also noted that the situation
 may become different would one resort to higher-order schemes.  
  Such an endeavor was undertaken by \cite{garcia2014exponential}
  in the context of three-dimensional rotating convection. A comparison
  was made between exponential integrators and multistep IMEX schemes
  that had been previously investigated by \cite{garcia2010comparison}. 
  Based on the investigation of moderately supercritical configurations, 
  the authors concluded on the superior accuracy of exponential
  integrators, at the expense of a larger cost. 
     

  In this study, the emphasis is set on the accuracy and 
  efficiency of multistage IMEX-RK methods, which are compared 
  with some multistep IMEX methods of order 2, 3 and 4.  IMEX-RK methods
  have been almost ignored by the core dynamics community. Interestingly, 
  however,   
 \cite{glatzmaier1996anelastic} implemented early on the IMEX-RK method 
proposed by \cite{spalart1991spectral} 
to three-dimensional dynamo modeling in spherical geometry. 
 This scheme, which will be evaluated in this study,  combines a third-order
explicit component with a second-order implicit component. 
 To our knowledge, no subsequent usage of that specific scheme
 was reported in the planetary core dynamics / dynamo community until a 
 few years ago. 
 It recently resurfaced in the study of magneto-convection
in Cartesian geometry by \cite{yan2019heat}. Aside from the scheme
by \cite{spalart1991spectral}, we note that \cite{hollerbach2000spectral}
defined a second order IMEX-RK scheme assembled from an explicit RK2 for its 
explicit part and a Crank-Nicolson for its implicit component. 
More recently, \cite{marti2016computationally} tested several of 
the IMEX-RK schemes
proposed by \cite{cavaglieri2015low} of second, third and fourth order 
applied to
standard core dynamics problems in a spherical shell geometry, with 
a focus on the impact of the implicit or explicit treatment of the
linear Coriolis force on the efficiency of their code. 
 \rmRone{The interest in} IMEX-RK methods \rmRone{was} \addRone{were} also investigated by one of us in the 
context of two-dimensional quasi-geostrophic spherical convection
\citep{gastine2019pizza}. It was noted that of the three 
third-order schemes that were tested, the multistep SBDF3
by \cite{ascher1995implicit}
and the IMEX-RK BPR353 by \cite{boscarino2013implicit}
 (both of which will also be evaluated
in this paper) had a similar efficiency in a rapidly-rotating, 
moderately turbulent configuration. The latter IMEX-RK scheme was 
recently employed by \cite{tassin2021geomagnetic} to obtain
 turbulent double diffusive geodynamo models
of higher accuracy.

\cite{grooms2011linearly} performed a thorough and inspiring 
comparison of IMEX-BDF, IMEX-RK and exponential integrators
applied to a variety a problems, including \addRone{the} two-dimensional
stratified Boussinesq \addRone{equations} \rmRone{convection} and \addRone{the}
quasigeostrophic \addRone{equation} \rmRone{convection}, both in a periodic Cartesian
domain. (The 6 IMEX-RK schemes that they analyzed
are part of the 22 IMEX-RK schemes analyzed in this work.)
Their conclusions can be tentatively summarized as follows:
 in the setups that they considered, exponential integrators
 are vastly superiors in terms of accuracy to multistep and
 multistage methods, even if  some  IMEX-RK 
 schemes, such as the BHR553 scheme of \cite{boscarino2009class}, may 
 display a convergence rate better than the nominal third order. 
 The exact treatment of linear
 diffusive terms enabled by exponential integrators is key 
 in the moderately nonlinear setups they considered, and 
 \cite{grooms2011linearly} stressed that 
 IMEX scheme could prove superior to exponential integrators
 when nonlinearities play 
 a more sizeable role in the dynamics. In passing, \cite{grooms2011linearly}
 also noted that for \addRone{the} 2D \addRone{stratified}
  Boussinesq \rmRone{convection} \addRone{equations}, IMEX-RK methods ``all displayed 
the 
 disturbing ability to produce stable but inaccurate results 
 at large step sizes'' (their Fig.~9). We shall come back to 
 this observation in our own analysis.  

Owing in part to the  ease of their implementation and interchangeability 
in modern computing frameworks \citep[e.g.][]{vos2011generic}
IMEX-RK recently received
attention in atmosphere and climate modeling \citep{giraldo2013implicit,gardner2018implicit,vogl2019evaluation}.
These studies looked into the possibility of
using such schemes to overcome the severe limitations of pure explicit
time marching that arise in nonhydrostatic models of the
atmosphere which host 
acoustic waves propagating in the vertical direction. 
 The smallness of the propagation time of those
waves compared with the characteristic time of convective transport makes the problem
stiff, and calls for an implicit treatment of those terms 
responsible for wave propagation. Using a testbed consisting
of a gravity wave test and a baroclinic instability test
from the 2012 dynamical core model intercomparison project \citep{ullrich2012dynamical}, 
\cite{gardner2018implicit} implemented an anisotropic splitting strategy, 
termed HEVI, for horizontally explicit -- vertically implicit, implemented
on  21 IMEX-RK schemes of order 2, 3, 4 and 5 (see their section 3.2.1
	for their description). Likewise, 
a comprehensive 
comparative study of 27 IMEX-RK schemes led \cite{vogl2019evaluation}
to recommend 5 schemes that consistently perform better 
than the rest of the pack for the same two test cases (their Table~6). 

With the aim of applying this systematic methodology to a problem
relevant to the dynamics of planetary interiors, we focus here on 2D Boussinesq convection in 
a cylindrical annulus, in the absence of background rotation. 
This setup is admittedly simpler than some problems recently reported in
the literature. Yet, 
its modest size allows us to cover a relatively broad range 
of behaviors including turbulent solutions
and to perform a comprehensive investigation
of 22 IMEX-RK schemes, in addition to 4 IMEX multistep schemes and two 
fully explicit schemes.

The outline of the paper is the following: we present the physical
model and its numerical approximation in section~\ref{sec:model}.
Results follow in section~\ref{sec:results}, where accuracy, order
reduction, and computational efficiency are investigated and
discussed for 11 different dynamical setups. 
We next summarize our findings and conclude in section~\ref{sec:conclusion} 
with 
some tentative recommendations for subsequent use of
IMEX-RK schemes in the context of three-dimensional dynamo 
simulations in spherical geometry.

\section{The model and its numerical approximation}
\label{sec:model}
\subsection{Governing equations}

Let us operate in cylindrical coordinates $(s,\varphi,z)$ with
local unit vectors $\hat{s}$, $\hat{\varphi}$ and $\hat{z}$. 
We consider a Newtonian fluid contained in a flat  
 annulus of outer radius $s_o$ and inner radius $s_i$. 
The fluid
is subjected to a uniform inward radial gravity field of amplitude
$g_0$. 
The outer and inner cylindrical walls are maintained at uniform temperatures
 $T_o$ and $T_i$, respectively. The temperature contrast $\Delta T \equiv T_i -T_o$
 is positive and gives rise to convective flow when it exceeds a critical value. 

  The primitive state variables describing the fluid are
   its velocity $\mathbf{u}=\left(u_s, u_\varphi \right)$, pressure $p$  and temperature $T$. 
 The material properties of the fluid relevant for the problem of interest
 here are its density $\rho$, 
 kinematic viscosity $\nu$, thermal diffusivity $\kappa$ and thermal expansion
 coefficient $\alpha$. The equation of state that relates  changes in temperature
 $\delta T$ to changes
  in density $\delta \rho$ is $$\delta \rho=-\alpha \rho \delta T.$$
 Under the Boussinesq approximation, the properties of the fluid are homogeneous, 
 save for density that can vary with the local temperature according to the previous
 law when (and only when) the gravitational force is computed. 
  The basic state about which convection can take place is the motionless hydrostatic
  conducting state. The conducting temperature profile is axisymmetric,
  \begin{equation}
  T_c(s,\varphi) = T_c(s) = \Delta T \frac{\log (s/s_i)}{\log(s_i/s_o)} + T_i. 
  \end{equation}
We scale length, time, and temperature by the gap width $(D=s_o-s_i)$, the viscous diffusion time $D^2/\nu$, and $\Delta T$, respectively. We also choose pressure, $p$, to be scaled by $\rho_o {\nu}^2/D^2$, where, $\rho_o$ is the background density. Conservation of mass, momentum and energy results in the following set of equations 
\begin{eqnarray}
\boldsymbol{\nabla} \cdot \mathbf{u} &=& 0, \label{eq:mass}  \\
\frac{\partial \mathbf{u}}{\partial t} &=& -  \boldsymbol{\nabla} \cdot
( \mathbf{u} \otimes \mathbf{u} ) - \boldsymbol{\nabla}p + \frac{\rayleigh}{\prandtl} T {\hat{s}} + {\nabla}^2 \mathbf{u}, \label{eq:momentum} \\
\frac{\partial T}{\partial t}&=& -  \boldsymbol{\nabla} \cdot ( \mathbf{u}  T) + \frac{1}{Pr}\nabla^{2} T, 
\end{eqnarray}
to be complemented with initial and boundary conditions (see below). 

The dimensionless control numbers are 
the Rayleigh number $\rayleigh$ and the Prandtl number $\prandtl$, defined by 
\begin{equation}
\rayleigh = \frac{g_o \alpha \Delta T D^3}{\nu \kappa},
\label{eq:rayleigh}
\end{equation}
and 
\begin{equation}
\prandtl=\frac{\nu}{\kappa}.
\label{eq:prandtl}
\end{equation}
The two-dimensional nature of the problem and the incompressibility constraint 
prompt us to resort to a vorticity-streamfunction formulation, see e.g. 
\cite{glatzmaier2013introduction}, \S 2.1 or \cite{peyret2002spectral}, \S 
II.6.
 Let 
 the symbol overbar denote
the azimuthal averaging operator, 
$$
\overline{f} = \frac{1}{2 \pi} \int_0^{2 \pi} f(\varphi) \mathrm{d} \varphi. 
$$
We introduce a streamfunction $\psi$ such that
\begin{subequations}
\begin{align}
\label{ur}
&\begin{aligned}
u_s=\frac{1}{s}\frac{\partial \psi}{\partial \varphi}, 
\end{aligned} \\
\label{uvarphi}
&\begin{aligned}
u_{\varphi}=\overline{u_{\varphi}}- \frac{\partial \psi}{\partial s}.
\end{aligned} 
\end{align}
\end{subequations}
Given the periodicity of the domain in the azimuthal direction, the decomposition of the
azimuthal flow into a mean component $\overline{u_{\varphi}}$ and a non-zonal component
$- \partial \psi/\partial s$ ensures the periodicity of pressure in the azimuthal
direction \citep[e. g.][\S 2]{plaut2002low}. The evolution of the mean component is governed by the
azimuthal average of Eq.~\eqref{eq:momentum}. 
The axial vorticity $\boldsymbol{\omega}=\omega \hat{z}$ is given by  
\begin{equation}
\omega=\frac{1}{s} \frac{\partial (s \overline{u_{\varphi}})}{\partial s}-{\nabla}^2 \psi,
\end{equation}
and its time-dependency is controlled by the axial component of the curl of the momentum equation 
 Eq.~\eqref{eq:momentum}. 
In summary, the set of dimensionless 
equations to solve in the vorticity-streamfunction formulation reads
\begin{subequations}
\label{eq:completeset}
\begin{eqnarray}
\frac{\partial \overline{u_{\varphi}}}{\partial t} & = & 
-\overline{u_s \omega } 
+ \tilde{\Delta} \overline{u_{\varphi}}, \label{eq:meanflow}\\
\frac{\partial \omega}{\partial t} &= & - \boldsymbol{\nabla}  \cdot \left( \mathbf{u}   \omega \right) + \nabla^{2} \omega - \frac{\rayleigh}{\prandtl} \frac{1}{s} \frac{\partial T}{\partial \varphi}, 
\label{eq:vort} \\
\frac{\partial T}{\partial t}&=& -  \boldsymbol{\nabla} \cdot ( \mathbf{u}  T) + \frac{1}{Pr}\nabla^{2} T, \label{eq:heat}  \\
\omega&=&\frac{1}{s} \frac{\partial (s \overline{u_{\varphi}})}{\partial s}-{\nabla}^2 \psi, 
\label{eq:vortpsi} \\
\mathbf{u} &=& \overline{u_{\varphi}} + \boldsymbol{\nabla} \times (\psi \hat{z}),
\label{eq:upsi}
\end{eqnarray}
\end{subequations}
where the modified Laplacian operator $\tilde{\Delta} = \partial_s(\partial_s+1/s)$.

\subsection{Boundary conditions}
Regarding the mechanical boundary conditions for the annulus, 
we shall assume throughout a no-slip boundary condition 
$u_\varphi=0$ along the curved inner and outer boundary walls, together with 
with the impermeable condition $u_s=0$. The latter condition implies that
\begin{equation}
\label{psibc1}
\psi = 0 \ \text{at} \ s=s_i, \ s_o.
\label{eq:bcpsi}
\end{equation}
 The no-slip boundary condition implies 
\begin{equation}
\label{bcns}
\frac{\partial \psi}{\partial s}= \overline{u_{\varphi}}=0 \text{ at} \ s=s_i, \ s_o,
\label{eq:bcmeanflow}
\end{equation}
which results in vorticity at the boundaries to be \begin{equation}
\omega=-\frac{\partial^2 \psi}{\partial s^2} 
\text{ at} \ s=s_i, \ s_o. 
\label{eq:bcvort}
\end{equation}
Thus, we have four boundary conditions on $\psi$ and two for $\overline{u_{\varphi}}$. 
For the temperature, the dimensionless boundary conditions are 
\begin{subequations}
\label{tempbc}
\begin{align}
&\begin{aligned}
T & = 1  \ \text{at} \ s=s_i,
\end{aligned} \\
&\begin{aligned}
T & = 0  \ \text{at} \ s=s_o.
\end{aligned}
\end{align}
\end{subequations}
\subsection{Diagnostics}
\label{sec:diagsection}
Before  dealing with the numerical approximation of the problem, let us introduce 
a few diagnostics 
 to obtain useful information and 
to check solution correctness. 
In the following, we denote the spatial average over the area $A$ of the annulus by a double overbar and time average by angular brackets $\langle \cdots \rangle$ . 
For a field $f(s,\varphi)$, 
\begin{equation}
 \overline{\overline{f}}  = \frac{1}{A} \iint_A f(s,\varphi) s \text{d}s \text{d}\varphi.
\end{equation}
where, $A = \pi (s_o^2 -s_i^2)$ is the area of the annulus. 
 We compute the 
kinetic energy $E_k$ at specified times of the simulation. 
At a given instant in time, it is given by
\begin{equation}
E_k(t) = \frac{1}{2}  \overline{\overline{ \left( u_s^2 + u_{\varphi}^2 \right)}}
(t).
\end{equation}
The Nusselt number $\nusselt$ quantifies 
the ratio of total heat flux to the reference heat flux carried 
by conduction alone. We define it at the inner and outer boundaries by considering
the time and azimuthal averages of the temperature, e.g.  
\begin{equation}
\nusselt_{o} =\left(\frac{\text{d} \langle \overline{T} \rangle }{\text{d} s}\right)_{s=s_{o}}/
\left({\frac{\text{d} {T_c}}{\text{d} s}}\right)_{s=s_{o}}
=   \left(\frac{\text{d} \langle \overline{T} \rangle }{\text{d} s}\right)_{s=s_{o}}   
 s_o \log\frac{s_o}{s_i}, 
 \label{eq:nusselt}
\end{equation}
for the outer boundary; the expression for $\nusselt_i$ being obtained upon substituting the $s_o$ factor 
by $s_i$ in this equation. 
The balance of inner and outer Nusselt numbers (incoming and outgoing heat 
fluxes) indicates that thermal relaxation has been reached and it is a good indicator for convergence of the solution. Now, we define the Reynolds number $\reynolds$ 
which measures the ratio of inertial to viscous forces. With our choice of scales, it is given as the time-averaged root-mean-square magnitude of the velocity 
\begin{equation}
\reynolds = 
\bigg \langle\left(\overline{\overline{u_s^2 + u_{\varphi}^2}}\right)^{1/2} \bigg \rangle
= \bigg \langle \sqrt{2 E_k}\bigg \rangle.
\label{eq:reynolds}
\end{equation}
The next diagnostic quantity we compute from the solution at specified time 
intervals is the power balance. 
From the solution, we check if heat loss by viscous dissipation balances on average
the buoyancy 
input power \citep[e.g.][]{king2012heat}. The expression for viscous dissipation is given as
\begin{equation}
D_\nu (t)= \overline{\overline{\mathbf{u} \cdot \nabla^2 \mathbf{u}}}(t).
\end{equation}
Using the vector identity 
$ \boldsymbol{\nabla} \times \boldsymbol{\nabla} \times \mathbf{u} = 
\boldsymbol{\nabla}(\boldsymbol{\nabla} \cdot \mathbf{u}) 
- \nabla^2 \mathbf{u}$, the definition of vorticity $\omega$  and the incompressibility i
constraint, the viscous dissipation term becomes
\begin{equation}
\label{visdis1}
D_\nu(t) =  -\overline{\overline{[\mathbf{u} \cdot (\boldsymbol{\nabla} \times \boldsymbol{\omega} )]}}(t).
\end{equation}
When no-slip boundary conditions are prescribed, it further simplifies to 
\begin{equation}
D_\nu(t) = 
 - \overline{\overline{ \omega^2}}(t). 
\label{eq:visdis2}
\end{equation}
The buoyancy input power $\power$ reads 
\begin{equation}
\power(t) = \frac{\rayleigh}{\prandtl} \overline{\overline{u_s T}}(t). 
\label{eq:buopower}
\end{equation}
On time average, we expect the solution to satisfy
\begin{equation}
\langle \power \rangle = -
\langle D_\nu  \rangle .
\end{equation}

\subsection{Spatial discretization}
We apply a Fourier-collocation approach to 
discretize 
Eqs.~(\ref{eq:meanflow}-\ref{eq:upsi}) in space.
The Fourier expansion is performed along the azimuthal direction which 
is naturally periodic and a Chebyshev collocation method is employed 
along the radial direction, see e.g. \citet{glatzmaier2013introduction}. 
The truncated Fourier expansions of field variables with a dependency on the azimuthal angle read
\begin{subequations}
\label{eq:fourierexp}
\begin{align}
&\begin{aligned}
{\omega (s,\varphi,t)} & \approx 2 \sideset{}{'}\sum^{N_m}_{m=0} \real{ {\omega}_m (s,t) e^{i m \varphi}},
\label{eq:vortf}
\end{aligned} \\
&\begin{aligned}
T(s,\varphi,t) & \approx 2 \sideset{}{'}\sum^{N_m}_{m=0} \real{ T_m (s,t) e^{i m \varphi}},
\label{eq:tempf}
\end{aligned} \\
&\begin{aligned}
{\psi (s,\varphi,t)} & \approx 2 \sideset{}{}\sum^{N_m}_{m=1} \real{ {\psi}_m (s,t) e^{i m \varphi}},
\label{eq:streamf}
\end{aligned} \\
&\begin{aligned}
{u_s} (s,\varphi,t)& \approx 2\sideset{}{}\sum^{N_m}_{m=1} \real{   u_{sm}(s,t) e^{i m \varphi}},
\label{eq:urm}
\end{aligned} \\
&\begin{aligned}
u_{\varphi} (s,\varphi,t)& \approx \overline{u_{\varphi}}(s,t) + 2\sideset{}{}\sum^{N_m}_{m=1}     
\real{ u_{\varphi m}(s,t) e^{i m \varphi}}.
\label{eq:uvarphim}
\end{aligned}
\end{align}
\end{subequations}
where $N_m$ is the maximum order of the truncation, and consequently 
the number of Fourier modes is $N_m+1$, $\real{f}$ 
is the real part of a complex-valued function $f$ and the single prime 
on the summation symbol means that the $m=0$ term  
in the series is multiplied by $1/2$. 

Substituting these expressions into Eq.~\eqref{eq:completeset} results 
in the following set of nondimensional equations 
\begin{subequations}
\label{eq:fcompleteset}
\begin{align}
&\begin{aligned}
\label{eq:fmeanflow}
\frac{\partial \overline{u_{\varphi}}}{\partial t}(s,t) & = &
-\overline{u_s \omega }(s,t)
+ \tilde{\Delta} \overline{u_{\varphi}}(s,t),
\end{aligned} \\
\label{eq:fvort}
&\begin{aligned}
{\frac{\partial \omega_m}{\partial t}} (s,t)& = - 
\boldsymbol{\nabla}_m \cdot 
(\mathbf{u}  \omega)_m (s,t) + \nabla_m^{2} \omega_m (s,t) - \frac{\rayleigh}{\prandtl} 
\frac{i m}{s} T_m (s,t),  \ \text{for} \ m > 0,
\end{aligned} \\
\label{eq:fheat}
&\begin{aligned}
{\frac{\partial T_m}{\partial t}} (s,t)& = -  \bm{\nabla}_m 
\cdot  (\mathbf{u}  T)_m(s,t) 
+ \frac{1}{\prandtl} \nabla^{2}_m T_m(s,t), \ \text{for} \ m \geq 0,
\end{aligned} \\
&\begin{aligned} 
\label{eq:fvortpsi}
\omega_0(s,t) = \frac{1}{s} \frac{\partial (s \overline{u_{\varphi}})}{\partial s}(s,t),  \quad 
{\omega_m }(s,t) & = -\nabla^{2}_m \psi_m(s,t), \ \text{for} \ m > 0,  
\end{aligned}\\
&\begin{aligned} 
\label{eq:fupsi}
u_{sm}(s,t) &= \frac{im}{s} \psi_m(s,t), \ \text{for} \ m > 0,  \quad \\
u_{\varphi 0} (s,t)  &=  \overline{u_{\varphi}} (s,t),  \quad 
u_{\varphi m} (s,t)  = -\frac{\partial \psi_m}{\partial s} (s,t), \ \text{for} \ m > 0,  
\end{aligned}
\end{align}
\end{subequations}
where the Fourier mode-dependent divergence and Laplacian operators read
$\bm{\nabla}_m 
\cdot \mathbf{a} = (1/s)\partial_s(sa_s) + (im / s)a_\varphi$ and 
$\nabla^{2}_m=\partial^2_{s} + (1/s) \partial_s - (m^2/s^2)$, respectively. 
The notation $(\cdots)_m$ refers to the
$m^{\text{th}}$ Fourier mode of the term inside the brackets. In the following, we shall refer to
the previous formulation as the $s-m$ formulation. 

We proceed with a radial approximation based 
on Chebyshev  polynomials $C_n$ up
to degree 
 $N_s-1$. 
Each field variable $g_m(s,t)$ appearing
in the previous system is expanded according to 
\begin{equation}
g_m(s,t) \approx  \left(\frac{2}{N_s-1}\right)^{1/2} \sideset{}{''}\sum^{N_s-1}_{n=0} \widehat{g}_{mn}(t) C_n[x(s)],
\label{eq:chebexp}
\end{equation}
where the double quote implies that the first and last terms are multiplied by $1/2$. The cylindrical 
radius $s$ is mapped into coordinate $x$ by
\begin{equation}
x = \frac{2s -s_o-s_i}{s_o-s_i} = 2s -s_o-s_i
\end{equation}
in order to use the Chebyshev--Gauss--Lobatto points defined by 
\begin{equation}
x_k = \cos \frac{k \pi}{N_s-1}
\end{equation}
with $k=0$ to $N_s-1$. The discrete Chebyshev expansion evaluates
$g_m(s=s_k,t)$ 
 such that
\begin{equation}
s_k = \frac{s_o-s_i}{2} x_k + \frac{s_o+s_i}{2}.
\end{equation}
Conversely, 
\begin{equation}
\widehat{g}_{mn}(t) = \left(\frac{2}{N_s-1}\right)^{1/2} \sideset{}{''}\sum^{N_s-1}_{k=0}
g(s_k,t) C_n(x_k),
\end{equation}
in which 
\[
C_n(x_k) = \cos({n \arccos x_k})=\cos\frac{n \pi k}{N_s-1}. 
\]

In practice, our unknowns consist of the $\widehat{g}_{mn}$. The radial
approximation of 
 Eqs.~\ref{eq:fcompleteset} leads to the following semi-discrete problem
\begin{subequations}
\label{eq:fccompleteset}
\begin{align}
\label{eq:fcmeanflow}
&\begin{aligned}
\frac{\mathrm{d}}{\mathrm{d}t} \mathbf{M}
\widehat{\mathsf{\overline{{u}_{\varphi}}}} = {\mathcal N}_{\overline{u_{\varphi}}} 
+ \mathbf{L}_{\overline{u_{\varphi}}} \widehat{\mathsf{\overline{{u}_{\varphi}}}}, &
\end{aligned} \\
&\begin{aligned}
\label{eq:fcvort}
\frac{\mathrm{d}}{\mathrm{d}t} \mathbf{M}
\widehat{{\omega}_m} 
= 
{\mathcal N}_{\omega,m}
+ \mathbf{L}_{\omega,m} \widehat{\mathsf{\omega}_m} 
+ \mathbf{B}_{m} \widehat{\mathsf{T}_m} & \quad \text{ for }
 m > 0,
\end{aligned}\\
&\begin{aligned}
\label{eq:fcheat}
\frac{\mathrm{d}}{\mathrm{d}t} \mathbf{M}
\widehat{\mathsf{T}_m} = 
{\mathcal N}_{T,m}
+ \mathbf{L}_{T,m} \widehat{\mathsf{T}_m} & \quad \text{ for } m \ge 0,
\end{aligned}\\
&\begin{aligned}
\label{eq:fcvortpsi}
\mathbf{M} \widehat{\mathsf{\omega}_0} = \mathbf{L}_0 \widehat{\mathsf{\overline{{u}_{\varphi}}}}
, \quad 
\mathbf{M} \widehat{\mathsf{\omega}_m} = 
 \mathbf{L}_{\omega\psi,m} 
 \widehat{\mathsf{\psi}_m}
 \quad \text{for } m \ge 0,
\end{aligned}\\
&\begin{aligned}
\label{eq:fcupsi}
\mathbf{M}\widehat{\mathsf{u}_{s,m}} &=  
\mathbf{L}_{u_s\psi,m} \ \widehat{\mathsf{\psi}_m}, 
\ \text{for} \ m > 0,  \quad \\
\mathbf{M}\widehat{\mathsf{u}_{\varphi,0}} &=
\widehat{\overline{\mathsf{u}_{\varphi}}},  \quad
\mathbf{M}\widehat{\mathsf{u}_{\varphi, m}}   = \mathbf{L}_{u_\varphi\psi,m} \ \widehat{\mathsf{\psi}_m}   , \ \text{for} \ m > 0. 
\end{aligned}
\end{align}
\end{subequations}
We have omitted the remaining dependency to time for the sake of conciseness. 
Bold capital letters refer to matrices acting upon column vectors of 
the 
kind $\widehat{\mathsf{g}_m}$, that contain the $N_s$ coefficients of the 
expansion of $g_m(s)$ on Chebyshev polynomials as given by 
Eq.~\eqref{eq:chebexp} above,
\begin{equation}
\widehat{\mathsf{g}_m} = \left[ g_{m0}, \dots, g_{m\ N_s-1}  \right] ^T,
\end{equation}
where $^T$ means transposition without conjugation. 
The $\mathbf{M}$ matrix  on the left-hand side 
of Eqs.~\ref{eq:fcmeanflow}--\ref{eq:fcupsi}
is a  $N_s\times N_s$ square matrix that
converts modal values to gridpoint values, since the equalities \ref{eq:fcmeanflow}--\ref{eq:fcupsi} are prescribed
at the $N_s$ collocation points $s_k$, with $k \in 0, \dots, N_s-1$. 
\text{ Its $k$-th row reads } 
\begin{equation}
\gamma \mbox{\dashline} 
C_0(x_k)/2, C_1(x_k),\dots,C_{N_s-2}(x_k), C_{N_s-1}(x_k)/2  
\mbox{\dashline}, 
\end{equation}
where the constant $\gamma= [ 2 / (N_s-1)]^{1/2}$.

The right-hand side of Eqs.~\ref{eq:fcmeanflow}--\ref{eq:fcheat} comprises 
nonlinear and linear terms. 
The nonlinear terms are vectors of size $N_s$ 
denoted by $\mathcal{N}$; 
$
{\mathcal N}_{\overline{u_{\varphi}}}, \ 
{\mathcal N}_{\omega,m}, 
$
and
$
{\mathcal N}_{T,m}
$
respectively 
contain the values of 
$ -\overline{u_s \omega }$, 
  $- \boldsymbol{\nabla}_m \cdot (\mathbf{u}  \omega)_m$
  and 
  $
  - \boldsymbol{\nabla}_m \cdot (\mathbf{u}  T)_m
  $
for each 
    collocation grid point $s_k$. For instance, 
\[ 
{\mathcal N}_{T,m} =  - \left[ \boldsymbol{\nabla}_m \cdot (\mathbf{u}  T)_m(s_0),
 \boldsymbol{\nabla}_m \cdot (\mathbf{u}  T)_m(s_1),
 \dots,
 \boldsymbol{\nabla}_m \cdot (\mathbf{u}  T)_m(s_{N_s-2}),
 \boldsymbol{\nabla}_m \cdot (\mathbf{u}  T)_m(s_{N_s-1})
 \right]^T.
\]
We resort to a pseudo-spectral approach: 
Nonlinear terms are evaluated 
on the physical grid, prior to being transformed back to spectral space. 
For instance, in the previous equation, the product $\mathbf{u}  T$ 
is computed on the $s-\varphi$ grid and transformed back to Fourier space using 
a Fast Fourier transform; 
a discrete cosine transform 
  \citep[][12.4.2]{press2007numerical} 
is next used to transform quantities from the 
 $s-m$ space to  the $n-m$ space.  
The divergence is 
computed using the appropriate three-term
recurrence relation for Chebyshev 
polynomials \citep[e.g.][\S 2.4.2]{chqz2006}. 

Linear terms in the system~\ref{eq:fccompleteset} are cast in terms of matrix-vector products; they
appear  in the algebraic equations 
Eqs.~\ref{eq:fcvortpsi}--\ref{eq:fcupsi}, in addition to the right-hand side
of differential Eqs.~\ref{eq:fcmeanflow}--\ref{eq:fcheat}. The matrices are 
all $N_s\times N_s$, and they possibly involve derivatives of the
Chebyshev polynomials, $C'$, $C''$, etc. For instance, the $k$-th  row of matrix
$\mathbf{L}_{\omega,m}$ reads

\begin{eqnarray*}
&\gamma&  
{\mbox{\scalebox{2}{\dashline}}}
\frac{1}{2}
\left[C''_0(x_k) + \frac{1}{s_k} C'_0(x_k) -\frac{m^2}{s_k^2}C_0(x_k)\right] ,
C''_1(x_k) + \frac{1}{s_k} C'_1(x_k) -\frac{m^2}{s_k^2}C_1(x_k),
\dots 
, \\
&&
C''_{N_s-2}(x_k) + \frac{1}{s_k} C'_{N_s-2}(x_k) -\frac{m^2}{s_k^2}C_{N_s-2}(x_k),
\frac{1}{2}\left[C''_{N_s-1}(x_k) + \frac{1}{s_k} C'_{N_s-1}(x_k) -\frac{m^2}{s_k^2}C_{N_s-1}(x_k)\right]
{\mbox{\scalebox{2}{\dashline}}}, 
\end{eqnarray*}
while the $k$-th row of the $\mathbf{B}_{m}$ buoyancy matrix reads
\[
-\frac{\rayleigh}{\prandtl}\frac{im \gamma}{s_k}
 {\mbox{\scalebox{1}{\dashline}}}  C_0(x_k)/2, C_1(x_k),\dots,C_{N_s-2}(x_k), C_{N_s-1}(x_k)/2  {\mbox{\scalebox{1}{\dashline}}}. 
\]
 Boundary conditions are prescribed through the appropriate modification of 
some of the matrices entering Eq.~\eqref{eq:fccompleteset}. 
We shall get back to this shortly. 

\subsection{Time discretization}

The Chebyshev--Fourier collocation method leads to a semi-discrete set of differential
algebraic equations (DAE) of the generic form
\begin{subequations}
\begin{align}
\frac{\mathrm{d}\statev}{\mathrm{d}t} &= \nonlinop(\statev,\algebraicv) + \linop \statev, \label{eq:ode}
\\
\mathbf{0} &= \constrainop(\statev,\algebraicv), \label{eq:constraint}
\end{align}
\label{eq:dae}
\end{subequations}
where $\nonlinop$ is the nonlinear operator
acting on the differential state vector $\statev=\left[\overline{u_{\varphi}},
\omega, T\right]$
and
the algebraic state vector $\algebraicv=\left[ \psi, \mathbf{u} \right]$, 
$\linop$ is the linear operator 
acting upon $\statev$ and 
$\constrainop$ is the linear operator relating $\statev$ and 
 $\algebraicv$. 
According to 
\cite{ascher1998computer} this defines a semi-explicit DAE of order $1$. 
We will solve it in time using methods developed for ordinary 
differential equations (ODEs), ensuring that constraint (\ref{eq:constraint})
is satisfied through a standard solution technique to be detailed below. 

We resort to implicit--explicit methods that
treat 
$\linop \statev$
and 
$\nonlinop(\statev,\algebraicv)$ implicitly and explicitly, respectively. 
There are different families of IMEX methods, and here we are interested in both IMEX multistep and IMEX multistage methods 
\citep{hairer1993solving,ascher1995implicit,hairer1996solving,ascher1997implicit,kennedy2003additive}. 

\subsubsection{Solution technique}
\label{sec:solution_technique}
Before \rmRone{to} \addRone{we} get to the details of the multistep and multistage
methods, let us describe  the backbone of our solution technique, in particular
how we deal with boundary conditions and the DAE when advancing in time. 
The former are enforced through the implicit solves, while
the latter is taken care of by means of a block-matrix solve. 

To update the field values and advance in time, our approach
is the following: the equations for 
 mean flow~(\ref{eq:fcmeanflow}) and
temperature~(\ref{eq:fcheat})
 are
solved first, as their implicit components are not 
coupled with the vorticity or the streamfunction equations 
(Eq.~\eqref{eq:fcvort}
and Eq.~\eqref{eq:fcvortpsi}). This comes down to inverting a linear system 
of the form 
\[
\left(\alpha \mathbf{M} - \beta \mathbf{L}\right) \mathtt{y} = \text{ r.h.s.}
\]
where the coefficients $\alpha$ and $\beta$ are both \rmRone{real} positive \addRone{real numbers} 
that
depend on the time integrator, $\mathtt{y}$
stands for the vector containing the $N_s$ coefficients
of the Chebyshev expansion of mean flow or temperature, 
and 
the right-hand side $\text{r.h.s}$ contains a mix of linear and nonlinear components. 
The first and last rows of the 
$N_s\times N_s$ matrix to invert, $\alpha \mathbf{M} - \beta \mathbf{L}$, are modified
in order to enforce the two boundary conditions that apply either on 
temperature or on the mean flow at $s=s_i$ and $s=s_o$, respectively
\citep[see e.g.][Table~4]{julien2009efficient}. The corresponding entries
 of the $\text{r.h.s}$ vector are modified accordingly and set to $0$ or $1$, depending
 on the boundary condition. 

Next, we update the vorticity and the non-zonal 
streamfunction,  Eq.~\eqref{eq:fcvort}
and Eq.~\eqref{eq:fcvortpsi}, 
 for each Fourier mode~$m>0$. 
     We do so by inverting the following $2N_s\times2N_s$
     block-matrix,
     \[
\left[
\begin{array}{cc} 
 \alpha \mathbf{M} -\beta \mathbf{L}_{\omega,m}  &  0   \\
 \mathbf{M}  & -\mathbf{L}_{\omega\psi,m}
\end{array}
\right]
\left[ 
\begin{array}{c}
\mathsf{y}_{\omega,m}        \\
\mathsf{y}_{\psi,m}
\end{array}
\right]
= 
\left[
\begin{array}{c}
 \text{r.h.s.}_{\omega}      \\
 0
\end{array}
\right]
     \]
The first
half of the vector $[\mathsf{y}_{\omega,m},\mathsf{y}_{\psi,m}]^T$, 
$\mathsf{y}_{\omega,m}$, is the updated vorticity, while its second
half is the updated streamfunction, $\mathsf{y}_{\psi,m}$. 
Each quadrant in the above equation has size $N_s\times N_s$. The top-left
quadrant originates from the time discretization of the vorticity
equation. 
 The right-hand side $\text{r.h.s}_\omega$ contains 
a mix of linear and nonlinear components, including the contribution
of the updated temperature field via the buoyancy term. 
The constraint (\ref{eq:fcvortpsi}) is enforced by means of the 
second row of the block-matrix system. The boundary conditions
are enforced for the streamfunction, by modifying the first, $N_s$-th, 
$N_s+1$-th and last row of the block matrix 
\citep[][\S 10.3.2]{glatzmaier2013introduction}. The first $N_s$ entries of these rows
are set to $0$, while the remaining $N_s$ entries are modified in order
enforce the homogeneous Dirichlet and Neumann conditions for the 
streamfunction, see e.g.  Table~4 in \cite{julien2009efficient}. Accordingly, 
the first, $N_s$-th,
$N_s+1$-th and last entry of the right-hand side vector  $[ 
\text{r.h.s.}_{\omega},0
]^T$ are set to 0.  
See also \cite[][his Fig.~1a,]{gastine2019pizza} for a graphical illustration 
of this implementation. 

Finally,  the updated mean flow and non-zonal streamfunction (which 
satisfy their respective 
boundary conditions, given by Eqs.~(\ref{eq:bcpsi}-\ref{eq:bcmeanflow}), 
  allow us to 
  evaluate the velocity components
  (Eq.~\eqref{eq:fcupsi}), thereby permitting 
 the evaluation of the nonlinear terms $\nonlinop(\statev)$ necessary for the
next update. 

We shall 
now describe the multistep and multistage time discretization methods.  
\subsubsection{IMEX multistep methods}
Multistep methods rely on a polynomial interpolation in time. 
Let  $K$ be the number of steps of an IMEX  multistep method, with $K \geq 1$. 
Let $\Delta t$ denote the timestep size and $\statev_i$ 
denote the approximate solution for the differential 
state vector at time $t_i=i \Delta t$, leaving aside the 
algebraic $\algebraicv$, which is updated alongside 
 $\statev$ following the solution technique detailed above. 
Then, for fixed $\Delta t$, following \cite{ascher1995implicit}, 
a general linear multistep IMEX method applied to Eq.~\eqref{eq:ode} can be 
written as
\begin{equation}
\label{imexgen}
\left(1 - \Delta tc_{-1} \linop \right) \statev_{i+1}
  = 
\sum^{K-1}_{j=0}\left[
a_j \statev_{i-j} + 
\Delta t b_j \nonlinop(\statev_{i-j}) 
+ \Delta t c_j \linop\statev_{i-j}\right],
\end{equation}
where $c_{-1} \neq 0$. It is noteworthy that a $K$-step IMEX method cannot have 
order of accuracy greater than $K$ \citep{ascher1995implicit}. 
The IMEX multistep methods we use for this study
have the same order as the number of steps $K$. 
%

In this work we shall consider multistep methods of order 2, 3 and 4: the 
popular Crank-Nicolson  Adams-Bashforth method of order 2 (CNAB2) 
already seen in the introduction,  and
the semi-implicit BDF (SBDF) schemes given e.g. in 
\cite{ascher1995implicit} of
order 2, 3 and 4. The three SBDF schemes apply a backward differentiation 
formula to the implicit part, and an extrapolation formula to the explicit 
part. 

In our 
convergence analysis, $\Delta t$ will remain fixed. We shall 
activate variable time-step 
for the equilibrated regime 
of the most turbulent of our reference cases to be reached 
(section~\ref{sec:pres_cases}), and 
for the stability analysis of section~\ref{sec:stab_analysis}. 
    The vectors of coefficients $\mathbf{a},$ $\mathbf{b}$ and $\mathbf{c}$
    for the four selected schemes are given in \ref{app:multistep}.

\subsubsection{IMEX multistage methods}
IMEX multistage methods rely on quadrature rules to evaluate
intermediate stages of $\statev$ (and $\algebraicv$) between
discrete times $t_i$ and $t_{i+1}$. The multistage methods of interest for this
work are often referred to as 
IMEX-RK (for IMEX Runge--Kutta) methods, which indicates that 
they involve a diagonally implicit Runge-Kutta (DIRK) and an explicit Runge-Kutta
(ERK) schemes  \citep{ascher1997implicit}. 
 Unlike multistep methods, their stability region can increase slightly 
 with their order. 

 Let $K$ denote  the number of internal 
 stages of an IMEX-RK method. 
  At each substage $k \in \{1,\dots,K\}$ of an IMEX-RK scheme applied to 
Eq.~\eqref{eq:ode}, one has 
  \begin{equation}
(1 - \Delta t a_{kk}^I \linop) \mathbf{y}_k = \statev_i + \Delta t \sum_{j=1}^{k-1} a^E_{kj} \mathcal{N}(\mathbf{y}_j) 
 + \Delta t \sum_{j=1}^{k-1} a^I_{kj} \mathcal{L}\mathbf{y}_j,
 \label{eq:substage}
 \end{equation}
 where the coefficients $a^I_{kj}$ and $a^E_{kj}$ define two matrices for the 
 DIRK and ERK schemes, 
 $\mathbf{A}^I$ and $\mathbf{A}^E$, respectively. 
 The first stage is defined by $\mathbf{y}_1=\statev_i$.  
 At each stage,  the algebraic variables $\algebraicv$ are updated alongside
 the differential variables following the strategy detailed in section~\ref{sec:solution_technique}. 

 \rmRtwo{For some, but not all, schemes, termed ``stiffly accurate", $\statev_{i+1}=\mathbf{y}_K$,
         meaning that the solution of the last stage corresponds to the updated state vector.}
 \rmRtwo{For the others, the updated differential state vector is assembled as}
 \[
\rmRtwo{\mbox{$\statev_{i+1} = \statev_{i} + \Delta t \sum_{j=1}^{K} b_j^E \mathcal{N}(\mathbf{y}_j)
                            + \Delta t \sum_{j=1}^{K} b_j^I \mathcal{L}\mathbf{y}_j.$}}
							\]
The DIRK and ERK components can be independently summarized using Butcher 
tableaux 
\citep[][\S~II.1]{hairer1993solving}
\begin{equation}
\begin{array}{c|cccc}
0 & 0  &  & & \\
c_2^E &  a^E_{21} & 0  &  & \\
\vdots & \vdots  & \ddots   & \ddots   & \\
c_K^E &  a^E_{K1} & \cdots &    a^E_{KK-1}& 0 \\ \hline
      &  b^E_{1} & \cdots  &  \cdots& b^E_{K}
\end{array},
\quad \quad 
\begin{array}{c|cccc}
0 &  0 &  & & \\
c_2^I &  a^I_{21} & a^I_{22}  &  & \\
\vdots & \vdots  & \ddots   & \ddots   & \\
c_K^I &  a^I_{K1} & \cdots &    a^I_{KK-1}& a^I_{KK} \\ \hline
      &  b^I_{1} & \cdots  &  \cdots& b^I_{K} 
\end{array}.
\end{equation}
The design of an IMEX-RK method 
implies a set of constraints on the coefficients 
$\mathbf{A}^I$,
 $\mathbf{b}^I$, 
 $\mathbf{c}^I$, 
$\mathbf{A}^E$,
 $\mathbf{b}^E$, 
 $\mathbf{c}^E$, 
  the number of which depends on the sought order of accuracy.  
Extra constraints 
  can be added \rmRone{i}
  to accommodate 
  discrete algebraic equations \citep{boscarino2009class}. 
  \addRtwo{According to the classification 
  by \cite{boscarino2007error}, the schemes considered in this study
  belong to the CK type, as the first row of the implicit matrix 
  $\mathbf{A}^I$ contains zeros. There exists schemes that have non-zero entries
  in this row, which 
  belong to the so-called type A \citep{boscarino2007error}, the first examples of which 
  were introduced by \cite{pareschi2005implicit}; we do not investigate such schemes
  in this study.}

    \addRtwo{In addition, a}ll the schemes considered in this work
    have $\mathbf{c}^E=\mathbf{c}^I$, which is why $c_1^E=c_1^I=a_{11}^I=0$ above. 
  Also, note that in practice,  $\mathbf{c}^E$ and $\mathbf{c}^I$ are not required
for thermal convection which has no explicit time-dependent forcing and
    is therefore an autonomous process. 

 In spite
of what Eq.~\eqref{eq:substage} suggests, 
 the IMEX-RK methods we analyze in this study do not necessarily 
comprise the same number of implicit and explicit stages, $K^I$ and $K^E$. 
  This is handled by setting the appropriate columns 
of $\mathbf{A}^I,$  and entries of $\mathbf{b}^I$ (or of $\mathbf{A}^E$ 
and $\mathbf{b}^E $) to zero. The number of linear inversions per time step, $
n^I$, is equal to $K^I$ provided $a_{k1}^I=0, \forall k \in {1,\dots,K^I}$.   

\addRtwo{DIRK schemes are called ``stiffly accurate" if
\begin{equation}
{\mathbf{b}^I}^T = \mathbf{e}_K^T \mathbf{A}^I,
\end{equation}
where $\mathbf{e}_K=[0,\dots,0,1]^T$. Following \cite{boscarino2013implicit}, we say that a IMEX-RK scheme
is globally stiffly accurate (GSA) if ${\mathbf{b}^I}^T = \mathbf{e}_K^T \mathbf{A}^I$ and
${\mathbf{b}^E}^T = \mathbf{e}_K^T \mathbf{A}^E$, and $c_K^I=c_K^E$, which implies that
the updated state vector is identical to the last internal stage value of the scheme. 
 For non-GSA schemes, the updated differential state vector is assembled as
\begin{equation}
\statev_{i+1} = \statev_{i} + \Delta t \sum_{j=1}^{K} b_j^E \mathcal{N}(\mathbf{y}_j)
                            + \Delta t \sum_{j=1}^{K} b_j^I \mathcal{L}\mathbf{y}_j. 
\label{eq:assembly}
\end{equation}}
\rmRtwo{For stiffly accurate schemes, 
the last row of $\mathbf{A}^E$ is equal to $\mathbf{b}^E$ 
and 
the last row of $\mathbf{A}^I$ is equal to $\mathbf{b}^I$. 
Some schemes 
have their DIRK component stiffly accurate while their ERK component is not.}
 If the chosen scheme requires \addRone{such} an assembly  \rmRtwo{(Eq.~\ref{eq:assembly})} then
 further work is needed to ensure that the updated vorticity meets
 condition~\addRone{\eqref{eq:bcvort}} \addRone{, see section~\ref{sec:assembly} below}. 

We follow e.g. \cite{boscarino2013implicit} and identify each IMEX-RK scheme 
by the initials of the authors (if they are no more than~3),
and three numbers ($K^I$, $K^E$, $r$) 
denoting, respectively, the number of implicit and explicit stages, 
and the theoretical order of
accuracy. Exceptions to this rule are DBM553 from \cite{vogl2019evaluation} 
and BHR553 from \cite{boscarino2009class} where we kept the initials of the 
original name; and PC432 which is a second order three stage 
predictor/corrector scheme constructed using the explicit scheme by 
\cite{jameson1981numerical} for its explicit component and a Crank-Nicolson 
for its implicit counterpart (see \ref{app:pc2} for details of its 
Butcher tableaux). 

In this work we investigate the properties of 22 IMEX-RK schemes, whose
properties are summarized in Table~\ref{tab:imexrk}.

\begin{table}[t!]
\caption{Multistage 
	IMEX-RK methods used in this study. The leftmost ``Scheme" column defines the scheme notation, as used in the main text, tables and figures.
	$K^I$ is the number of stages of the diagonally implicit Runge--Kutta (DIRK) component. 
	$K^E$ is the number of stages of the explicit Runge--Kutta (ERK) component. 
	The next column contains the expected order of accuracy $o$ of the combined IMEX-RK scheme. 
	$n^I$ is the number of linear solves for each time-step, which differs 
	from $K^I$ if $a_{k1}^I \neq 0\ \forall k$. 
	 The star indicates that the scheme involves several matrices 
because of the changes on the diagonal of the implicit Butcher table (non 
S-DIRK schemes). 
	S. A. indicates whether the ERK or DIRK part of the method is stiffly accurate, and $\mathbf{b}^I=\mathbf{b}^E$
	indicates if the DIRK and ERK methods have the same solution weights to 
compute the assembly. Storage denotes the number of state vectors that need 
to be stored simultaneously for the time advance of one physical quantity.
	The last column provides the relevant 
    reference augmented with a section or paragraph number, and possibly the name
	of the scheme as it appears in the reference.}
\label{tab:imexrk}
\resizebox{\textwidth}{!}{\begin{tabular}{lcccl|ccccl}
  \toprule
  \multirow{2}{*}{\raisebox{-\heavyrulewidth}{Scheme}} 
  & \multirow{2}{*}{\raisebox{-\heavyrulewidth}{$K^I$}} 
  & \multirow{2}{*}{\raisebox{-\heavyrulewidth}{$K^E$}} 
  & \multirow{2}{*}{\raisebox{-\heavyrulewidth}{$o$ }} 
  & \multirow{2}{*}{\raisebox{-\heavyrulewidth}{$n^I$}} 
  & \multicolumn{1}{c}{S. A.} 
  & \multicolumn{1}{c}{S. A.} & \multirow{2}{*}{\raisebox{-\heavyrulewidth}{$\mathbf{b}^I=\mathbf{b}^E$}} 
  & \multirow{2}{*}{\raisebox{-\heavyrulewidth}{storage}} 
  & \multirow{2}{*}{\raisebox{-\heavyrulewidth}{Reference}}\\
  & & &  &  & DIRK & ERK & \\
  \midrule
  ARS222 & 2 & 2 & 2 & 2    & \checkmark & 
\checkmark & X & 4 & \cite{ascher1997implicit}, \S2.6\\
  ARS232 & 2 & 3 & 2 & 2    & \checkmark & X & 
\checkmark  & 6 & \cite{ascher1997implicit}, \S2.5 \\
  BPR442& 4 & 4 & 2 & 4   & 
\checkmark & \checkmark & X  & 8 & \cite{boscarino2017unified}, Eq.~(76) \\
  PC432    & 4 & 3 & 2 &  3  & \checkmark & \checkmark & X &  7
&\cite{jameson1981numerical}, Eq.~(4.18);  Schaeffer (priv. comm.) \\
    SMR432 & 4 & 3 & 2 & 3$^\star$  &  \checkmark 
& \checkmark & X & 7 &\cite{spalart1991spectral}, App.~A \\
  \hline
  ARS233 & 2 & 3 & 3 & 2          & X          & X & 
\checkmark  & 6 &\cite{ascher1997implicit}, \S2.4 \\
  ARS343 & 3 & 4 & 3 & 3   & \checkmark & X & 
\checkmark  & 8 &\cite{ascher1997implicit}, \S2.7\\
ARS443 & 4 & 4 & 3 & 4   & \checkmark & 
\checkmark & X  & 8 &\cite{ascher1997implicit}, \S2.8\\
BHR553 & 5 & 5 & 3 & 4   & \checkmark & X & \checkmark 
 & 11 & \cite{boscarino2009class}, App.~1, BHR(5,5,3) \\
  BPR533 & 5 & 3 & 3 & 4  & \checkmark & 
\checkmark & X  & 8 & \cite{boscarino2013implicit}, \S8.3, BPR(3,5,3) \\
  BR343 & 3 & 4 & 3 & 3  & \checkmark & X & \checkmark 
 & 8
& \cite{boscarino2007uniform}, \S3, MARS(3,4,3) \\
  CB443 & 4 & 4 & 3 & 3$^\star$   & \checkmark & X 
& \checkmark  & 9 & \cite{cavaglieri2015low}, \S4, IMEXRKCB3f \\
  CFN343 & 3 & 4 & 3 & 3   & \checkmark & X & 
\checkmark  & 8 & \cite{calvo2001linearly}, Eq.~(8) and (10) \\
 DBM553 & 5 & 5 & 3 & 4   & \checkmark & X & 
\checkmark  & 11
& \cite{vogl2019evaluation}, App.~A, DBM453; \cite{kinnmark1984one}  \\
KC443 & 4 & 4 & 3 & 3   & \checkmark & X & 
\checkmark  & 9 &\cite{kennedy2003additive}, App.~C, ARK3(2)4L[2]SA \\
  LZ543 & 5 & 4 & 3 & 4$^\star$   & \checkmark & 
\checkmark & X  & 9 &\cite{liu2006additive}, \S6, RK.3.L.1 \\
  \hline
CB664 & 6 & 6 & 4 & 5$^\star$   & \checkmark & X & 
\checkmark 
 & 13 &\cite{cavaglieri2015low}, \S5,  IMEXRKCB4 \\
  CFN564 & 5 & 6 & 4 & 5   & \checkmark & X & 
\checkmark  & 12 &\cite{calvo2001linearly}, Eq.~(14);   
\cite{hairer1996solving}, Eq.~(6.16)
\\
  KC664 & 6 & 6 & 4 & 5    & \checkmark & X & 
\checkmark  & 13 &\cite{kennedy2003additive}, App.~C, ARK4(3)6L[2]SA \\
KC774 & 7 & 7 & 4 & 6    & \checkmark & X & 
\checkmark  & 15 &\cite{kennedy2019higher}, App.~A,  ARK4(3)7L[2]SA$_1$ \\
  LZ764 & 7 & 6 & 4 & 6$^\star$   & \checkmark & 
\checkmark & \checkmark  & 13 &\cite{liu2006additive}, \S6, RK.4.A.1 \\
\hline
KC885 & 8 & 8 & 5 & 7    & \checkmark & X & 
\checkmark  & 17 &\cite{kennedy2019higher}, App.~A, ARK5(4)8L[2]SA$_2$ \\
  \bottomrule
\end{tabular}} 
\end{table}

\subsubsection{Treatment of IMEX-RK methods with an assembly stage}
\label{sec:assembly}
IMEX-RK methods which are not stiffly accurate require the assembly stage 
Eq.~\eqref{eq:assembly} to be performed.
The assembly of the temperature and mean flow is straightforward, as it is a linear combination of
nonlinear and linear terms evaluated at the substages. Applying the same linear combination for 
the vorticity, however, does not guarantee that the boundary conditions \eqref{eq:bcvort}
are properly enforced. Indeed, the assembly stage 
does not lend itself to the solution method outlined 
in section \ref{sec:solution_technique} for the vorticity and streamfunction, which allows one to 
bypass the explicit enforcement of the boundary conditions on vorticity. 

 To make sure that the vorticity built at the assembly stage is consistent 
 with the boundary conditions, we follow the strategy outlined by \cite{johnston2009comparison}. 

We begin by assembling a first guess of the final vorticity 
$\mathsf{y}^\star_{\omega,m}$ for each mode $m>0$ by means
of Eq.~\eqref{eq:assembly}. Using that intermediate value, we compute 
$\mathsf{y}_{\psi,m}$ by 
inverting
\begin{equation}
\mathbf{L}_{\omega\psi,m} \mathsf{y}_{\psi,m}  = \mathbf{M} \mathsf{y}^\star_{\omega,m}, 
\end{equation}
having modified the first and last lines of $\mathbf{L}_{\omega\psi,m}$  so that $\psi_m =0$ at $s=s_i$ and $s=s_o$. 

The knowledge of $\mathsf{y}_{\psi,m}$ makes it possible to construct a local interpolant in the 
vicinity of the two walls, $L_w$, that is constrained by the values of $\psi_m$ at the first $J+1$, say,  
Chebyshev-Gauss-Lobatto points, $\psi_m^j$ (that include the point on the wall) 
and the extra requirement that
$\partial \psi_m /\partial s =0$ on the wall as well. For the inner wall, $s=s_i$, 
the interpolant reads
\begin{equation}
L_w(s) = \sum_{j=0}^J \psi_m^j \ell_j(s) - (s-s_i) \ell_0(s) \left[ \sum_{j=0}^J \psi_m^j \ell_j'(s_i)\right],
\end{equation}
with a similar expression for the outer wall. In this expression, $\ell_j$ is the Lagrange polynomial attached
to the $j$-th point away from the wall, and $\ell_j'$ is its first derivative. 
This local interpolant allows us to compute the vorticity on the inner and outer walls, 
\begin{eqnarray}
\omega_m(s_i) &=& \left. -\frac{\mathrm{d}^2 L_w}{\mathrm{d}s^2} \right|_{s=s_i} = -\sum_{j=0}^J \psi_m^j \left[\ell_j''(s_i) 
                - 2 \ell_j'(s_i) \ell_0'(s_i)\right], \\
\omega_m(s_o) &=& \left. -\frac{\mathrm{d}^2 L_w}{\mathrm{d}s^2} \right|_{s=s_o} = -\sum_{j=0}^J \psi_m^j \left[\ell_j''(s_o) 
                - 2 \ell_j'(s_o) \ell_0'(s_o)\right],
\end{eqnarray}
where $\ell''_j$ denotes the second derivative of $\ell_j$. 
We form a vector of nodal values of vorticity whose interior values are based on $\mathsf{y}^\star_{\omega,m}$
and whose boundary values are the ones we just computed based on the local interpolant $L_w$. 
We finally determine $\mathsf{y}_{\omega,m}$ by applying the inverse of $\mathbf{M}$ to this vector of nodal 
values. In our experiments, we set the value of $J$ to $14$. 
\subsubsection{Fully explicit RK methods}
\label{sec:rk}
In addition to the IMEX multistep and IMEX-RK multistage techniques detailed above, we found
it useful sometimes to consider two well-known fully explicit methods, the explicit Runge-Kutta methods
of order 2 and 4, RK2 and RK4 \citep[e.g.][Eqs.~D.2.15 and D.2.17]{chqz2006}. 
The solution technique that these
methods imply is based on the technique laid out for IMEX-RK methods 
(no linear solve, except at the assembly stage). 

\subsection{Implementation and validation}

The code for solving the problem using the aforementioned pseudospectral methods and time-stepping strategies was written 
from scratch in the Fortran programming language. The code contains several modules and subroutines where each module has specific dependencies.
The fast Fourier and discrete cosine transforms resort to the FFTW3 library \citep{fftw3}. 
The matrix equations are solved using standard matrix solvers available in the LAPACK routines 
\texttt{dgetrf} and \texttt{dgetrs} \citep{anderson1999lapack}. The \texttt{dgetrf} 
routine is used for computing the LU factorization and the \texttt{dgetrs} 
routine is used for solving the system using the factored matrix 
obtained by using the \texttt{dgetrf} routine.

To benchmark the code against peer-reviewed results, we compare it 
with a reference solution obtained by \cite{alonso2000transition}.
They performed their numerical simulations using spectral methods with a fixed radius ratio $s_i/s_o = 0.3$ 
and Prandtl number $\prandtl = 0.025$ (which corresponds to liquid Mercury Hg), and 
the second-order stiffly stable time integrator by \cite{karniadakis1991high}. 
 In table \ref{tab:nu_bench} we list the dependency of the equilibrated 
Nusselt number $\nusselt=\nusselt_o=\nusselt_i$ 
 to the Rayleigh number reported by \cite{alonso2000transition} and obtained here using
 the ARS443 IMEX-RK time integrator, together with $N_s=32$ and $N_m=192$. 
 Furthermore, for the range of Rayleigh numbers shown in table \ref{tab:nu_bench}, 
 we observe an oscillation of the solution about the periodic azimuthal direction, which is 
 a characteristic of low Prandtl number fluids. For $\rayleigh = 6500$, 
 the frequency of oscillation we find is $f = 5.15$ 
 which exactly matches value published by \cite{alonso2000transition}. 
 Thus we ascertain that the code was benchmarked and ready to be used for the 
 study of various time integration methods. 

\begin{table}
\begin{minipage}{0.4\linewidth}
\begin{center}
\begin{tabular}{lcc}
\hline
$\rayleigh$ & $\nusselt-1$ & $\nusselt-1$ \\
& (ref.) & (this study) \\
\hline
1892 & 0.005 & 0.005  \\ 
2510 & 0.163 & 0.162  \\ 
3268 & 0.383 & 0.383  \\ 
4013 & 0.544 & 0.544  \\ 
4106 & 0.562 & 0.561  \\ 
4500 & 0.617 & 0.618  \\ 
5000 & 0.679 & 0.678  \\ 
5500 & 0.733 & 0.733  \\ 
6000 & 0.783 & 0.783  \\ 
6500 & 0.827 & 0.827  \\ 
7000 & 0.871 & 0.869  \\ \hline
\end{tabular}
\end{center}
\end{minipage}
\hfill
\begin{minipage}{0.57\linewidth}
\caption{Nusselt number $\nusselt$ obtained for $\prandtl=0.025$, $s_i/s_o=0.3$ 
         and an increasing Rayleigh number $\rayleigh$, by \cite{alonso2000transition} (the reference) and with the code 
		 developed for this study. 
		 }
\label{tab:nu_bench}
\end{minipage}
\end{table}

\section{Results}
\label{sec:results}
We begin by a presentation of the 11 cases studied in this work, followed by the analysis
of the convergence properties of the time schemes we investigated. We investigate
the likely causes of the order reduction observed for some configurations, and finally
weigh these findings against a more practical estimate of the computational efficiency. 
\subsection{Cases studied}
\label{sec:pres_cases}
All cases considered  have a radius ratio $s_i/s_o$ set to 0.35. 
They are initialized with a temperature perturbation of localized 
compact support as introduced by \cite{gaspari1999construction}, 
of width $0.1/\sqrt{2}$ and amplitude $10^{-4}$. 
The properties of the cases are summarized in table~\ref{tab:summ_cases}. 
 This table comprises the input control parameters
$\prandtl$ and $\rayleigh$, the Reynolds
number $\reynolds$ (Eq.~\eqref{eq:reynolds}), Nusselt number at the outer boundary (Eq.~\eqref{eq:nusselt})
  and the temporal averages of the buoyancy input power (Eq.~\eqref{eq:buopower}), 
  and heat loss by viscous dissipation (Eq.~\eqref{eq:visdis2}). 
  In addition, we provide
 in this table the spatial
discretization parameters $N_s$ and $N_m$ introduced in the previous section. Unless otherwise stated, 
 a given 
case was always run for the same $(N_s,N_m)$ pair. That pair was chosen to make spatial discretization
error negligible against temporal discretization error.  
We chose to run 3 cases with a Prandtl number equal to $0.025$, which corresponds to liquid metals,  
7 cases with $\prandtl$ equal to 1, which corresponds to a commonly 
taken value in numerical simulations,  and 
one case with $\prandtl=40$, to have at least one situation in the large $\prandtl$ limit. The numbering
 in table~\ref{tab:summ_cases}
was adopted to follow the increase of the Reynolds number.  
Case 0 is extremely laminar, while case 10 is our most turbulent case with $\reynolds > 10^4$. 
Our goal is to  
exercise the time schemes
over a broad range of regimes. 

\begin{table}
\caption{Properties of the 11 convection cases investigated in this study. From left to right: 
Case number, Prandtl number (input), Rayleigh number (input), 
Reynolds number (output), 
 Nusselt number at the outer boundary (output), time average buoyancy input power (output), time average heat loss by
 viscous dissipation (output), and spatial resolution used.}
\label{tab:summ_cases}
\begin{center}
\begin{tabular}{cllrrccr}
\hline
Case & $\prandtl$ & $\rayleigh$ & $ \reynolds $  
& $ \nusselt_o $  & $\langle \power \rangle$ & $\langle D_\nu \rangle$    & $(N_s,N_m)$ \\
\hline
0 & 1 & 2 $\times 10^3$ &  $2.87$ & $1.16$ & $2.03\times 10^3$ & $-2.03\times 10^3$ & $(36,36)$  \\ 
1 & 1&  1 $\times 10^4$ & $18.85$  &$2.51$ & $9.29\times 10^4$ &  $-9.29\times 10^4$ & $(48,48)$\\ 
2 & 40& 1 $\times 10^7$ & $26.00$ & $12.63$  & $4.39\times 10^5 $ & $-4.39 \times 10^5 $ & $(256,256)$\\ 
3 & 1& 1 $\times 10^5$  &  $77.33$ & $4.64$     & $2.22\times 10^6$ & $-2.22\times 10^6$ & $(64,64)$\\
4 & 1& 1 $\times 10^6$ & $279.76$  & $7.70$  & $4.06\times 10^7 $ & $-4.06\times 10^7$ & $(96,128)$ \\ 
5 & 0.025& 1 $\times 10^4$ & $513.44 $ & $2.07$  & $1.07\times 10^8 $ &$-1.07\times 10^8$ &$(64,192)$ \\ 
6 & 1& 1 $\times 10^7$ & $943.12$ & $13.17$ & $7.33\times 10^8 $ & $-7.33\times 10^8$& $(128,160)$ \\ 
7 & 0.025& 1 $\times 10^5$ &$2023.52$ &$3.97$ & $2.89\times 10^9 $  & $-2.89\times 10^9 $ & $(128,320)$\\ 
8 & 1& 1 $\times 10^8$ & $3462.47$& $23.30$  & $1.34\times 10^{10}$ & $-1.34\times 10^{10}$ &$(256,256)$\\ 
9  & 0.025& 1 $\times 10^6$ &$6835.69$ & $6.56$  & $5.36\times 10^{10}$ & $-5.37\times 10^{10}$  & $(160,384)$\\ 
10  & 1 & 1 $\times 10^9$ & $13320.12$  & $ 44.22$ & $2.60\times 10^{11} $ & $-2.60\times 10^{11}$ & $(384,384)$ \\
\hline
\end{tabular}
\end{center}
\end{table}

\begin{figure}
\begin{minipage}{0.6\linewidth}
\centerline{\includegraphics[width=.9\linewidth]{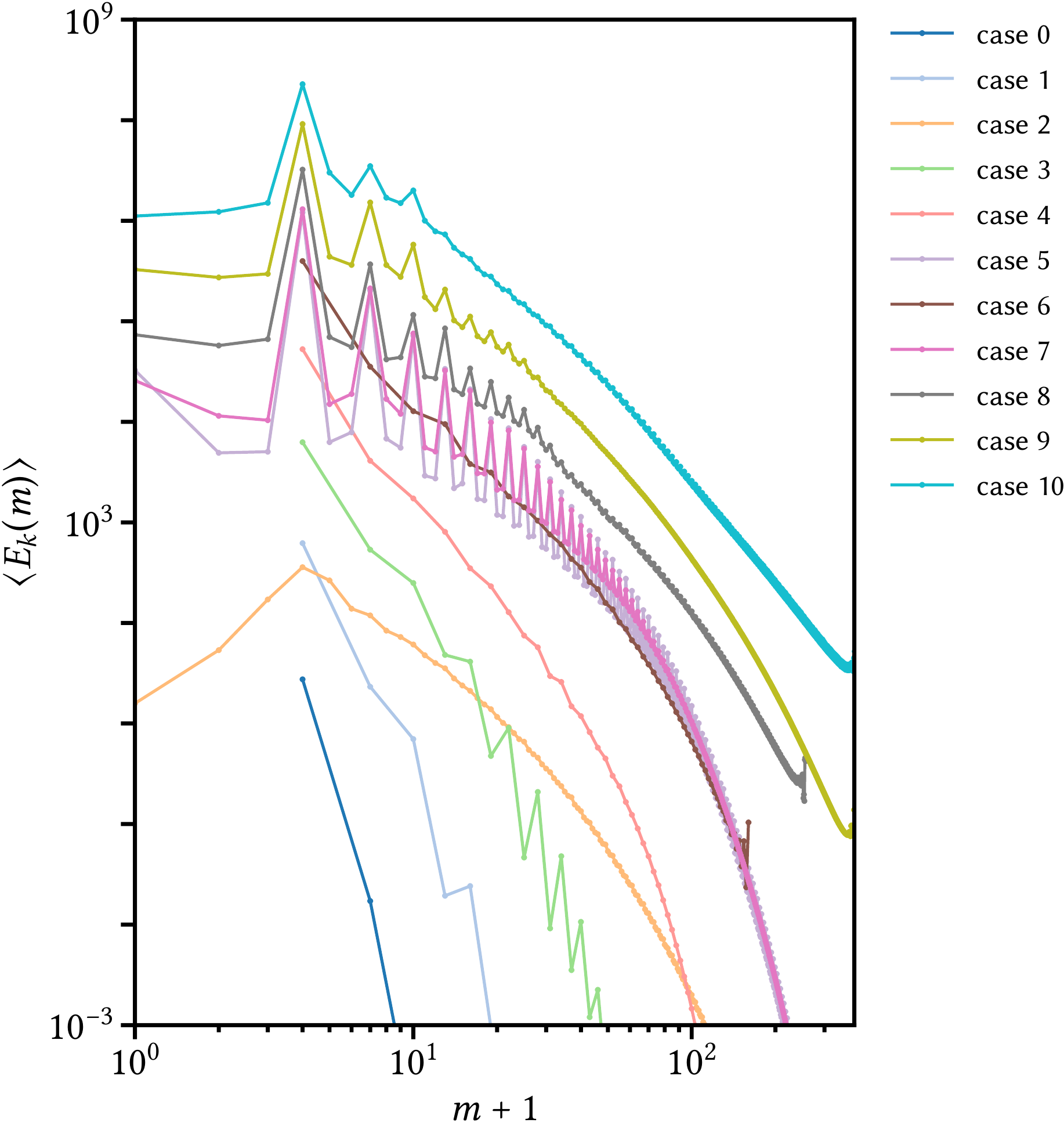}}
\end{minipage}
\hfill
\begin{minipage}{0.39\linewidth}
\caption{Time averaged kinetic energy versus azimuthal wavenumber $m$ for the 11 configurations
considered in this study. Every third mode shown for clarity ($m=0,3,6,\dots$) for cases 
0, 1, 3, 4 and 6.  The scale on both axes
is logarithmic. Note that cases 0, 1 and 3 have zero energy in the axisymmetric $m=0$ mode.}
\label{fig:spectra}
\end{minipage}
\begin{minipage}{.48\linewidth}
(a) Case~2, vorticity $\omega$\\
\centerline{\includegraphics[width=\linewidth]{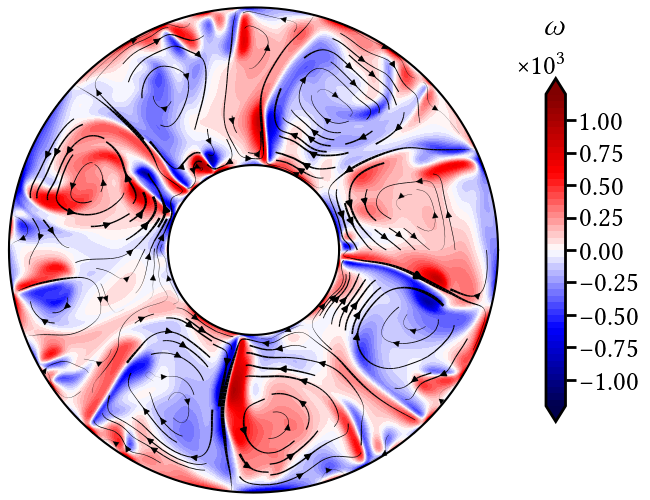}}
\end{minipage}
\hfill
\begin{minipage}{.48\linewidth}
(b) Case~2, temperature $T$\\
\centerline{\includegraphics[width=\linewidth]{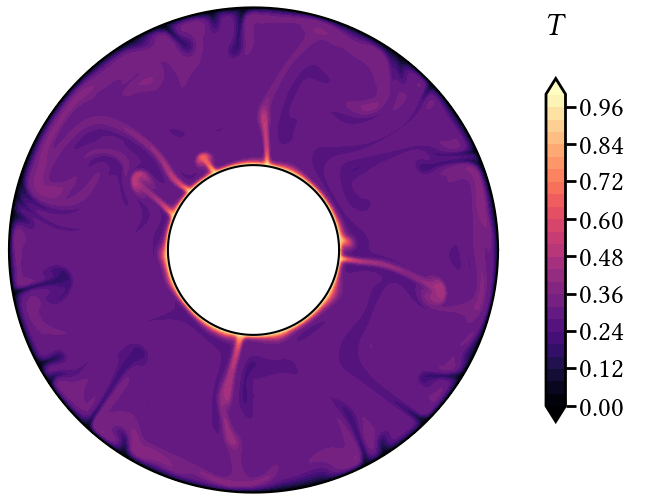}}
\end{minipage}
\begin{minipage}{.48\linewidth}
(c) Case~10, vorticity $\omega$\\
\centerline{\includegraphics[width=\linewidth]{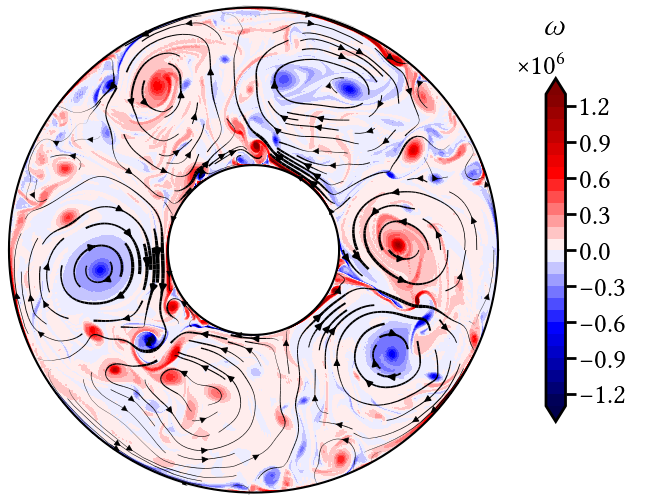}}
\end{minipage}
\hfill
\begin{minipage}{.48\linewidth}
(d) Case~10, temperature $T$\\
\centerline{\includegraphics[width=\linewidth]{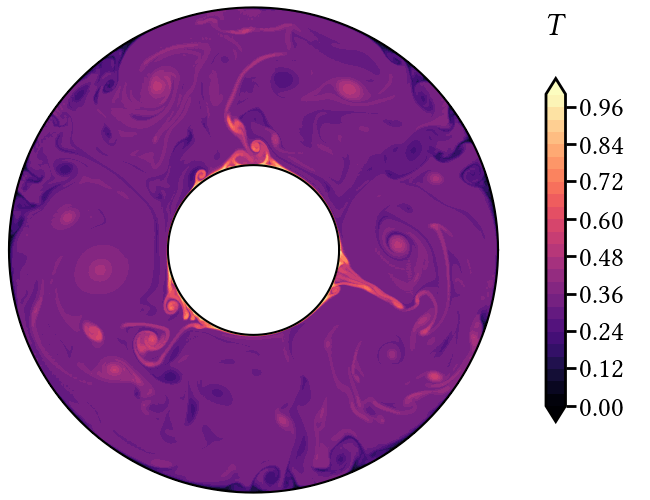}}
\end{minipage}
\caption{Solution snapshots. 
a: vorticity field, with superimposed velocity streamlines, for
case~2 ($\rayleigh=10^{7}$, $\prandtl=40$). b: temperature field, for case~2 and at the same
discrete time. 
c: vorticity field, with superimposed velocity streamlines, for
case~10 ($\rayleigh=10^{9}$, $\prandtl=1$). d: temperature field, for case~10 and at the same
discrete time. 
}
\label{fig:cont}
\end{figure}

For this radius ratio, and regardless of the value of the Prandtl number considered
 ($0.025, 1, 40$), the most unstable convective 
mode has a threefold symmetry in the periodic azimuthal direction, that is to say
that the value of the critical wavenumber $m_{\mbox{\footnotesize crit}}=3$. 
The corresponding value of the critical Rayleigh number is 
$\rayleigh_{\mbox{\footnotesize crit}}=1768$. 
In the range of forcing that we
cover, the threefold symmetry is a persisting feature. 
When increasing the level of turbulence, 
the energy found in other wavenumbers increases, by virtue of the larger importance 
taken by turbulent transport of momentum and heat. This is shown in Figure~\ref{fig:spectra}, 
which displays the time averaged kinetic energy spectra of the 11 cases. 
\addRone{ The kinetic energy in each Fourier mode $m$ is given by 
\begin{eqnarray*}
 E_k(m=0)  & = & \pi \int_{s_i}^{s_o} \overline{u_{\varphi}}^2 s \mathrm{d}s,  \\
 E_k(m>0)  & = & 2 \pi \int_{s_i}^{s_o} \left( |u_{sm}|^2 + |u_{\varphi m}|^2 \right) s \mathrm{d}s. 
\end{eqnarray*}
	}
Note that
for case 10 (the most turbulent case) the azimuthal 
truncation (the value of $N_m$) chosen enables a $10^6$ factor
to be achieved 
between the highest energy level (for $m=3$)  and the lowest energy level (around $m=N_m$). 
We use a similar criterion to set the truncation in radius, $N_s$. 
 \addRone{How stiff are these cases
 numerically? We shall see below
  that stiffness, as measured by the disparity between
 linear and nonlinear time scales, remains moderate
  across the region of parameter space explored by the cases,
    with a stiffness parameter that varies
  between $10^{-3}$ and $10^{-5}$ (see
  section~\ref{sec:stiffness} and Table~\ref{tab:stif}
  below for more details).}

We now consider in Fig.~\ref{fig:cont} a snapshot of the solution obtained for cases~2 and~10. 
In the latter case, the temperature field (Fig.~\ref{fig:cont}d)  shows three major plumes originating 
from the hot inner boundary, which reflect the maximum energy at $m=3$ shown 
in Fig.~\ref{fig:spectra}. Accordingly, the vorticity in Fig.~\ref{fig:cont}c exhibits a large
scale $m=3$ overturning circulation, with pockets of intense vorticity found in the eyes of
the large-scale circulation. 
Note that in this set-up the plumes are anchored at the inner boundary, and that time-dependency
appears mostly in the form of undulations occurring at their tip. 
In contrast, case~2, which corresponds to a more viscous fluid, and a lower level of forcing, 
is more laminar; its vorticity is notably concentrated along the edges of the large-scale
convective cells (Fig.~\ref{fig:cont}a). The temperature field shown in Fig.~\ref{fig:cont}b
appears symmetrical with regard to the top and bottom boundary layers, which are destabilized
by similar cold or hot plumes displaying a mushroom head on top of a thin conduit.

\subsection{Convergence analysis}
\label{sec:accuracy}
Our analysis of the convergence of the 26 schemes of interest in this study follows this procedure: 
 for each case, we compute a reference simulation using the 4th order SBDF4 IMEX multistep scheme, 
using a time-step size $\Delta t^r$ small enough in order to enable a convergence analysis 
that spans two orders of magnitude in terms of $\Delta t$. To equilibrate the solution prior 
to using $\Delta t^r$, we activated  the possibility of a variable $\Delta t$ for the high-resolution cases. 
We select a time window $[t_s,t_e]$  
that typically cover\addRone{s} a sizeable fraction of a convective turnover time. The reference state vector
at $t=t_s$ is taken as the initial condition for the forward 
integration up to $t=t_e$, 
performed with each of the 26 schemes. The accuracy of the 
solution at $t=t_e$ is assessed using the $\mathcal{L}^2$ norm. For instance, the error 
in $\omega$ is given by
\[
e_\omega = \sqrt{\iint_A [\omega(s,\varphi,t_e) - \omega^r(s,\varphi,t_e)]^2 s \mathrm{d}s\mathrm{d}\varphi},
\]
where the superscript $r$ corresponds to the 
reference solution. In the following, unless otherwise stated, we will 
systematically use  
this absolute definition of error. 

\begin{figure}
\centerline{\includegraphics[width=\linewidth]{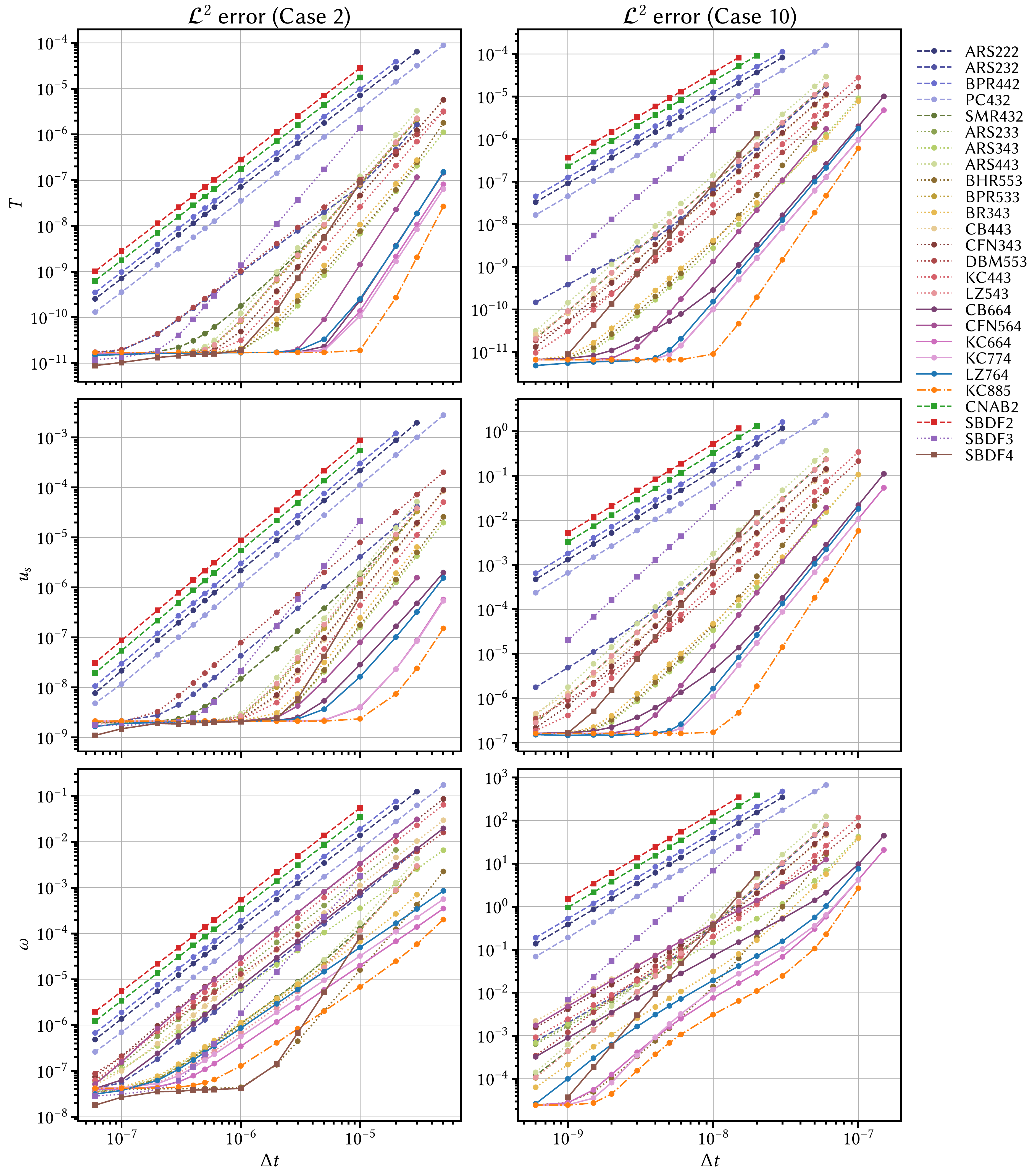}}
\caption{Convergence of the $\mathcal{L}^2$ error for the temperature field 
(top panels), the cylindrical radial velocity  $u_s$ (middle panels) and the 
vorticity $\omega$ (bottom panels) for Case 2 (left column) and Case 10 (right 
column) as a function of the timestep size $\Delta t$. The markers correspond to 
the class of IMEX, with squares denoting IMEX multistep and circles 
IMEX-RK multistage schemes. The linestyles highlight the theoretical order with dashed 
lines for second order, dotted lines for third order, solid lines for fourth 
order and dash-dotted lines for fifth order.}
\label{fig:er_2_10} 
\end{figure}
We begin by a global inspection of the error behavior for the two cases we already looked at, cases~2 and cases~10, whose convergence results are shown in Figure~\ref{fig:er_2_10}. 
As explained above, we tried to assess the convergence properties by having at least two orders of magnitude in the
range of $\Delta t$; this is sometimes barely achieved, in particular for IMEX multistep methods SBDF3 and SBDF4 whose
stability domain is narrower than IMEX-RK schemes of the same order. 
Schemes of nominal order~2 systematically display a higher error level than schemes of order 3 and beyond, and
never reach the plateau of numerical roundoff error in the range of time step sizes that we considered. 
Two exceptions are the ARS232 scheme by \cite{ascher1997implicit} and SMR432 scheme by \cite{spalart1991spectral} that exhibit a higher convergence rate; their convergence
curves are in fact mixed with those of the schemes of nominal order~3. These schemes aside, we also note that at any 
given $\Delta t$ there can be a factor of $10$ difference between the worst (in this sense) order 2 scheme and the 
best one - PC432 is more accurate by one order of magnitude than SBDF2 or CNAB2. 
Order 3 schemes find themselves sandwiched between order 2 and order 4 schemes. Their convergence rate is such 
that the roundoff error plateau is reached for some schemes, for all fields
($T$, $u_s$ and $\omega$) for case~2 and all fields but the vorticity 
for case~10. BR343, BHR553 and ARS343 appear as the most accurate third-order IMEX-RK schemes, especially in the
	turbulent case~10. For the latter (ARS343) this makes sense as its explicit component
	is designed to match the stability properties of the RK4 scheme \citep[][\S 2.7]{ascher1997implicit}; see also
	\ref{app:stab}.  
	For case 2, comparison of the behavior of $e_{u_s}$ of IMEX-RK schemes of theoretical order $3$
	with that of SBDF3 highlights that some 
	do not exhibit third-order accuracy.   
	The 4th-order IMEX-RK schemes display overall similar error levels, below the lowest
	level attained by third-order schemes, this being marginally true for CFN564. For a given $\Delta t$, if one
	considers the temperature $T$ (Figure~\ref{fig:er_2_10}, top panel) 
	 4th-order IMEX-RK schemes are more accurate by two orders of magnitude than the SBDF4 multistep scheme. The situation is not so clear
when one considers the error in the vorticity. There SBDF4 displays a high convergence rate towards the plateau, 
whereas $4$th order IMEX-RK schemes do not show a clear trend. In fact, the sole scheme that appears to compete 
with SBDF4 is BHR553. We will return to this later. For now, we complete this preliminary overview by 
noting that our sole $5$th order scheme, KC885, is as expected more accurate than any other scheme considered, with
the exception of SBDF4 and BHR553 being more accurate with regard to the vorticity for case~$2$, 
over a limited range of $\Delta t$. 
Note finally that the fully explicit schemes that we have at our disposal (RK2 and RK4) are unstable
over the range of $\Delta t$ investigated here for cases~2 and 10. This is due to the stricter limits
imposed on $\Delta t$, which are such that stability coincides with reaching the numerical 
roundoff error plateau. Case~3 provides a configuration for which 
fully explicit schemes can be studied; the corresponding convergence curves are
provided in \ref{app:rk}). 

\begin{figure}
\centerline{\includegraphics[width=\linewidth]{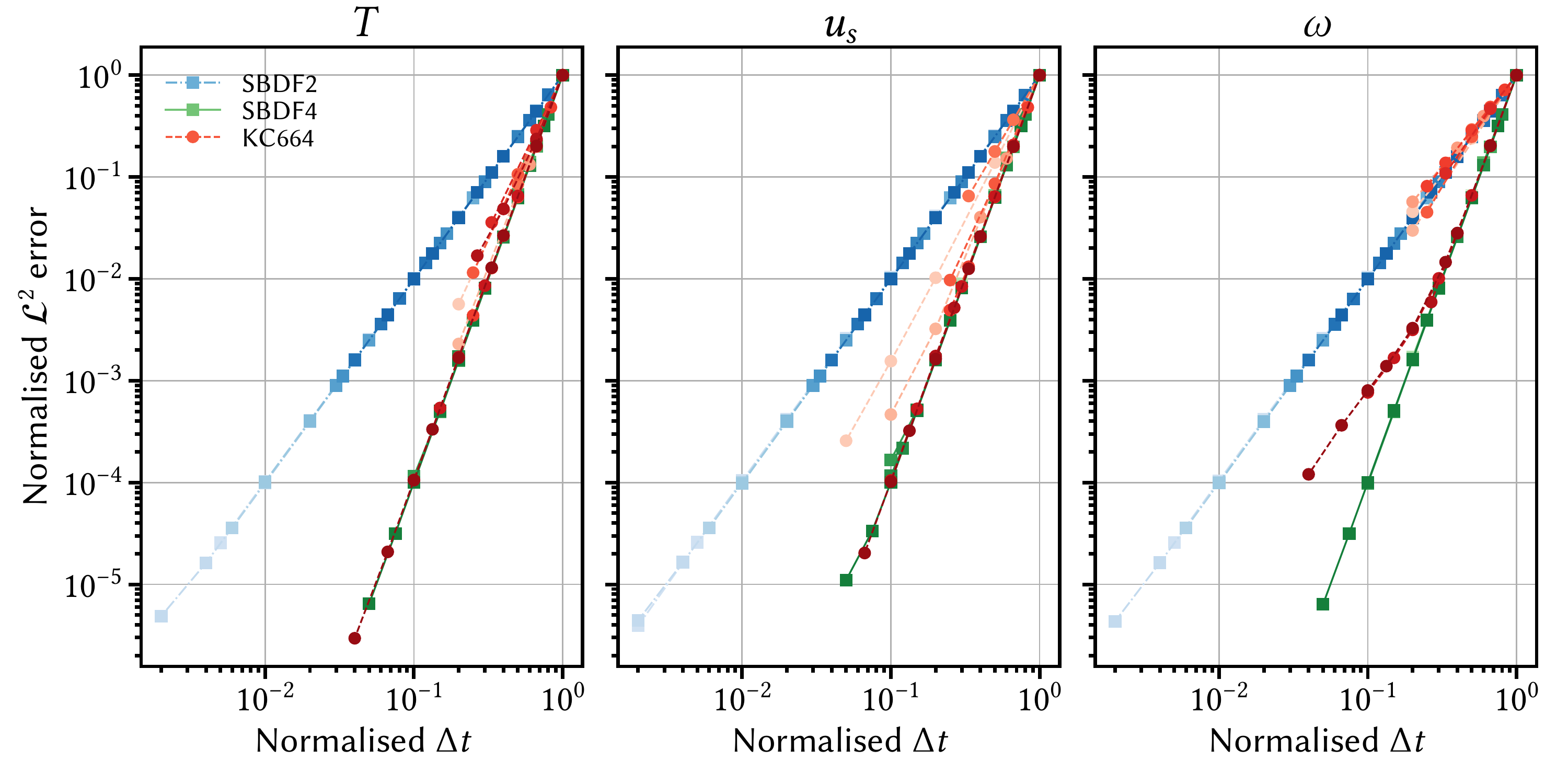}}
\caption{$\mathcal{L}^2$ error for $(T,u_s,\omega)$ normalized by its maximum 
value (for a given case) as a function of $\Delta t$ normalized by its maximum value (for a given case too). Three 
schemes are considered: SBDF2, SBDF4 and KC664. The darker the color of a symbol, the higher the case number.
}
\label{fig:collapse}
\end{figure}

In summary, 
 multistep schemes display the expected convergence rate, and an error level overall higher
 than IMEX-RK
schemes of the same order. 	
To get a better understanding of the error behavior across the 11 cases, 
we now show in Fig~\ref{fig:collapse} how  it evolves for the
IMEX-RK KC664 scheme of \cite{kennedy2003additive} and the two 
multistep schemes SBDF2 and SBDF4. 
Despite the fact that the error level and admissible time step values 
vary across the 11 cases, we collapse the information
by normalizing the error of a given scheme for a given case
by its maximum value, and the timestep $\Delta t$ by the value it has
when this maximum value is obtained. In addition symbols are represented such that
the darker the symbol, the larger the case number. 
Note also that prior to collapsing the data, we got rid of those unwanted  points 
located on a plateau, such as those 
present in Fig.~\ref{fig:er_2_10} for cases 2 and 10. 
 In the log-log representation of Fig.~\ref{fig:collapse}, we first observe 
that the behavior of SBDF2 and SBDF4 is well captured across the cases by two straight lines, 
of slopes 2 and 4, respectively, for the three fields of interest, $T$, $u_s$ and $\omega$. 
This illustrates nicely that they indeed conform to their expected convergence rate over the range
of regimes studied.  
 KC664, on the contrary, displays 
some scatter, lighter symbols (laminar cases) being overall further apart from the 4th order reference defined
by SBDF4 than darker symbols (turbulent cases). This is particularly striking for the vorticity field, more
moderate for $u_s$, and even less pronounced for $T$. For cases 9 and~10 (darker symbols) 
the vorticity appears
to transition from 4th order to 2nd order as the value of the normalized time step decreases. 

This is evidence of order reduction, to an extent that depends strongly on the regime considered. 

\subsection{Order reduction}
\begin{figure}
\centerline{\includegraphics[width=\linewidth]{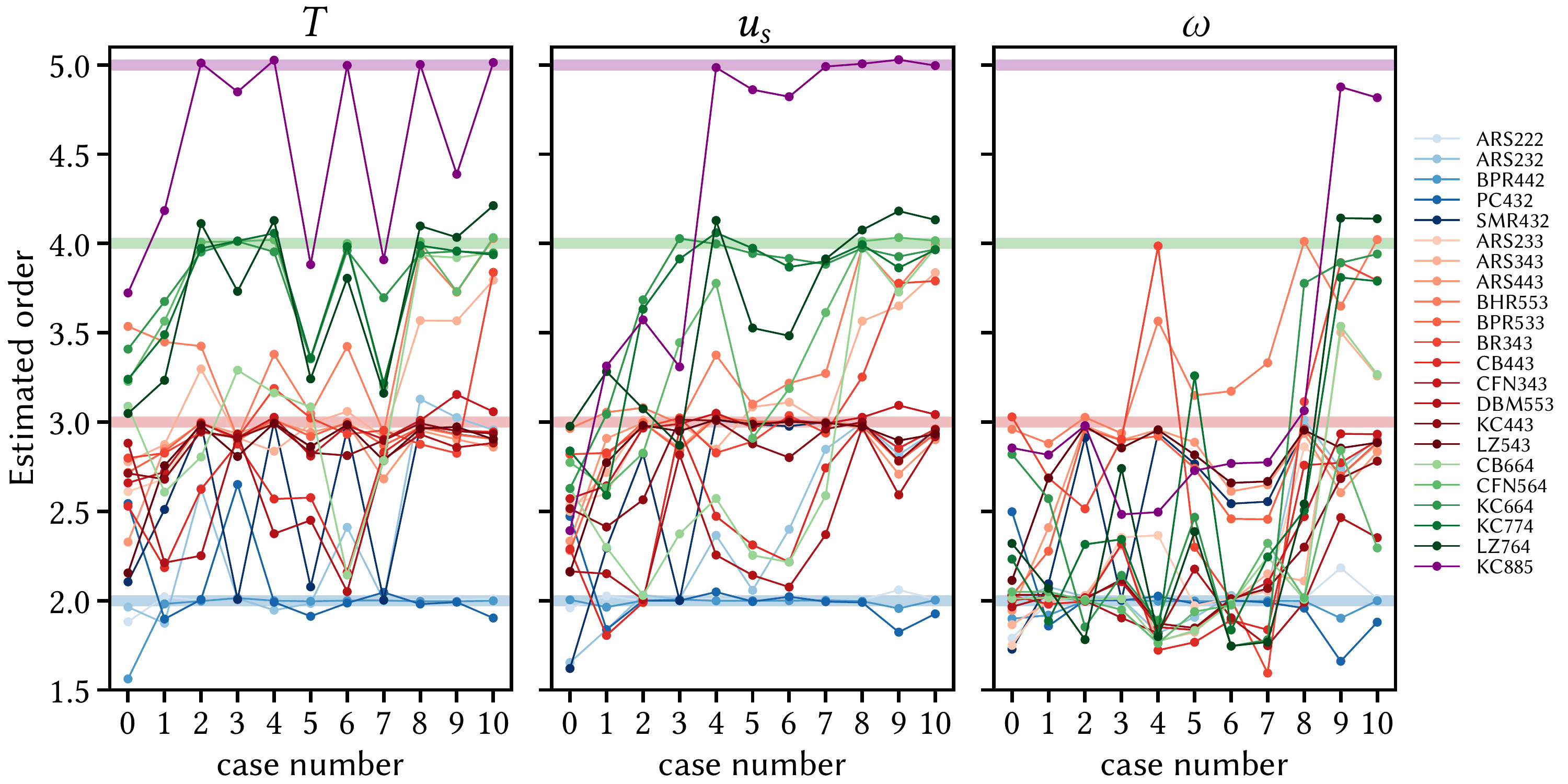}}
\caption{Measured order of convergence for the 11 cases and 22 IMEX-RK time integrators, based on the 
error impacting the temperature $T$ (left panel), cylindrical radial velocity $u_s$ (middle panel)
and vorticity $\omega$ (right panel). Thick 
 horizontal lines highlight integer values of $2$, $3$, $4$ 
and $5$. The 22 schemes are listed to the right, and their color reflects the theoretical
order of convergence.}
\label{fig:order_est}
\centerline{\includegraphics[width=\linewidth]{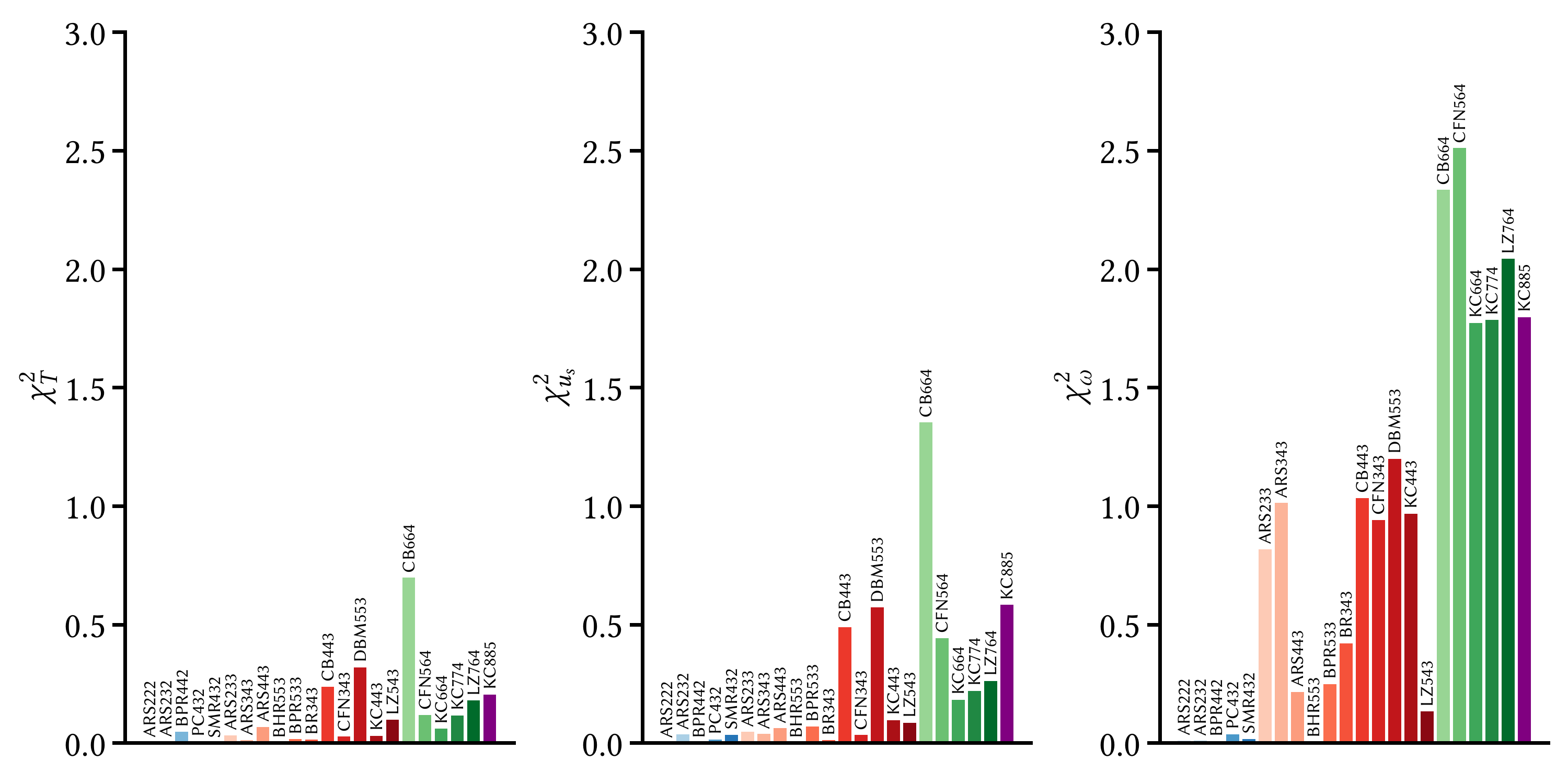}}
\caption{$\chi^2$ measure of order reduction for the 22 IMEX-RK schemes, for the temperature $T$ (left panel),
cylindrical radial velocity $u_s$ (middle panel), and vorticity $\omega$ (right panel). A value of $0$ means
no order reduction over all cases considered in this study.}
\label{fig:chi2_full}
\end{figure}

We now quantify order reduction for all cases and the 22 IMEX-RK schemes considered. 
To assess the order of convergence of a scheme $\mu$, we consider the two largest values of
$\Delta t$ used in the convergence analysis and seek a fit for the error of the form
\[
e (\Delta t) \propto \Delta t^{\,\mu}, 
\]
with $\mu$ the sought order, that depends on the scheme and the case, and the field
of interest ($T$, $u_s$ and $\omega$). 
\addRone{ The two largest values of $\Delta t$ are admittedly 
          not representative of the asymptotic behavior of a scheme
	  when $\Delta t \rightarrow 0$. 
	  Several methods display
	  more than one scaling through the range of tested time step sizes
	   (recall Fig.~\ref{fig:collapse} above), 
	  which makes it difficult to correctly define the order of convergence. 
	   This definition of the order 
	  is not perfect, as 
	  it is biased towards runs that favor
	   the largest integration length over accuracy, for a given 
	   number of time steps performed.} 
The $726$ values of $\mu$ that we estimated 
are displayed in Figure~\ref{fig:order_est}. 
To try and synthesize the information, we additionally introduce a $\chi^2$ measure of order reduction, defined by
\[
\chi^2 = \sum_{\mbox{\scriptsize cases}} \left(\frac{\mu^m -\mu^t}{\mu^t}\right)^2,
\]
where the measured order $\mu^m$ is set to the theoretical order $\mu^t$ when superconvergence is observed, i.e.
when the measured order exceeds the theoretical order. 
Values of $\chi^2$ are reported in Figure~\ref{fig:chi2_full}. 
Figure~\ref{fig:chi2_full} reveals that order reduction, when it affects a scheme, is more severe 
for $\omega$  than it is for $u_s$, which is itself more severe than the order reduction that impacts $T$, if it
	impacts $T$ at all. 

Overall, we observe in Fig.~\ref{fig:order_est}
 that
IMEX-RK order 2 schemes are immune to order reduction, with SMR432 and ARS232 showing superconvergence, in particular
for high Reynolds number cases. 
	Third-order schemes show a variety of behaviors. 
Two schemes stand out as being particularly impacted by order reduction: CB443 and DBM553. On the contrary, 
BHR553 is immune to order reduction, and occasionally superconverges, as also  
reported by e.g. \cite{grooms2011linearly}. Fourth-order schemes are all prone to order reduction, especially CB664. 
Order reduction manifests itself mostly for our most laminar cases, from $0$ to $5$, and appears to be
stronger in the most laminar cases. 
This phenomenon is a well-known issue that can arise due to two factors: the 
discrete algebraic equation (Eq.~\eqref{eq:dae}), and the stiffness of the problem
\addRtwo{\citep[consult][for a thorough theoretical investigation of this issue]{boscarino2007error}}. 
As previously noticed by e.g. \cite{kennedy2003additive} in their analysis of order reduction
in a convection-diffusion-reaction problem,  
both differential variables ($T$, $\omega$ here) and algebraic variables ($u_s$ here) are impacted. 

\subsection{An attempt to assess the impact of the DAE on order reduction}
\begin{figure}
\centerline{\includegraphics[width=0.6\linewidth]{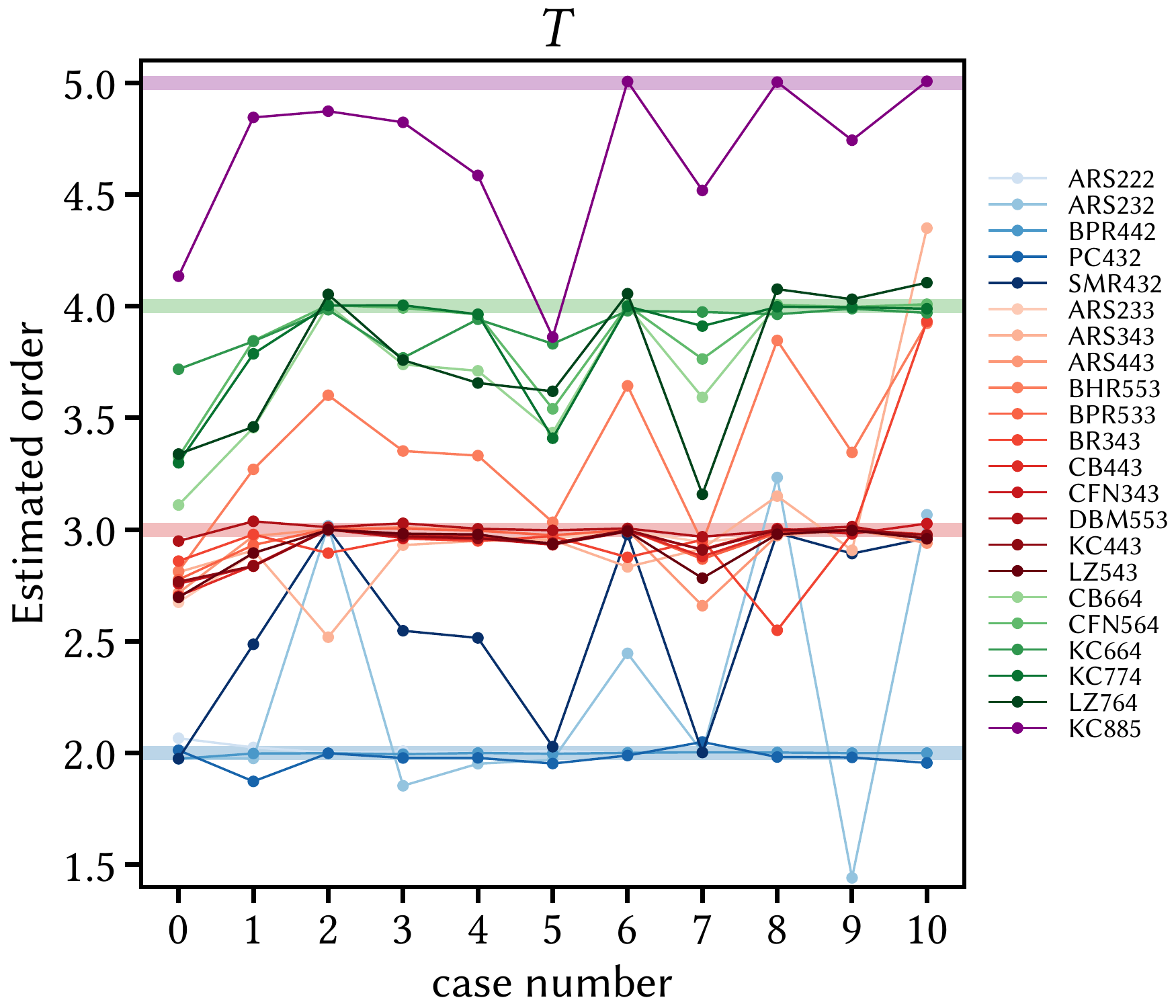}}
\caption{
	Measured order of convergence for the 11 cases and 22 IMEX-RK time integrators, 
	for the simplified advection-diffusion problem whose only time-dependent variable is temperature $T$. 
		Thick horizontal lines highlight integer values of $2$, $3$, $4$
and $5$. The 22 schemes are listed to the right, and their color reflects the theoretical
order of convergence.
		}
\label{fig:order_est_temp}
\end{figure}
We try to estimate to which extent order reduction can be ascribed to the DAE by considering the reduced
advection-diffusion problem
 \begin{equation}
	\frac{\partial T}{\partial t}= -  \boldsymbol{\nabla} \cdot ( \mathbf{u}  T) + \frac{1}{Pr}\nabla^{2} T,
\end{equation}
subject to the same boundary conditions for temperature, and the same procedure for spatial
discretization as detailed above. 
To cover the 11 cases investigated, 
we specify $\mathbf{u}$ by extracting the velocity from a random 
snapshot of the full problem for each case, 
 and take the temperature field
from that snapshot as the  initial condition. 
We use this strategy in order to retain the physical properties of the solution to the full problem (in particular 
the level of turbulent transport) while getting rid of its algebraic component. 
Reference solutions to this reduced problem 
 are
produced via its time integration with the SBDF4 scheme using, again, a time step size small enough
to allow convergence properties to be determined over two orders of magnitude, for each case. 
The temperature field $T$ is the only field that remains to evaluate accuracy and convergence 
properties. 
Estimated orders are now presented in Figure~\ref{fig:order_est_temp}, 
noting that the findings that we present are not 
sensitive to the randomly chosen snapshot.   
 Inspection of Figures~\ref{fig:order_est_temp} reveals
that order reduction persists, but  
to a much lesser extent;  there is less scatter in Fig.~\ref{fig:order_est_temp} than
in the left panel of Fig.~\ref{fig:order_est}. In fact order reduction
is now restricted to well-identified
cases, namely cases 0, 5 and 7, for which the curves of Fig.~\ref{fig:order_est_temp} 
tend to globally dip. Schemes that exhibit occasional superconvergence (SMR432, BHR553, ARS232)
do not superconverge for those cases. Schemes of order 3 and 4 that underperform for the full
	problem 
	 (most notably CB443, DBM553 and CB664)
	are now on par with other schemes of order 3 and 4. 
In summary, using this heuristic method of comparing orders of convergence estimated
for a simplified, DAE-free problem against the full problem, we find that a significant
	fraction of the order reduction that 
	impacts both differential and algebraic variables
	can be ascribed to the DAE. 

\subsection{An attempt to assess stiffness and its relationship with the observed order reduction}
\label{sec:stiffness}
We now try to assess the level of stiffness of the problem we are interested
in. 
As opposed to standard systems of ODEs for which the stiffness is set 
by means of an input control parameter, such as those analyzed e.g. by 
\cite{boscarino2009class}, in our case it is the combination
of the physical control parameters ($\rayleigh$,$\prandtl$) and the spatial properties
of the 2D mesh that sets the level of stiffness, and makes its
definition less straightforward. 

As seen above, the differential component of the problem at hand reads schematically
\begin{equation}
\frac{\mathrm{d}\statev}{\mathrm{d}t} = \nonlinop(\statev,\algebraicv) + \linop \statev,
\end{equation}
where $\statev$ is the differential state vector,
and $\nonlinop$ and  $\linop$ are the nonlinear and linear operators, respectively. 
To estimate the stiffness we consider the
ratio 
\begin{equation}
\rmRone{\sigma(t)} \addRone{\epsilon(t)} = \frac{\taul}{\taunl(t)}, 
\end{equation}
where $\taunl$ and $\taul$ are the smallest time scales associated with the
nonlinear and linear terms, respectively. For Boussinesq thermal convection, nonlinearities
reflect the transport of momentum and heat by fluid flow, while
linearities arise from the diffusion of those same fields. The nonlinear time scale
 $\taunl$
 varies along the dynamical trajectory $\statev(t)$.
 On the contrary, the linear time scale $\taul$ corresponds to the most negative
 real eigenvalue of $\linop$, and is set once and for all upon prescription of 
 the Prandtl number 
 and the grid properties. A stiff situation occurs when 
  $\taul \ll \taunl $, or \rmRone{$\sigma \gg 1$} \addRone{$\epsilon \ll 1$}, which is a strong incentive for an implicit treatment
  of the linear term~$\linop \statev$. 

A first option to estimate $\taunl$ and $\taul$ is to follow the logic of e.g. \cite{grooms2011linearly}
by writing in dimensionless form 
\begin{eqnarray}
\taunl &=& \min_{\mbox{\scriptsize grid}} \left\{ \frac{h_s}{|u_s|}, \frac{h_\varphi}{|u_\varphi|} \right\},  \\
\taul &=& \frac{\prandtl}{1+\prandtl}  \min_{\mbox{\scriptsize grid}} \left\{ h_ s^2, h_\varphi ^2 \right\},
\end{eqnarray}
where $\min_{\mbox{\scriptsize grid}}$  is the minimum over the physical $s-\varphi$ grid, 
and $h_s$ and $h_\varphi$ are the space-varying grid spacings in the $s$ and $\varphi$
directions, respectively.  \addRone{The nonlinear time scale is estimated based on a local measure of transport in the two directions of space, 
while the linear time scale assumes that 
    the most negative eigenvalues of the linear operator 
    correspond to an effective diffusivity equal to $\kappa +\nu$, as suggested 
    by Eq.(24) of \cite{grooms2011linearly}.}

 We list the nonlinear and linear time scales determined for 
 the 11 cases in the leftmost two columns of Table~\ref{tab:stif}, noting that
 the value of $\taunl$ is, for each case, the average value
 of $\taunl(t)$ found for $5$ independent snapshots. 
 \rmRone{If we introduce a small} \addRone{The corresponding 
 stiffness parameter, $\epstau$, }
 \rmRone{$=1/\sigma$, we see that it}
 spans a moderate range of two decades. The increase of turbulence 
 and reduction of $\taunl$ goes alongside an increase in the resolution that induces a concomitant decrease
 of $\taul$.  We are in what authors currently refer to 
 as an intermediate range of stiffness \citep{kennedy2019higher}, that can indeed be detrimental to the order of convergence 
 of some IMEX-RK methods. 
\begin{table}
\centering
\begin{tabular}{c|llrrrr} 
Case & $\taunl$ & $\taul$ & $\epstau$ & $\epsdt$ & $\epslambda$ & 
$\epslambda$ (red.) \\ \hline
0 & 9.14 $10^{-3}$ & 2.06 $10^{-6}$ & 2.22 $10^{-4}$ & 1.97 $10^{-4}$ & 2.84 $10^{-4}$ &{\bfseries  2.92 $\mathbf{10^{-4}}$}  \\
1 & 8.92 $10^{-4}$ & 6.23 $10^{-7}$ & 6.99 $10^{-4}$ & 8.14 $10^{-4}$ & 1.69 $10^{-3}$ & 1.05 $10^{-3}$ \\
2 & 5.99 $10^{-5}$ & 1.40 $10^{-9}$ & {\bfseries 2.34 $\mathbf{10^{-5}}$} & {\bfseries 3.66 $\mathbf{10^{-5}}$ } & {\bfseries 3.50 $\mathbf{10^{-5}}$} & 1.40 $10^{-3}$\\
3 & 1.55 $10^{-4}$ & 1.93 $10^{-7}$ & 1.25 $10^{-3}$ & 2.99 $10^{-3}$ & 4.28 $10^{-3}$& 2.65 $10^{-3}$\\
4 & 1.85 $10^{-5}$ & 3.74 $10^{-8}$ & 2.02 $10^{-3}$ & 3.59 $10^{-3}$  & 5.93 $10^{-3}$ & 4.97 $10^{-3}$ \\
5 & 1.07 $10^{-5}$ & 9.42 $10^{-9}$ & 8.78 $10^{-4}$ & 7.94 $10^{-4}$  & 1.36 $10^{-3}$ & 4.27 $10^{-4}$\\
6 & 4.10 $10^{-6}$ & 1.17 $10^{-8}$ & {\bfseries 2.86 $\mathbf{10^{-3}}$} & {\bfseries 6.13 $\mathbf{10^{-3}}$} & {\bfseries 9.18 $\mathbf{10^{-3}}$} & {\bfseries 7.99 $\mathbf{10^{-3}}$}\\
7 & 1.39 $10^{-6}$ & 5.71 $10^{-10}$ & 4.10 $10^{-4}$ & 6.87 $10^{-4}$ & 6.98 $10^{-4}$ & 2.80 $10^{-4}$ \\
8 & 7.42 $10^{-7}$ & 7.20 $10^{-10}$ & 9.71 $10^{-4}$ & 3.04  $10^{-3}$ & 3.54 $10^{-3}$ & 3.47 $10^{-3}$\\
9 & 2.26 $10^{-7}$ & 2.32 $10^{-10}$ & 1.03 $10^{-3}$ & 2.05 $10^{-3}$& 1.78 $10^{-3}$ & 1.06 $10^{-3}$\\
10 &1.35 $10^{-7}$ & 1.41 $10^{-10}$ & 1.05 $10^{-3}$ & 1.57 $10^{-3}$ & 3.82 $10^{-3}$ & 4.14 $10^{-3}$   
\end{tabular}
\caption{Three different estimates of the stiffness 
of the problem at hand, for all
cases considered in this study. The rightmost column gives an estimate
of stiffness for the problem reduced to the advection--diffusion equation
for temperature. Bold face fonts used for largest and smallest values. See text for details.}
\label{tab:stif}
\end{table}

Our second estimate of stiffness 
compares the maximum time-step allowed for stable computation 
using the ARS343 IMEX-RK method, $\Delta t _{\mbox{\scriptsize max}}^{{\mbox{\scriptsize imex}}}$,
 with the maximum time step
$\Delta t _{\mbox{\scriptsize max}}^{{\mbox{\scriptsize expl}}}$
allowed if one resorts to the fully explicit RK4 integrator. By considering
the ratio 
\[
\epsdt = 
\frac{\Delta t _{\mbox{\scriptsize max}}^{{\mbox{\scriptsize expl}}}}%
{\Delta t _{\mbox{\scriptsize max}}^{{\mbox{\scriptsize imex}}}},
\]
we have an estimate of the stiffness based on an indirect probing
of the stability regions of the timeschemes considered. 
 As a matter of fact, the IMEX-RK scheme chosen here 
is ARS343 because the stability region of its explicit component matches by
design that of RK4 \citep[][\S 2.7]{ascher1997implicit}; therefore we should not expect an offset of
$\epsdt$ by an unwanted factor. Values of $\epsdt$ are tabulated alongside 
values of
$\epstau$ in Table~\ref{tab:stif}. They fall within a factor of $3$ within
the values of $\epstau$, in a non-systematic way. 

\begin{figure}
\centerline{\includegraphics[width=\linewidth,draft=false]{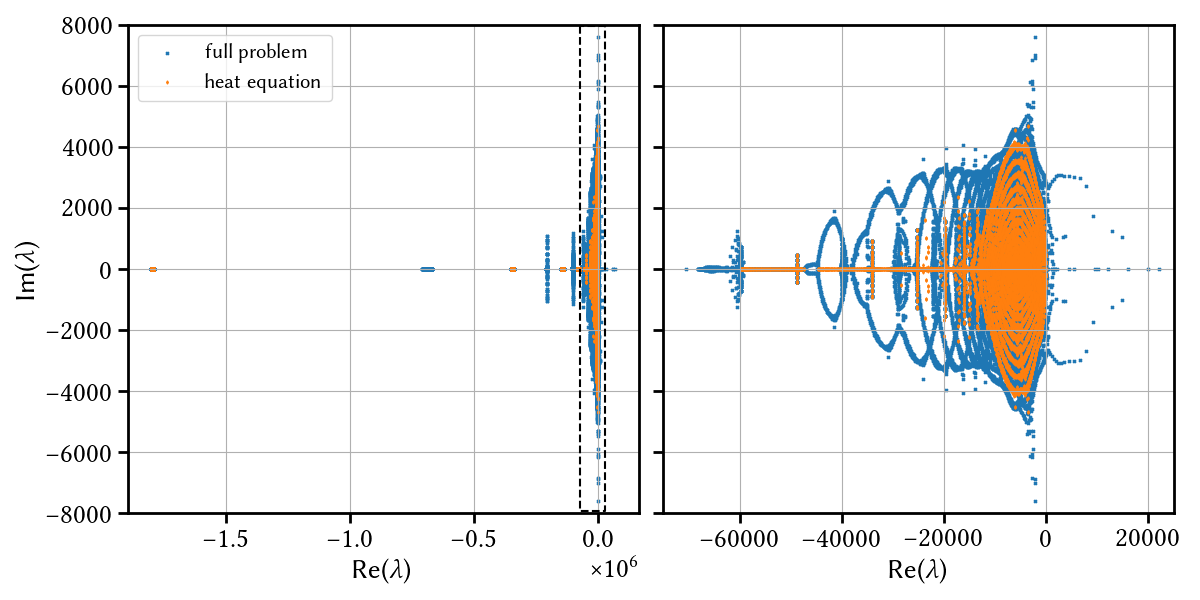}}
\caption{Eigenvalues $\lambda$ of the tangent linear operator computed for case~3, for the full 
problem and the reduced problem restricted to the heat equation considered alone. Left: full spectrum. Right: zoom in the vicinity of the origin, in the region of the complex plane defined by the dashed rectangle in the left panel.
}
\label{fig:full_spec}
\end{figure}

The third and final
option we consider is to investigate directly the eigenvalues of the operators at hand. 
In the vicinity of a point $\statev^\star(t^\star)$, we approximate $\nonlinop$ by
its tangent linear operator $\tlnonlinop$ such that
\begin{equation}
\nonlinop(\statev^\star + \delta \statev) = \nonlinop(\statev^\star) + \tlnonlinop(\statev^\star) \delta\statev. 
\end{equation}
Under these circumstances, if we define
$\totalop(\statev^\star)=\tlnonlinop(\statev^\star) + \linop$,
the behavior of the
solution in the vicinity of $\statev^\star$ obeys
\begin{equation}
\frac{\mathrm{d}\delta \statev}{\mathrm{d}t} = \totalop(\statev^\star) \delta \statev. 
\end{equation}
For each case, we constructed a two-dimensional, 
second-order finite-difference approximation of $\totalop(\statev^\star)$, upon the Chebyshev--Fourier
grid used in our spatial approximation. We did so in order to obtain a sparse, 2D, operator 
amenable to eigenvalue analysis 
for both full and reduced problems. The full
problem, though, was approximated using a formulation based on the streamfunction alone
\citep[see e.g.][\S 1.4]{chqz2007}, in order 
 to facilitate 
 its implementation. 
We benchmarked our finite difference approximation by computing the critical parameters for 
convection, with a critical wavenumber $m_{\mbox{\footnotesize crit}}=3$ and 
Rayleign number $\rayleigh_{{\mbox{\footnotesize crit}}}=1768$ (recall section~\ref{sec:pres_cases}). 
To obtain the eigenvalues of $\totalop(\statev^\star) $, we resorted to the sparse library
of the scientific python package Scipy \citep{virtanen2020scipy}, in conjunction 
with the SLEPc toolbox for python \citep{hernandez2005slepc,dalcin2011parallel}. 
Cases of modest resolution lend themselves to the full calculation of the spectrum, and 
the example of case~3 is given in Figure~\ref{fig:full_spec}. 
We observe that the distribution of eigenvalues  is symmetrical with respect to the real axis, 
and that eigenvalues are concentrated in the vicinity of the origin, at the exception 
of a few purely real values that are quite distant, and correspond to the inverse value
of the diffusive time scale on the smallest grid spacing, of non-dimensional value 
${\pi^2/N_s^4}$. \addRone{These negative eigenvalues of large magnitude are due to the linear 
component of the problem at hand, and the ones responsible for stiffness.} 
We associate the eigenvalues of largest imaginary part with the 
advective component of $\totalop$, and therefore the reciprocal value of $\taunl$. 

For cases of larger size (for case 10~the size of the 
sparse matrix to deal with is $878206\times878206$) 
we computed only the 100~eigenvalues of largest negative real parts, $\lambda^r$, and 100~eigenvalues 
of largest imaginary parts, $\lambda^i$. Our estimate of stiffness is given by 
\begin{equation}
\epslambda = \frac{\max{\Im({\lambda^i})}}{\max{|\Re({\lambda^r})|}},  
\end{equation} 
in which $\Im()$ is the imaginary part. 
For this calculation, we used the same 5 independent snapshots for each case as the ones used 
to estimate $\epstau$. Results are listed in Table~\ref{tab:stif}, and are in 
agreement, 
again within a factor of $3$ with estimates based on $\epstau$ and $\epsdt$. 
The 
three estimates of $\epsilon$ point to case~6 as being the least stiff, even though the stiffness parameter
does not exceed $10^{-2}$. Case~6 had a Prandtl number of unity, which implies that
the diffusivities of heat and momentum are equal. 
 The stiffest case is case~2, which has $\prandtl=40$. For case~2, a relatively large resolution is needed
 to resolve the small-scale temperature anomalies within a pretty laminar flow (recall 
 Figures~\ref{fig:cont}a and~\ref{fig:cont}b). Table~\ref{tab:stif} also comprises
 values of $\epslambda$ computed for the reduced problem (advection-diffusion 
equation of temperature with a prescribed flow), that we used previously 
 to try and assess the impact of the DAE on the observed order reductions. 
 The values of $\epslambda$ for the reduced problem are consistent with 
 those found for the complete problem, with one exception. 
 In this simplified system, case~6 remains the least stiff case, and it is now 
 the barely supercritical case~0 that is the stiffest case. Case~2 ceases to be the 
 stiffest case, since its most negative eigenvalues are associated with the diffusion
 of momentum, not heat. Accordingly, the value of $\epslambda$ we find for the 
reduced case~2 is
 precisely multiplied by a factor of $\prandtl=40$, compared with the value found for the 
 full problem. This increase is not seen in the three
 cases that have $\prandtl$ below unity (5, 7 and 9), for which the largest negative eigenvalues
 remain connected with the diffusion of heat. 
\begin{figure}
\centerline{\includegraphics[width=\linewidth]{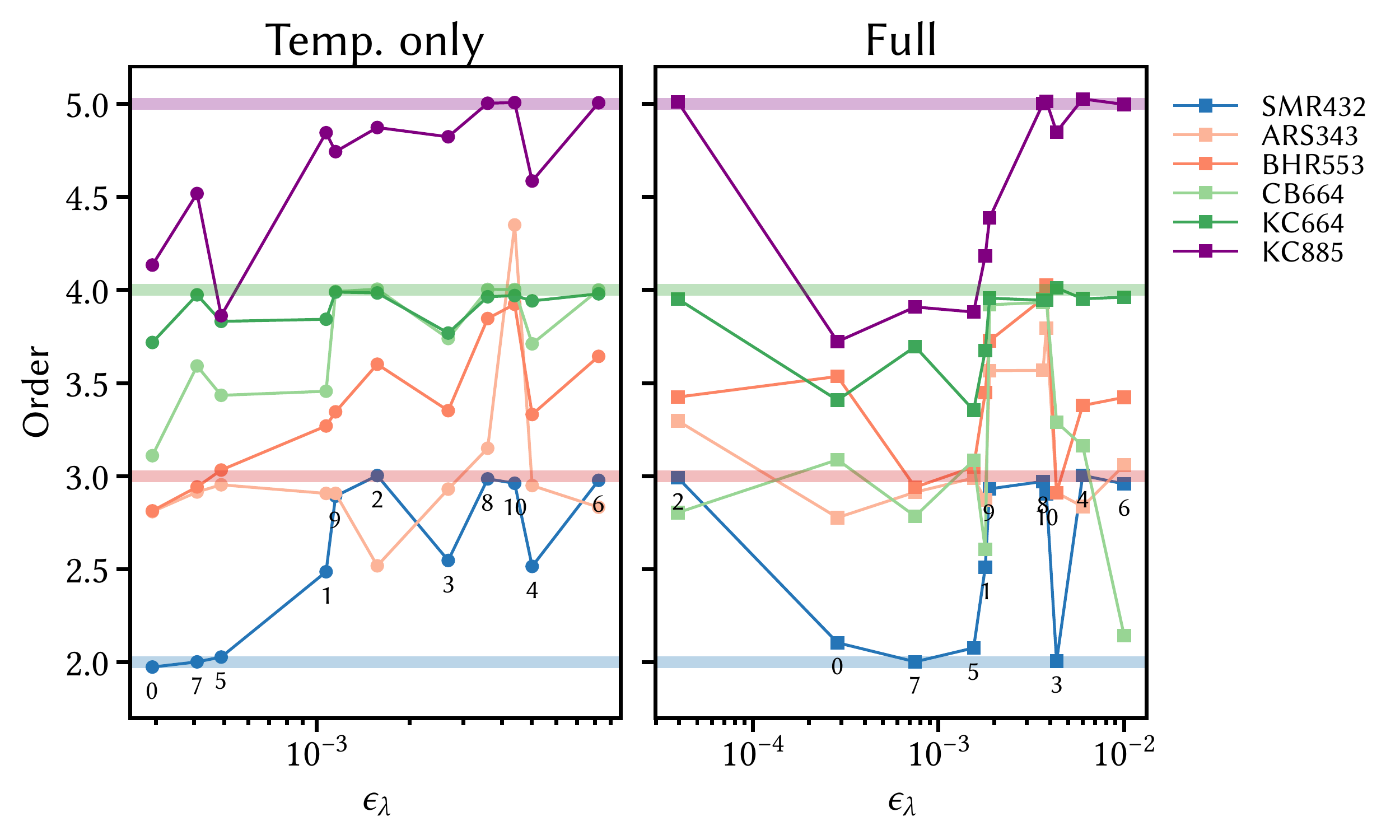}}
\caption{Order of convergence for the temperature field $T$, in the reduced problem set-up (left panel), and for the full problem (right panel), as a function
of the stiffness parameter $\epslambda$ estimated for each problem for all 11 
cases
and for some selected time schemes. See text for details.
$\epslambda$ sur
}
\label{fig:order_vs_stif}
\end{figure}
 We conclude the analysis of the impact of stiffness by showing the measured 
 order of convergence for the reduced, DAE-free,  and full problems against the corresponding values
 of $\epslambda$ in
 figure~\ref{fig:order_vs_stif}, with the hope that this representation will 
 get rid of the jaggedness of figures~\ref{fig:order_est} (left panel) and ~\ref{fig:order_est_temp}.  
 Within the modest range of $\epslambda$ that our investigations enabled, 
 we observe in Fig.~\ref{fig:order_vs_stif}, left panel, that for the 
 reduced problem, KC774, CB664 and KC885 are close to meeting  their
 expected order of convergence for $\epslambda \gtrsim 10^{-3}$. 
 SMR432 exhibits an order of convergence larger than 2 for $\epslambda \gtrsim 
10^{-3}$
 as well.  ARS343 superconverges only for case~10, while 
 BHR553 superconverges for all cases but case~0. 
 This scheme is by design supposed to be immune to order reductions caused by 
 stiffness and the DAE \citep{boscarino2009class}. For the full problem, the 
range of $\epslambda$ covered
 is a bit broader, and we find it noteworthy
 that for the stiffest case 2, order is restored. In fact, 
 the curves we obtain in particular for schemes KC774 and KC885 show a reasonable
 similarity with the curves obtained for those same schemes applied to Kap's problem
 by their genitors \cite{kennedy2019higher}.  

 In summary, we think that the stiffness of the cases we considered in this study
 covers a moderate, intermediate range of values spanning slightly less than two decades for the reduced
 problem and two and a half decades for the full Boussinesq convection problem. 
We find that even if stiffness has an impact
 on the convergence of some schemes over a limited range, typically
 $10^{\addRone{-4}} \lesssim \epslambda \lesssim 10^{-3}$ for the stiffness parameter we 
determined, 
 it is mostly the DAE that causes the degradation of convergence. 
 The reduction in order affects both differential and algebraic variables, probably
 by virtue of the coupling induced by the equations of the problem at hand. 
Yet, schemes of theoretical order 2 are not affected by order reduction. 
 Higher-order time integrators that are by design immune to such problems, such as BHR553 or the 
 IMEX multistage methods, may appear as the schemes of choice. 
 
This statement remains to be 
 weighted against a measure of the stability and computational efficiency of those
 schemes, which is the topic of the next subsection. 
\subsection{Stability and computational efficiency}
\label{sec:stab_analysis}
\begin{figure}
\centerline{\includegraphics[width=\linewidth]{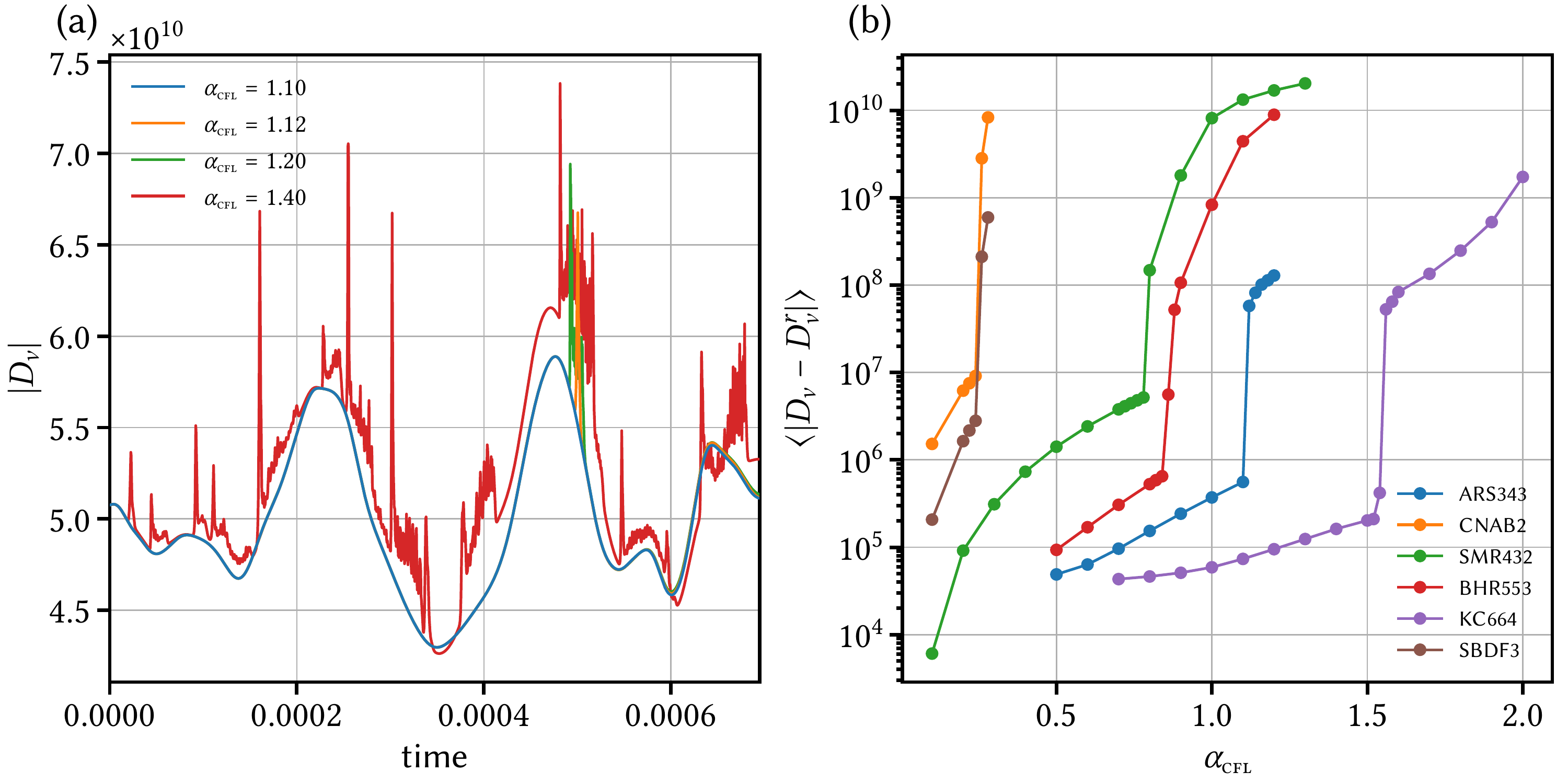}}
\caption{a: Time series of viscous dissipation $|D_\nu|$ for case~9
advanced with the ARS343 IMEX-RK scheme, 
 considering various values of $\alpha_{\mbox{\scriptsize \sc cfl}}$. The highest admissible
 value we find in this setup is $\alpha_{\mbox{\scriptsize \sc cfl}}=1.10$. 
  b: Time average of $|D_\nu - D_\nu^r|$ as a function of
  $\alpha_{\mbox{\scriptsize \sc cfl}}$ 
  for case 9 and 6 time integrators. 
	  $D_\nu^r(t)$ is a reference time series for viscous dissipation obtained with
	  the SBDF4 time scheme. 
	  Note that the scale on the $y$-axis is logarithmic.
	  For a scheme, the maximum admissible value of $\alpha_{\mbox{\scriptsize \sc cfl}}$
	  is the largest one that precedes the steep increase in the curve. 
	   }
\label{fig:getalphamax}
\end{figure}

The explicit treatment of nonlinearities imposes a restriction on the available
time step that is subject to a time-dependent Courant-Friedrich-Levy condition 
\[
\Delta t \leq \alpha_{\mbox{\scriptsize \sc cfl}} \min_{\mbox{\scriptsize grid}} 
\left\{ \frac{h_s}{|u_s|}, \frac{h_\varphi}{|u_\varphi|} \right\}, 
\]
where the $\mathcal{O}(1)$ prefactor $\alpha_{\mbox{\scriptsize \sc cfl}}$  
depends on the time integrator and the case considered. In 
order to determine empirically the maximum admissible value of 
$\alpha_{\mbox{\scriptsize \sc cfl}}$, 
$\alpha^{\mbox{\scriptsize \sc max}}_{\mbox{\scriptsize \sc cfl}}$, 
for the 
$11 \times 26$ combinations of this study, we followed \cite{gastine2019pizza}
and inspected the timeseries 
of viscous dissipation 
$D_\nu(t)$ 
and its fluctuations for different values 
of $\alpha_{\mbox{\scriptsize \sc cfl}}$. 
The maximum admissible value of 
$\alpha_{\mbox{\scriptsize \sc cfl}}$, 
$\alpha^{\mbox{\scriptsize \sc max}}_{\mbox{\scriptsize \sc cfl}}$
is determined to $0.02$ accuracy 
by requesting that the timeseries 
of $D_\nu(t)$ does not exhibit any flagrant spike, 
over a 
time window of width roughly equal to 5 convective turnover times; the time window
is case-dependent, but for a given case, it is the same for all time integrators.   
Figure~\ref{fig:getalphamax}a illustrates this methodology for 
case~9 and the ARS343 scheme, a configuration for which we find 
$\alpha^{\mbox{\scriptsize \sc max}}_{\mbox{\scriptsize \sc cfl}}=1.10$. This arguably tedious methodology 
is meant at preserving the accuracy of the solution, and can not
lead to the disturbing
occurrence  
of
stable yet inaccurate IMEX-RK schemes
reported by \cite{grooms2011linearly}, and mentioned
in the introduction. In this regard, our estimate
of $\alpha^{\mbox{\scriptsize \sc max}}_{\mbox{\scriptsize \sc cfl}}$ is conservative. \addRone{We refer readers interested in a more standard assessment of stability and efficiency to appendix~\ref{sec:std_eff}, where we report accuracy versus runtime for cases 2 and 10.}

We now propose an automated way of reaching the same conclusions. 
We begin by establishing a master curve for
$D_\nu(t)$ over the interval of interest using the SBDF4 time scheme and the
smallest $\alpha_{\mbox{\scriptsize \sc cfl}}$. 
 This master curve is denoted by $D_\nu^r(t)$ where again, 
the superscript $r$ stands for reference. 
 Given the $D_\nu(t)$ computed for an integrator and a value of  $\alpha_{\mbox{\scriptsize \sc cfl}}$,
 we evaluate the time average of $| D_\nu - D_\nu^r|$ over the window of interest using splines 
 for the numerical 
 integration. 
 As an example, we 
 show  $\langle |D_\nu - D_\nu^r| \rangle$  in Figure~\ref{fig:getalphamax}b  for case~9 and 6~schemes.
 We observe a sharp transition in the behavior of this quantity, for relatively low values 
 of $\alpha_{\mbox{\scriptsize \sc cfl}}$ for the two multistep schemes (CNAB2 and SBDF3) 
 and larger values for the IMEX-RK schemes. The largest value of $\alpha_{\mbox{\scriptsize \sc cfl}}$
 before the transition matches the $\alpha^{\mbox{\scriptsize \sc max}}_{\mbox{\scriptsize \sc cfl}}$
 obtained by visual inspection of the timeseries of $D_\nu(t)$.  

\begin{figure}
\centerline{\includegraphics[width=0.5\linewidth]{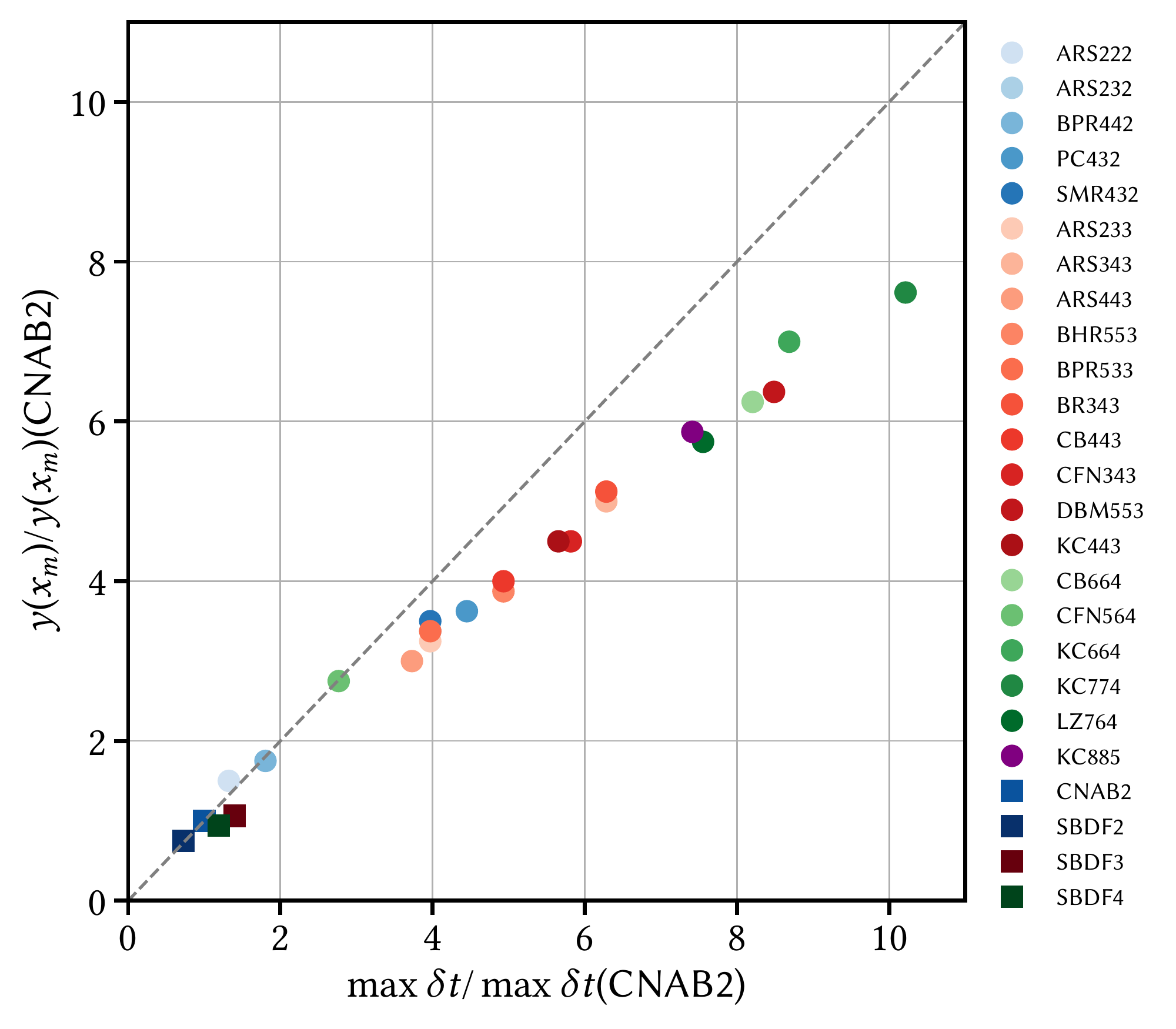}}
\caption{Ratio of the maximum ordinate of the stability curve of the explicit component of
a time integrator to that of the CNAB2 scheme as a function of the ratio of the maximum timestep 
size admissible to the maximum time step admissible for CNAB2. 
The case considered is case~10 ($\reynolds=13320$, stiffness parameter $\epsilon \sim 3\times 10^{-3}$).
The dashed grey line is the prediction based on Eq.~\eqref{eq:tg_guess}.} 
\label{fig:dtmax}
\end{figure}
The value of $\alpha^{\mbox{\scriptsize \sc max}}_{\mbox{\scriptsize \sc cfl}}$ can 
be converted into a maximum attainable value of the time step size, $\Delta t ^{\mbox{\scriptsize \sc max}}$: 
we take it to be the average  $\Delta t$ obtained in the same setups that led to the determination
of $\alpha^{\mbox{\scriptsize \sc max}}_{\mbox{\scriptsize \sc cfl}}$, meaning that
we find $26$ values of $\Delta t ^{\mbox{\scriptsize \sc max}}$ (one per scheme) per case. 
 In those cases where nonlinearities dominate, with stiffness parameters $\epsilon$ of the order
of $10^{-2}$ (recall~Table~\ref{tab:stif} above),  
it is actually possible to make a decent guess of  $\Delta t ^{\mbox{\scriptsize \sc max}}$
based on the boundary of the stability region of the explicit component of the scheme 
under scrutiny. This stability region is bounded
by a curve that mostly lies in the left hand side of the complex plane, 
that of negative real parts. 
We anticipate that in moderately stiff situations where nonlinearities play a major part in the dynamics, 
it is this curve that will control the stability of the implicit explicit scheme, even if it does 
not correspond to the stability curve of the combined scheme, as studied by e.g. \cite{karniadakis1991high} and
\cite{izzo2017highly} for test problems. As discussed above in our analysis of stiffness, 
transport will effectively probe the eigenvalues
of the semi-discrete tangent linear 
 operator with the largest imaginary parts (in absolute value). Let  $\lambda_m$ denote the eigenvalue
of largest imaginary part. As a rule of thumb
we want  $\lambda_m \Delta t$ to be close 
to the stability curve. 
 For transport-dominated
physics, we thus make the tentative prediction that for any scheme 
\begin{equation}
\Delta t^{\mbox{\scriptsize \sc max}} \mbox{ (scheme) } = f \times  y(x_m) \mbox{ (scheme) },
\end{equation}
where the factor $f$ is a function of the spatial discretization only, $x_m=\Re({\lambda_m} \Delta t)$  
 and the ordinate $y(x_m)$ on the stability curve can be determined
numerically. These curves are provided in \ref{app:stab} for 
completeness. If this equality holds, provided we know $\Delta t^{\mbox{\scriptsize \sc max}}$ of a case
for one scheme (CNAB2, say)  we may expect that
\begin{equation}
\Delta t^{\mbox{\scriptsize \sc max}} \mbox{ (scheme) } = \frac{y(x_m) \mbox{(scheme)}}{{y(x_m) \mbox{(CNAB2)}}}
\, \, 
\Delta t^{\mbox{\scriptsize \sc max}} \mbox{ (CNAB2) }
.  
\label{eq:tg_guess}
\end{equation}
The relevance of this line of reasoning is shown in Figure~\ref{fig:dtmax} where this empirical prediction
is compared with the measured value for case 10, our most turbulent case. We find that the overall trend is that
the prediction slightly overestimates the actual value, typically by within 10 to 20 \%. 
Figure~\ref{fig:dtmax} highlights the fact that the most stable order~3 scheme is DBM553, 
a result that can be understood by inspection of the stability region of its explicit
component given in Figure~\ref{fig:stability_regions}, middle panel. Of all the third order schemes
we considered, DBM553 has the most elongated region of stability in the vicinity of the $y$-axis
of the complex plane. In fact, it was designed for the very purpose of accommodating the
constraint on the available time step size arising from the location of eigenvalues
of the explicit component of a model being located along the 
imaginary axis \citep[][]{kinnmark1984one}.
A closer inspection of the IMEX-RK schemes 
which share the same stability domain for their explicit component reveals 
very similar CFL coefficients for case 10 with for instance 
$\alpha^{\mbox{\scriptsize \sc max}}_{\mbox{\scriptsize \sc cfl}}(\text{ARS232})=0.54$ and 
$\alpha^{\mbox{\scriptsize \sc max}}_{\mbox{\scriptsize \sc cfl}}(\text{SMR432})=0.56$; or 
$\alpha^{\mbox{\scriptsize \sc max}}_{\mbox{\scriptsize \sc cfl}}(\text{ARS343})=0.8$ and 
$\alpha^{\mbox{\scriptsize \sc max}}_{\mbox{\scriptsize \sc cfl}}(\text{BR343})=0.82$. This is another 
indication that the stability domain of the explicit part only provides a 
decent estimate of the actual stability of an IMEX-RK scheme in the limit of 
advection-dominated flows.
 To conclude this paragraph, we note that 
for cases of more dramatic stiffness,  one should probably consider the stability region of
the complete IMEX scheme, an endeavor that we did not pursue. 



\begin{figure}[t!]
 \centerline{\includegraphics[width=\linewidth]{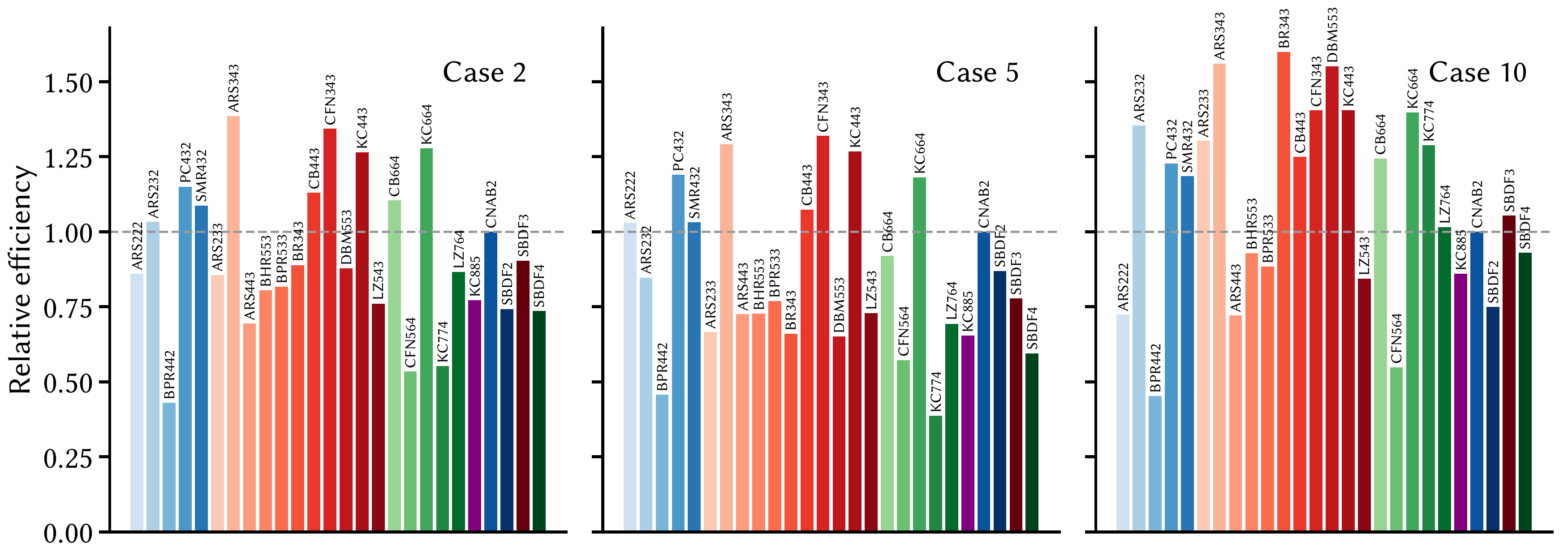}}
 \caption{Efficiency of the time integrators relative to the efficiency of CNAB2 for cases 2, 5 and 10 from left to 
 right. 
 Horizontal dashed lines correspond to a value of unity. 
 See text for details. }
 \label{fig:perf}
 \centerline{\includegraphics[width=\linewidth]{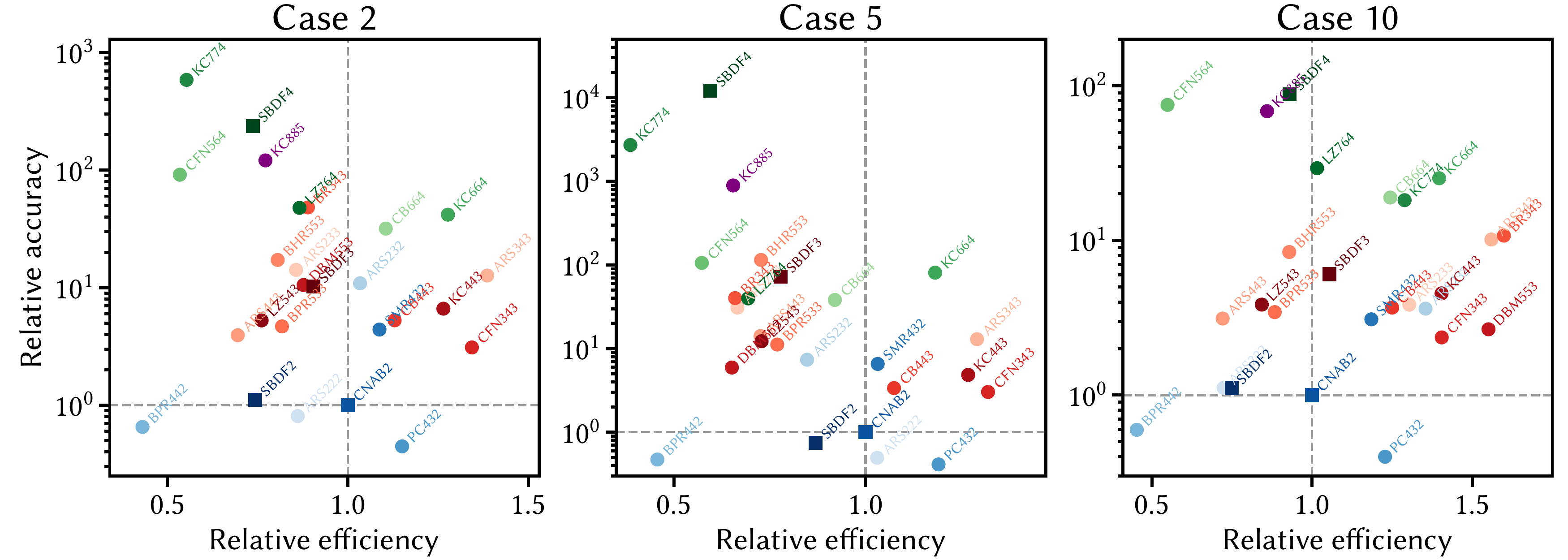}}
 \caption{Efficiency and accuracy of the time integrators relative to the efficiency and accuracy 
 of CNAB2 for cases 2, 5 and 10 from left to right. The scale on the $y$-axis is logarithmic. 
 See text for details. 
 Horizontal and vertical dashed lines correspond to a value of unity. 
 }
 \label{fig:perfacc}
 \end{figure}

To evaluate the efficiency of a given scheme,  we consider the following ratio
\begin{equation}
\eff = \frac{\alpha^{\mbox{\scriptsize \sc max}}_{\mbox{\scriptsize \sc cfl}}}{\cost},
\end{equation}
where $\cost$ refers to the amount of work required to advance the solution by one time step $\Delta t$. 
In practice $\cost$ is the average cpu time measured over $1000$ iterations using reproducible
runtime conditions (same compute nodes, exclusive access to the compute node, one single
OpenMP thread). Note that the linear matrix solves amount for the majority of the walltime, 
and hence the number of solves per iteration $n^I$ in Table~\ref{tab:imexrk} provides
a decent estimate of the actual relative walltime. 
We evaluated the efficiency of the 26 schemes considering the 11~cases. We investigate 
 how the gain one may obtain 
in terms of a larger $\alpha^{\mbox{\scriptsize \sc max}}_{\mbox{\scriptsize \sc cfl}}$  using an IMEX-RK scheme 
trades off with the extra operations that are needed. (Again, at this stage, we do not consider the benefit in terms
of accuracy.)
 Given the popularity of the CNAB2 integrator in our community, we normalize the efficiency by the efficiency 
 of CNAB2. Results are displayed in Figure~\ref{fig:perf} for cases, 2, 5 and 10 that have $\reynolds=26$, $513$ and $13320$, respectively, 
 and whose stiffness parameter $\epsilon$ we estimated in section~\ref{sec:stiffness}
 to be $\sim 3 \times 10^{-5}$, $\sim10^{-3}$ and $\sim3 \times 10^{-3}$, respectively. 
 Relative efficiency is in all cases bounded between $0.5$ and $1.50$. 
 We observe that the number of IMEX-RK schemes that outperform CNAB2 increases dramatically with the Reynolds number from 5 for Case~2
 to 14 for case~10. For the latter, the 14 schemes comprise 3, 7 and 4 schemes of order 2, 3 and 4, respectively. 
 This indicates that the gain in stability in transport-dominated regimes outweigh the increase in the
 number of operations as the resolution increases. 
  For case~10, 
 we note a relative grouping of multistep schemes around 1, 
  with SBDF3 being slight more efficient than CNAB2, in agreement with its more elongated stability
  domain close to the imaginary axis (\ref{app:stab}). 
 A scheme that appears to be consistently efficient across the 3 cases is ARS343. The DBM553 scheme by 
 \cite{vogl2019evaluation} and the BR343 scheme by \cite{boscarino2007uniform} 
 transition from a poor efficiency in the laminar case~2 to an excellent one in the turbulent case~10. 

\subsection{Trade-off between accuracy and efficiency}
We now weigh efficiency and accuracy for recommendations to be made to the practitioner. The accuracy is characterized
by the error one expects for a simulation performed at the maximum available time step size, $\Delta t^{\mbox{\scriptsize \sc max}}$, 
as defined in the previous section. This error is computed based on the scaling of 
the error on temperature with $\Delta t$  that 
can be obtained for the various convergence curves introduced 
in section~\ref{sec:accuracy} above, by taking $\Delta t = \Delta t^{\mbox{\scriptsize \sc max}}$.  
Therefore, our target practitioner
wishes to integrate the solution for the longest time span possible given 
his/her computing resources with the hope that the error will remain as small as possible. We normalize the error
with the error estimated for CNAB2 and show in Figure~\ref{fig:perfacc} where the various schemes considered in this work are
located in the (relative efficiency, relative error) plane for cases 2, 5 and 10 again.

Schemes located in the top right quadrant of each panel are more accurate and yet more efficient
than CNAB2. Six 
IMEX-RK schemes are systematically located in this quadrant, 
one scheme of order~2, SMR432 \citep{spalart1991spectral}, 
four schemes 
of order 3, ARS343 \citep{ascher1997implicit}, CB443 \citep{cavaglieri2015low}, 
CFN343 \citep{calvo2001linearly}, 
KC443 \citep{kennedy2003additive} and one scheme of order 4, KC664 \citep{kennedy2003additive}. At the operational\addRone{, dissipation-based} limit of stability, it is remarkable to note that ARS343 and
KC664 enable a significant improvement of accuracy at a lower cost, regardless of the configuration. In the turbulent regime, where the explicit components
of time integrators prevail, CNAB2 is outperformed by no less than 14 schemes, which 
include 2 schemes of order 2, 8 schemes of order 3 (including SBDF3), and 4 schemes of order 4. 


\section{Summary and recommendations}
\label{sec:conclusion} 
We have applied 
22 multistage, 4 multistep implicit-explicit,  and 2 fully explicit
 integrators 
to the problem of Boussinesq
thermal convection in a two-dimensional cylindrical annulus, over a broad range
of regimes whose exploration was made possible by the
definition of 11~different physical setups. Our spatial discretization rests 
on  a pseudo-spectral collocation method applied to the vorticity
 streamfunction formulation of the problem.  
We summarize our findings as follows: 
\begin{itemize}
\item Over the range of cases considered, IMEX multistep methods  
      exhibit their expected order of convergence. IMEX-RK methods of second order also show  
	  the expected convergence. Some IMEX-RK methods of order 3 and 4 show order reduction, 
	  that affect both the algebraic and differential variables of the chosen formulation. 
\item We have attempted to evaluate the stiffness of the cases using 
      three different strategies that lead to the same conclusion: 
	  the small parameter that quantifies stiffness, 
	  $\epsilon$, covers a moderate and intermediate range of values 
	  spanning two and a half decades for the full Boussinesq convection problem, between
	  $3\times 10^{-5}$ and $10^{-2}$. 
\item  We find that even if stiffness has an impact on the convergence of some schemes 
        over a limited range of~$\epsilon$, 
       typically $10^{-4} \lesssim \epsilon \lesssim 10^{-3}$, it is mostly the discrete 
	   algebraic equation that causes the degradation of convergence. 
	   The reduction in order affects both differential and algebraic variables, probably by virtue of the coupling induced by the equations of the problem at hand. 
\item  IMEX-RK time integrators that are by design immune to such problems, 
       such as BHR553 \citep{boscarino2009class}, exhibit nominal convergence
       properties, or even better-than-nominal properties in the least stiff (transport-dominated) situations. 
\item  We have defined the efficiency of a scheme 
      by the ratio of the maximum admissible
	  value of the Courant number, $\alpha^{\mbox{\scriptsize \sc max}}_{\mbox{\scriptsize \sc cfl}}$, divided
	  by the cost of performing one time step. 
	  By maximum admissible value we understand a value that generates a smooth timeseries
	  of viscous dissipation within the system. Viscous dissipation is a demanding 
	  quantity, whose behavior help us define an acceptable (or not acceptable)  solution. 
	  With this at hand,  we have reported the efficiency of the schemes relative to the popular CNAB2 
       for 3 cases that we consider representative. We found that the 
	   relative efficiency was bounded between $0.5$ and $1.5$. Also, 
	   CNAB2 is not the method of choice when going to transport-dominated cases, as no less 
	    than 14 schemes are more efficient than CNAB2 for our most turbulent case. 
\item This last statement becomes even stronger when relative accuracy is added to the  analysis. 
      By relative accuracy here we mean the ratio of 
	  the error anticipated for a scheme running at the 
	  operational\addRone{, dissipation-based} limit of stability, whose expected error can be estimated
	  based on the convergence analysis, to the same error 
	  anticipated for CNAB2 at its limit of stability (and that for each case).
	  For the same three representative cases, 
	  we find that $6$ schemes combine the assets of being
	  less expensive and more accurate than CNAB2: SMR432 \citep{spalart1991spectral}, 
	  CFN343 \citep{calvo2001linearly}, ARS343 \citep{ascher1997implicit}, 
	  CB443 \citep{cavaglieri2015low}, 
	  KC443 and KC664 \citep{kennedy2003additive}. 
\item For the problem of thermal convection in 
      a cylindrical annulus with a spectral discretization in space, it appears
	  that the default integrator should be the third-order ARS343, or possibly KC664 
	  if higher accuracy is sought.  
\end{itemize}

 For turbulent cases, or in more general terms, transport-dominated cases, 
      the performance of an implicit-explicit scheme is dictated by 
	  its explicit component\addRone{, as the part treated explicitly here 
	  is purely advective}. The previous general recommendation 
	  can be amended on a case-by-case basis: for instance the third-order scheme 
	  ~DBM553 proposed by \cite{vogl2019evaluation} may well prove superior
	  to any third-order scheme for turbulent transport-dominated 2D problems reaching 
	  $\reynolds \sim 10^5$ and beyond. In fact, on this path, 
	  provided $\epsilon\sim1$ is reached, RK4 may well prove competitive, even more so
	  if a regridding is performed to alleviate the stringent time step 
	      limitations due to the clustering of grid points near
	      the boundaries, as done e.g. by \cite{johnston2009comparison} using
	 the mapping proposed by \cite{kosloff1993modified}. 

 \addRone{Before discussing a tentative extrapolation 
          of our results to three-dimensional geometry, we should 
	  add the following important caveat: stiffness in our setup is
	  controlled by the most negative real eigenvalues 
	   of the linear part that is treated implicitly, and 
	  our recommendations may not be suited for those
	   problems where stiffness comes from fast waves, i.e. large
	   imaginary eigenvalues of the linear terms.}

 Three-dimensional problems concerned with the modeling of planetary interiors
 in global spherical geometry 
 are nowhere near reaching values of $\reynolds$ as extreme as~$10^6$. Recent
 parametric studies of convection-driven dynamo in a spherical shell geometry by 
 \cite{schwaiger2019force}, 
\cite{gastine2020dynamo} 
 and  
\cite{tassin2021geomagnetic}
 have values of $\reynolds$ in the range $100-3000$, while single-case, 
 rapidly-rotating trophy studies by 
 \cite{schaeffer2017turbulent,sheyko2018scale} (DNS) and \cite{aubert2019approaching} (LES)
 report $\reynolds \sim 5000$ and $\reynolds \sim 20000$ for the DNS and LES, respectively. 
 We may wonder how our findings can carry over to these simulations. In \ref{sec:pred3d}, 
 we attempt to convert our 2D tradeoff diagrams into their 3D counterparts, by making the
 strong 
 assumptions that accuracy and stability remain the same, i.e. that the convergence properties
 and values of $\alpha^{\mbox{\scriptsize \sc max}}_{\mbox{\scriptsize \sc cfl}}$ are not affected. 
 The sole impact of the change of geometry that factor in the analysis lies in the cost, where we acknowledge that in 
 a three-dimensional pseudo-spectral code of planetary core dynamics, explicit stages are the most
 expensive steps in a calculation, as there is no efficient equivalent of the fast Fourier transform
 for the Legendre transform. Taking this into account, non \addRtwo{globally} stiffly accurate IMEX-RK
	 schemes, that require an extra assembly stage, are penalized compared with 
 stiffly accurate ones. 

For the range of $\reynolds$ typical of recent parametric studies, and under the strong assumptions that we made, 
 we predict that the schemes of choice for $\reynolds \sim 1000$ in three dimensions
 should be PC432 (order 2), ARS343 and BPR533 (order~3) and KC664 (order~4). Relative to the classic CNAB2, 
 these schemes should yield
 a reduction of the time to solution while enabling more accurate solutions. 
 This analysis ignores the memory imprint of the schemes 
 (recall Tab.~\ref{tab:imexrk}) 
 , as the codes
 in question are massively parallel and should be immune to this issue, at least for the level of $\reynolds$
 considered. If the level of turbulence should increase for 3D simulations, then we anticipate 
 again that it is the robustness and efficiency of the explicit component of IMEX schemes that
 will matter. Also, depending on the problem studied, hybrid splitting 
 strategies, that would treat explicitly some linear terms with the
 aim of increasing the efficiency of the calculation, look like an interesting
 avenue for investigation, especially in turbulent situations. 
  Also worth investigating are the IMEX general
linear method (GLM) that have recently come to the fore. 
 They appear to overcome some of the inherent
limitations of IMEX-RK schemes, such as order reduction. \cite{zhang16high}
for instance explored several IMEX strategies
to compute 2-D and 3-D simulations of thermal rising bubbles \citep{giraldo2008study} 
and showed that
the IMEX-GLM were immune to order reduction and exhibited the best
accuracy/efficiency tradeoff among the tested schemes (see their Fig. 10). 
  We conclude by hoping that our findings will serve as an incentive for the community 
  of stars and planetary fluid interiors modelers to transition 
 towards IMEX-RK integrators, which possess the extra convenient property of being self-restarting (their
 initialization only 
 requires knowledge of the current state vector), regardless
 of their order of accuracy.



\section*{Acknowledgments}

We thank the two anonymous reviewers whose comments and suggestions helped improve this paper. 
This work was made possible by support from
Indo-French Centre for the Promotion of Advanced Research
CEFIPRA/IFCPAR (project 5307-1). Near the completion of the writing of this manuscript we were
saddened to learn the passing of Ms. A. Sathidevi,  senior scientific officer at CEFIPRA that had accompanied us
 over the course of this project, 
 and we would like
to dedicate this work to her memory.
We also acknowledge fruitful interactions over the long course of this work
with Prof. H.~Johnston,
Dr.~P.~Livermore,
Dr.~L.~M\'etivier,
Prof. D.~Reynolds,
Dr.~N.~Schaeffer
and Prof.~J.-P.~Vilotte.
Numerical computations were performed on the S-CAPAD platform, IPGP, France.

\bibliographystyle{model1-num-names}

\newcommand{\noop}[1]{}
\begin{thebibliography}{81}
\expandafter\ifx\csname natexlab\endcsname\relax\def\natexlab#1{#1}\fi
\providecommand{\url}[1]{\texttt{#1}}
\providecommand{\href}[2]{#2}
\providecommand{\path}[1]{#1}
\providecommand{\DOIprefix}{doi:}
\providecommand{\ArXivprefix}{arXiv:}
\providecommand{\URLprefix}{URL: }
\providecommand{\Pubmedprefix}{pmid:}
\providecommand{\doi}[1]{\href{http://dx.doi.org/#1}{\path{#1}}}
\providecommand{\Pubmed}[1]{\href{pmid:#1}{\path{#1}}}
\providecommand{\bibinfo}[2]{#2}
\ifx\xfnm\relax \def\xfnm[#1]{\unskip,\space#1}\fi
\bibitem[{McKenzie et~al.(1974)McKenzie, Roberts, and
  Weiss}]{mckenzie1974convection}
\bibinfo{author}{D.~P. McKenzie}, \bibinfo{author}{J.~M. Roberts},
  \bibinfo{author}{N.~O. Weiss},
\newblock \bibinfo{title}{Convection in the {E}arth's mantle: towards a
  numerical simulation},
\newblock \bibinfo{journal}{Journal of Fluid Mechanics} \bibinfo{volume}{62}
  (\bibinfo{year}{1974}) \bibinfo{pages}{465--538}.
\bibitem[{Sato and Thompson(1976)}]{sato1976finite}
\bibinfo{author}{A.~Sato}, \bibinfo{author}{E.~G. Thompson},
\newblock \bibinfo{title}{Finite element models for creeping convection},
\newblock \bibinfo{journal}{Journal of Computational Physics}
  \bibinfo{volume}{22} (\bibinfo{year}{1976}) \bibinfo{pages}{229--244}.
\bibitem[{Kopitzke(1979)}]{kopitzke1979finite}
\bibinfo{author}{U.~Kopitzke},
\newblock \bibinfo{title}{Finite element convection models: comparison of
  shallow and deep mantle convection, and temperatures in the mantle},
\newblock \bibinfo{journal}{Journal of Geophysics} \bibinfo{volume}{46}
  (\bibinfo{year}{1979}) \bibinfo{pages}{97--121}.
\bibitem[{Jarvis(1984)}]{jarvis1984time}
\bibinfo{author}{G.~T. Jarvis},
\newblock \bibinfo{title}{Time-dependent convection in the {E}arth's mantle},
\newblock \bibinfo{journal}{Physics of the Earth and Planetary Interiors}
  \bibinfo{volume}{36} (\bibinfo{year}{1984}) \bibinfo{pages}{305 -- 327}.
\bibitem[{Zhong et~al.(2015)Zhong, Yuen, Moresi, and
  Knepley}]{zhong2015numerical}
\bibinfo{author}{S.~J. Zhong}, \bibinfo{author}{D.~A. Yuen},
  \bibinfo{author}{L.~N. Moresi}, \bibinfo{author}{M.~G. Knepley},
\newblock \bibinfo{title}{Numerical methods for mantle convection},
\newblock in: \bibinfo{editor}{G.~Schubert} (Ed.), \bibinfo{booktitle}{Treatise
  on Geophysics}, \bibinfo{edition}{second edition} ed.,
  \bibinfo{publisher}{Elsevier}, \bibinfo{address}{Oxford},
  \bibinfo{year}{2015}, pp. \bibinfo{pages}{197--222}. \URLprefix
  \url{http://www.sciencedirect.com/science/article/pii/B9780444538024001305}.
  \DOIprefix\doi{10.1016/B978-0-444-53802-4.00130-5}.
\bibitem[{Davies et~al.(2011)Davies, Wilson, and Kramer}]{davies2011fluidity}
\bibinfo{author}{D.~R. Davies}, \bibinfo{author}{C.~R. Wilson},
  \bibinfo{author}{S.~C. Kramer},
\newblock \bibinfo{title}{Fluidity: A fully unstructured anisotropic adaptive
  mesh computational modeling framework for geodynamics},
\newblock \bibinfo{journal}{Geochemistry, Geophysics, Geosystems}
  \bibinfo{volume}{12} (\bibinfo{year}{2011}) \bibinfo{pages}{Q06001}.
\bibitem[{Kronbichler et~al.(2012)Kronbichler, Heister, and
  Bangerth}]{kronbichler2012high}
\bibinfo{author}{M.~Kronbichler}, \bibinfo{author}{T.~Heister},
  \bibinfo{author}{W.~Bangerth},
\newblock \bibinfo{title}{High accuracy mantle convection simulation through
  modern numerical methods},
\newblock \bibinfo{journal}{Geophysical Journal International}
  \bibinfo{volume}{191} (\bibinfo{year}{2012}) \bibinfo{pages}{12–29}.
\bibitem[{Canuto et~al.(2006)Canuto, Hussaini, Quarteroni, and Zang}]{chqz2006}
\bibinfo{author}{C.~Canuto}, \bibinfo{author}{M.~Y. Hussaini},
  \bibinfo{author}{A.~Quarteroni}, \bibinfo{author}{T.~A. Zang},
  \bibinfo{title}{Spectral methods: Fundamentals in Single Domains}, Scientific
  Computation, \bibinfo{publisher}{Springer}, \bibinfo{address}{Berlin
  Heidelberg}, \bibinfo{year}{2006}.
\bibitem[{Glatzmaier and Roberts(1995)}]{glatzmaier1995three}
\bibinfo{author}{G.~A. Glatzmaier}, \bibinfo{author}{P.~H. Roberts},
\newblock \bibinfo{title}{A three-dimensional convective dynamo solution with
  rotating and finitely conducting inner core and mantle},
\newblock \bibinfo{journal}{Physics of the Earth and Planetary Interiors}
  \bibinfo{volume}{91} (\bibinfo{year}{1995}) \bibinfo{pages}{63--75}.
\bibitem[{Kageyama and Sato(1995)}]{kageyama1995computer}
\bibinfo{author}{A.~Kageyama}, \bibinfo{author}{T.~Sato},
\newblock \bibinfo{title}{Computer simulation of a magnetohydrodynamic dynamo.
  {II}},
\newblock \bibinfo{journal}{Physics of Plasmas} \bibinfo{volume}{2}
  (\bibinfo{year}{1995}) \bibinfo{pages}{1421--1431}.
\bibitem[{Glatzmaier(1984)}]{glatzmaier1984numerical}
\bibinfo{author}{G.~A. Glatzmaier},
\newblock \bibinfo{title}{{Numerical simulations of stellar convective dynamos.
  I- The model and method}},
\newblock \bibinfo{journal}{Journal of Computational Physics}
  \bibinfo{volume}{55} (\bibinfo{year}{1984}) \bibinfo{pages}{461--484}.
\bibitem[{Clune et~al.(1999)Clune, Elliott, Miesch, Toomre, and
  Glatzmaier}]{clune1999computational}
\bibinfo{author}{T.~Clune}, \bibinfo{author}{J.~Elliott},
  \bibinfo{author}{M.~Miesch}, \bibinfo{author}{J.~Toomre},
  \bibinfo{author}{G.~Glatzmaier},
\newblock \bibinfo{title}{Computational aspects of a code to study rotating
  turbulent convection in spherical shells},
\newblock \bibinfo{journal}{Parallel Computing} \bibinfo{volume}{25}
  (\bibinfo{year}{1999}) \bibinfo{pages}{361–380}.
\bibitem[{Schaeffer(2013)}]{schaeffer2013efficient}
\bibinfo{author}{N.~Schaeffer},
\newblock \bibinfo{title}{Efficient spherical harmonic transforms aimed at
  pseudospectral numerical simulations},
\newblock \bibinfo{journal}{Geochemistry, Geophysics, Geosystems}
  \bibinfo{volume}{14} (\bibinfo{year}{2013}) \bibinfo{pages}{751--758}.
\bibitem[{Matsui et~al.(2016)Matsui, Heien, Aubert, Aurnou, Avery, Brown,
  Buffett, Busse, Christensen, Davies, Featherstone, Gastine, Glatzmaier,
  Gubbins, Guermond, Hayashi, Hollerbach, Hwang, Jackson, Jones, Jiang,
  Kellogg, Kuang, Landeau, Marti, Olson, Ribeiro, Sasaki, Schaeffer, Simitev,
  Sheyko, Silva, Stanley, Takahashi, Takehiro, Wicht, and
  Willis}]{matsui2016performance}
\bibinfo{author}{H.~Matsui}, \bibinfo{author}{E.~Heien},
  \bibinfo{author}{J.~Aubert}, \bibinfo{author}{J.~M. Aurnou},
  \bibinfo{author}{M.~Avery}, \bibinfo{author}{B.~Brown},
  \bibinfo{author}{B.~A. Buffett}, \bibinfo{author}{F.~Busse},
  \bibinfo{author}{U.~R. Christensen}, \bibinfo{author}{C.~J. Davies},
  \bibinfo{author}{N.~Featherstone}, \bibinfo{author}{T.~Gastine},
  \bibinfo{author}{G.~A. Glatzmaier}, \bibinfo{author}{D.~Gubbins},
  \bibinfo{author}{J.-L. Guermond}, \bibinfo{author}{Y.-Y. Hayashi},
  \bibinfo{author}{R.~Hollerbach}, \bibinfo{author}{L.~J. Hwang},
  \bibinfo{author}{A.~Jackson}, \bibinfo{author}{C.~A. Jones},
  \bibinfo{author}{W.~Jiang}, \bibinfo{author}{L.~H. Kellogg},
  \bibinfo{author}{W.~Kuang}, \bibinfo{author}{M.~Landeau},
  \bibinfo{author}{P.~Marti}, \bibinfo{author}{P.~Olson},
  \bibinfo{author}{A.~Ribeiro}, \bibinfo{author}{Y.~Sasaki},
  \bibinfo{author}{N.~Schaeffer}, \bibinfo{author}{R.~D. Simitev},
  \bibinfo{author}{A.~Sheyko}, \bibinfo{author}{L.~Silva},
  \bibinfo{author}{S.~Stanley}, \bibinfo{author}{F.~Takahashi},
  \bibinfo{author}{S.-i. Takehiro}, \bibinfo{author}{J.~Wicht},
  \bibinfo{author}{A.~P. Willis},
\newblock \bibinfo{title}{Performance benchmarks for a next generation
  numerical dynamo model},
\newblock \bibinfo{journal}{Geochemistry, Geophysics, Geosystems}
  \bibinfo{volume}{17} (\bibinfo{year}{2016}) \bibinfo{pages}{1586--1607}.
\bibitem[{{Hollerbach}(2000)}]{hollerbach2000spectral}
\bibinfo{author}{R.~{Hollerbach}},
\newblock \bibinfo{title}{{A spectral solution of the magneto-convection
  equations in spherical geometry}},
\newblock \bibinfo{journal}{International Journal for Numerical Methods in
  Fluids} \bibinfo{volume}{32} (\bibinfo{year}{2000})
  \bibinfo{pages}{773--797}.
\bibitem[{Tilgner(1999)}]{tilgner1999spectral}
\bibinfo{author}{A.~Tilgner},
\newblock \bibinfo{title}{Spectral methods for the simulation of incompressible
  flows in spherical shells},
\newblock \bibinfo{journal}{International Journal for Numerical Methods in
  Fluids} \bibinfo{volume}{30} (\bibinfo{year}{1999})
  \bibinfo{pages}{713--724}.
\bibitem[{Ascher et~al.(1997)Ascher, Ruuth, and Spiteri}]{ascher1997implicit}
\bibinfo{author}{U.~M. Ascher}, \bibinfo{author}{S.~J. Ruuth},
  \bibinfo{author}{R.~J. Spiteri},
\newblock \bibinfo{title}{Implicit-explicit {R}unge--{K}utta methods for
  time-dependent partial differential equations},
\newblock \bibinfo{journal}{Applied Numerical Mathematics} \bibinfo{volume}{25}
  (\bibinfo{year}{1997}) \bibinfo{pages}{151--167}.
\bibitem[{Ascher et~al.(1995)Ascher, Ruuth, and Wetton}]{ascher1995implicit}
\bibinfo{author}{U.~M. Ascher}, \bibinfo{author}{S.~J. Ruuth},
  \bibinfo{author}{B.~T.~R. Wetton},
\newblock \bibinfo{title}{Implicit-explicit methods for time-dependent partial
  differential equations},
\newblock \bibinfo{journal}{SIAM Journal on Numerical Analysis}
  \bibinfo{volume}{32} (\bibinfo{year}{1995}) \bibinfo{pages}{797--823}.
\bibitem[{Willis et~al.(2007)Willis, Sreenivasan, and
  Gubbins}]{willis2007thermal}
\bibinfo{author}{A.~P. Willis}, \bibinfo{author}{B.~Sreenivasan},
  \bibinfo{author}{D.~Gubbins},
\newblock \bibinfo{title}{Thermal core–mantle interaction: Exploring regimes
  for “locked” dynamo action},
\newblock \bibinfo{journal}{Physics of the Earth and Planetary Interiors}
  \bibinfo{volume}{165} (\bibinfo{year}{2007}) \bibinfo{pages}{83–92}.
\bibitem[{{Fournier} et~al.(2005){Fournier}, {Bunge}, {Hollerbach}, and
  {Vilotte}}]{fournier2005fourier}
\bibinfo{author}{A.~{Fournier}}, \bibinfo{author}{H.-P. {Bunge}},
  \bibinfo{author}{R.~{Hollerbach}}, \bibinfo{author}{J.-P. {Vilotte}},
\newblock \bibinfo{title}{{A Fourier-spectral element algorithm for thermal
  convection in rotating axisymmetric containers}},
\newblock \bibinfo{journal}{Journal of Computational Physics}
  \bibinfo{volume}{204} (\bibinfo{year}{2005}) \bibinfo{pages}{462--489}.
\bibitem[{Stellmach and Hansen(2008)}]{stellmach2008efficient}
\bibinfo{author}{S.~Stellmach}, \bibinfo{author}{U.~Hansen},
\newblock \bibinfo{title}{An efficient spectral method for the simulation of
  dynamos in {C}artesian geometry and its implementation on massively parallel
  computers},
\newblock \bibinfo{journal}{Geochemistry, Geophysics, Geosystems}
  \bibinfo{volume}{9} (\bibinfo{year}{2008}) \bibinfo{pages}{Q05003}.
\bibitem[{Verhoeven and Stellmach(2014)}]{verhoeven2014compressional}
\bibinfo{author}{J.~Verhoeven}, \bibinfo{author}{S.~Stellmach},
\newblock \bibinfo{title}{The compressional beta effect: A source of zonal
  winds in planets?},
\newblock \bibinfo{journal}{Icarus} \bibinfo{volume}{237}
  (\bibinfo{year}{2014}) \bibinfo{pages}{143--158}.
\bibitem[{Lecoanet et~al.(2019)Lecoanet, Vasil, Burns, Brown, and
  Oishi}]{lecoanet2019tensor}
\bibinfo{author}{D.~Lecoanet}, \bibinfo{author}{G.~M. Vasil},
  \bibinfo{author}{K.~J. Burns}, \bibinfo{author}{B.~P. Brown},
  \bibinfo{author}{J.~S. Oishi},
\newblock \bibinfo{title}{Tensor calculus in spherical coordinates using
  {J}acobi polynomials. part-{II}: {I}mplementation and examples},
\newblock \bibinfo{journal}{Journal of Computational Physics: X}
  \bibinfo{volume}{3} (\bibinfo{year}{2019}) \bibinfo{pages}{100012}.
\bibitem[{{Marti} et~al.(2014){Marti}, {Schaeffer}, {Hollerbach}, {C{\'e}bron},
  {Nore}, {Luddens}, {Guermond}, {Aubert}, {Takehiro}, {Sasaki}, {Hayashi},
  {Simitev}, {Busse}, {Vantieghem}, and {Jackson}}]{marti2014full}
\bibinfo{author}{P.~{Marti}}, \bibinfo{author}{N.~{Schaeffer}},
  \bibinfo{author}{R.~{Hollerbach}}, \bibinfo{author}{D.~{C{\'e}bron}},
  \bibinfo{author}{C.~{Nore}}, \bibinfo{author}{F.~{Luddens}},
  \bibinfo{author}{J.~L. {Guermond}}, \bibinfo{author}{J.~{Aubert}},
  \bibinfo{author}{S.~{Takehiro}}, \bibinfo{author}{Y.~{Sasaki}},
  \bibinfo{author}{Y.~Y. {Hayashi}}, \bibinfo{author}{R.~{Simitev}},
  \bibinfo{author}{F.~{Busse}}, \bibinfo{author}{S.~{Vantieghem}},
  \bibinfo{author}{A.~{Jackson}},
\newblock \bibinfo{title}{{Full sphere hydrodynamic and dynamo benchmarks}},
\newblock \bibinfo{journal}{Geophysical Journal International}
  \bibinfo{volume}{197} (\bibinfo{year}{2014}) \bibinfo{pages}{119--134}.
\bibitem[{Livermore(2007)}]{livermore2007implementation}
\bibinfo{author}{P.~W. Livermore},
\newblock \bibinfo{title}{An implementation of the exponential time
  differencing scheme to the magnetohydrodynamic equations in a spherical
  shell},
\newblock \bibinfo{journal}{Journal of Computational Physics}
  \bibinfo{volume}{220} (\bibinfo{year}{2007}) \bibinfo{pages}{824--838}.
\bibitem[{Garcia et~al.(2014)Garcia, Bonaventura, Net, and
  S{\'a}nchez}]{garcia2014exponential}
\bibinfo{author}{F.~Garcia}, \bibinfo{author}{L.~Bonaventura},
  \bibinfo{author}{M.~Net}, \bibinfo{author}{J.~S{\'a}nchez},
\newblock \bibinfo{title}{Exponential versus {IMEX} high-order time integrators
  for thermal convection in rotating spherical shells},
\newblock \bibinfo{journal}{Journal of Computational Physics}
  \bibinfo{volume}{264} (\bibinfo{year}{2014}) \bibinfo{pages}{41--54}.
\bibitem[{Garcia et~al.(2010)Garcia, Net, Garc\'{i}a-Archilla, and
  S{\'a}nchez}]{garcia2010comparison}
\bibinfo{author}{F.~Garcia}, \bibinfo{author}{M.~Net},
  \bibinfo{author}{B.~Garc\'{i}a-Archilla}, \bibinfo{author}{J.~S{\'a}nchez},
\newblock \bibinfo{title}{{A comparison of high-order time integrators for
  thermal convection in rotating spherical shells}},
\newblock \bibinfo{journal}{Journal of Computational Physics}
  \bibinfo{volume}{129} (\bibinfo{year}{2010}) \bibinfo{pages}{7997--8010}.
\bibitem[{Glatzmaier and Roberts(1996)}]{glatzmaier1996anelastic}
\bibinfo{author}{G.~A. Glatzmaier}, \bibinfo{author}{P.~H. Roberts},
\newblock \bibinfo{title}{An anelastic evolutionary geodynamo simulation driven
  by compositional and thermal convection},
\newblock \bibinfo{journal}{Physica D: Nonlinear Phenomena}
  \bibinfo{volume}{97} (\bibinfo{year}{1996}) \bibinfo{pages}{81--94}.
\bibitem[{Spalart et~al.(1991)Spalart, Moser, and Rogers}]{spalart1991spectral}
\bibinfo{author}{P.~R. Spalart}, \bibinfo{author}{R.~D. Moser},
  \bibinfo{author}{M.~M. Rogers},
\newblock \bibinfo{title}{Spectral methods for the {N}avier-{S}tokes equations
  with one infinite and two periodic directions},
\newblock \bibinfo{journal}{Journal of Computational Physics}
  \bibinfo{volume}{96} (\bibinfo{year}{1991}) \bibinfo{pages}{297--324}.
\bibitem[{Yan et~al.(2019)Yan, Calkins, Maffei, Julien, Tobias, and
  Marti}]{yan2019heat}
\bibinfo{author}{M.~Yan}, \bibinfo{author}{M.~A. Calkins},
  \bibinfo{author}{S.~Maffei}, \bibinfo{author}{K.~Julien},
  \bibinfo{author}{S.~M. Tobias}, \bibinfo{author}{P.~Marti},
\newblock \bibinfo{title}{Heat transfer and flow regimes in quasi-static
  magnetoconvection with a vertical magnetic field},
\newblock \bibinfo{journal}{Journal of Fluid Mechanics} \bibinfo{volume}{877}
  (\bibinfo{year}{2019}) \bibinfo{pages}{1186–1206}.
\bibitem[{Marti et~al.(2016)Marti, Calkins, and
  Julien}]{marti2016computationally}
\bibinfo{author}{P.~Marti}, \bibinfo{author}{M.~A. Calkins},
  \bibinfo{author}{K.~Julien},
\newblock \bibinfo{title}{A computationally efficient spectral method for
  modeling core dynamics},
\newblock \bibinfo{journal}{Geochemistry, Geophysics, Geosystems}
  \bibinfo{volume}{17} (\bibinfo{year}{2016}) \bibinfo{pages}{3031--3053}.
\bibitem[{{Cavaglieri} and {Bewley}(2015)}]{cavaglieri2015low}
\bibinfo{author}{D.~{Cavaglieri}}, \bibinfo{author}{T.~{Bewley}},
\newblock \bibinfo{title}{{Low-storage implicit/explicit Runge-Kutta schemes
  for the simulation of stiff high-dimensional ODE systems}},
\newblock \bibinfo{journal}{Journal of Computational Physics}
  \bibinfo{volume}{286} (\bibinfo{year}{2015}) \bibinfo{pages}{172--193}.
\bibitem[{Gastine(2019)}]{gastine2019pizza}
\bibinfo{author}{T.~Gastine},
\newblock \bibinfo{title}{pizza: an open-source pseudo-spectral code for
  spherical quasi-geostrophic convection},
\newblock \bibinfo{journal}{Geophysical Journal International}
  \bibinfo{volume}{217} (\bibinfo{year}{2019}) \bibinfo{pages}{1558--1576}.
\bibitem[{Boscarino et~al.(2013)Boscarino, Pareschi, and
  Russo}]{boscarino2013implicit}
\bibinfo{author}{S.~Boscarino}, \bibinfo{author}{L.~Pareschi},
  \bibinfo{author}{G.~Russo},
\newblock \bibinfo{title}{Implicit-explicit {R}unge--{K}utta schemes for
  hyperbolic systems and kinetic equations in the diffusion limit},
\newblock \bibinfo{journal}{{SIAM} Journal on Scientific Computing}
  \bibinfo{volume}{35} (\bibinfo{year}{2013}) \bibinfo{pages}{A22--A51}.
\bibitem[{Tassin et~al.(2021)Tassin, Gastine, and
  Fournier}]{tassin2021geomagnetic}
\bibinfo{author}{T.~Tassin}, \bibinfo{author}{T.~Gastine},
  \bibinfo{author}{A.~Fournier},
\newblock \bibinfo{title}{Geomagnetic semblance and dipolar-multipolar
  transition in top-heavy double-diffusive geodynamo models},
\newblock \bibinfo{journal}{Geophysical Journal International}
  \bibinfo{volume}{226} (\bibinfo{year}{2021}) \bibinfo{pages}{1897--1919}.
\bibitem[{Grooms and Julien(2011)}]{grooms2011linearly}
\bibinfo{author}{I.~Grooms}, \bibinfo{author}{K.~Julien},
\newblock \bibinfo{title}{Linearly implicit methods for nonlinear {PDE}s with
  linear dispersion and dissipation},
\newblock \bibinfo{journal}{Journal of Computational Physics}
  \bibinfo{volume}{230} (\bibinfo{year}{2011}) \bibinfo{pages}{3630--3650}.
\bibitem[{Boscarino and Russo(2009)}]{boscarino2009class}
\bibinfo{author}{S.~Boscarino}, \bibinfo{author}{G.~Russo},
\newblock \bibinfo{title}{On a class of uniformly accurate {IMEX}
  {R}unge--{K}utta schemes and applications to hyperbolic systems with
  relaxation},
\newblock \bibinfo{journal}{{SIAM} Journal on Scientific Computing}
  \bibinfo{volume}{31} (\bibinfo{year}{2009}) \bibinfo{pages}{1926--1945}.
\bibitem[{Vos et~al.(2011)Vos, Eskilsson, Bolis, Chun, Kirby, and
  Sherwin}]{vos2011generic}
\bibinfo{author}{P.~E. Vos}, \bibinfo{author}{C.~Eskilsson},
  \bibinfo{author}{A.~Bolis}, \bibinfo{author}{S.~Chun}, \bibinfo{author}{R.~M.
  Kirby}, \bibinfo{author}{S.~J. Sherwin},
\newblock \bibinfo{title}{A generic framework for time-stepping partial
  differential equations ({PDE}s): general linear methods, object-oriented
  implementation and application to fluid problems},
\newblock \bibinfo{journal}{International Journal of Computational Fluid
  Dynamics} \bibinfo{volume}{25} (\bibinfo{year}{2011})
  \bibinfo{pages}{107–125}.
\bibitem[{Giraldo et~al.(2013)Giraldo, Kelly, and
  Constantinescu}]{giraldo2013implicit}
\bibinfo{author}{F.~X. Giraldo}, \bibinfo{author}{J.~F. Kelly},
  \bibinfo{author}{E.~M. Constantinescu},
\newblock \bibinfo{title}{Implicit-explicit formulations of a three-dimensional
  nonhydrostatic unified model of the atmosphere ({NUMA})},
\newblock \bibinfo{journal}{SIAM Journal on Scientific Computing}
  \bibinfo{volume}{35} (\bibinfo{year}{2013}) \bibinfo{pages}{B1162–B1194}.
\bibitem[{Gardner et~al.(2018)Gardner, Guerra, Hamon, Reynolds, Ullrich, and
  Woodward}]{gardner2018implicit}
\bibinfo{author}{D.~J. Gardner}, \bibinfo{author}{J.~E. Guerra},
  \bibinfo{author}{F.~P. Hamon}, \bibinfo{author}{D.~R. Reynolds},
  \bibinfo{author}{P.~A. Ullrich}, \bibinfo{author}{C.~S. Woodward},
\newblock \bibinfo{title}{Implicit--explicit ({IMEX}) {R}unge--{K}utta methods
  for non-hydrostatic atmospheric models},
\newblock \bibinfo{journal}{Geoscientific Model Development}
  \bibinfo{volume}{11} (\bibinfo{year}{2018}) \bibinfo{pages}{1497--1515}.
\bibitem[{Vogl et~al.(2019)Vogl, Steyer, Reynolds, Ullrich, and
  Woodward}]{vogl2019evaluation}
\bibinfo{author}{C.~J. Vogl}, \bibinfo{author}{A.~Steyer},
  \bibinfo{author}{D.~R. Reynolds}, \bibinfo{author}{P.~A. Ullrich},
  \bibinfo{author}{C.~S. Woodward},
\newblock \bibinfo{title}{Evaluation of implicit-explicit additive
  {R}unge-{K}utta integrators for the {HOMME-NH} dynamical core},
\newblock \bibinfo{journal}{Journal of Advances in Modeling Earth Systems}
  \bibinfo{volume}{11} (\bibinfo{year}{2019}) \bibinfo{pages}{4228--4244}.
\bibitem[{Ullrich et~al.(2012)Ullrich, Jablonowski, Kent, Lauritzen, Nair, and
  Taylor}]{ullrich2012dynamical}
\bibinfo{author}{P.~A. Ullrich}, \bibinfo{author}{C.~Jablonowski},
  \bibinfo{author}{J.~Kent}, \bibinfo{author}{P.~H. Lauritzen},
  \bibinfo{author}{R.~D. Nair}, \bibinfo{author}{M.~A. Taylor},
\newblock \bibinfo{title}{Dynamical core model intercomparison project
  ({DCMIP}) test case document},
\newblock \bibinfo{journal}{DCMIP Summer School} \bibinfo{volume}{83}
  (\bibinfo{year}{2012}).
\bibitem[{Glatzmaier(2013)}]{glatzmaier2013introduction}
\bibinfo{author}{G.~A. Glatzmaier}, \bibinfo{title}{Introduction to Modeling
  Convection in Planets and Stars: Magnetic Field, Density Stratification,
  Rotation}, \bibinfo{publisher}{Princeton University Press},
  \bibinfo{year}{2013}.
\bibitem[{Peyret(2002)}]{peyret2002spectral}
\bibinfo{author}{R.~Peyret}, \bibinfo{title}{Spectral Methods for
  Incompressible Viscous Flow}, \bibinfo{publisher}{Springer New York},
  \bibinfo{year}{2002}. \URLprefix
  \url{http://dx.doi.org/10.1007/978-1-4757-6557-1}.
  \DOIprefix\doi{10.1007/978-1-4757-6557-1}.
\bibitem[{Plaut and Busse(2002)}]{plaut2002low}
\bibinfo{author}{E.~Plaut}, \bibinfo{author}{F.~H. Busse},
\newblock \bibinfo{title}{Low-{Prandtl}-number convection in a rotating
  cylindrical annulus},
\newblock \bibinfo{journal}{Journal of Fluid Mechanics} \bibinfo{volume}{464}
  (\bibinfo{year}{2002}) \bibinfo{pages}{345--363}.
\bibitem[{King et~al.(2012)King, Stellmach, and Aurnou}]{king2012heat}
\bibinfo{author}{E.~M. King}, \bibinfo{author}{S.~Stellmach},
  \bibinfo{author}{J.~M. Aurnou},
\newblock \bibinfo{title}{Heat transfer by rapidly rotating
  {Rayleigh–Bénard} convection},
\newblock \bibinfo{journal}{Journal of Fluid Mechanics} \bibinfo{volume}{691}
  (\bibinfo{year}{2012}) \bibinfo{pages}{568–582}.
\bibitem[{Press et~al.(2007)Press, Teukolsky, Vetterling, and
  Flannery}]{press2007numerical}
\bibinfo{author}{W.~H. Press}, \bibinfo{author}{S.~A. Teukolsky},
  \bibinfo{author}{W.~T. Vetterling}, \bibinfo{author}{B.~P. Flannery},
  \bibinfo{title}{Numerical recipes: The art of scientific computing},
  \bibinfo{edition}{3} ed., \bibinfo{publisher}{Cambridge university press},
  \bibinfo{year}{2007}.
\bibitem[{Ascher and Petzold(1998)}]{ascher1998computer}
\bibinfo{author}{U.~M. Ascher}, \bibinfo{author}{L.~R. Petzold},
  \bibinfo{title}{Computer Methods for Ordinary Differential Equations and
  Differential-Algebraic Equations}, \bibinfo{edition}{1st} ed.,
  \bibinfo{publisher}{Society for Industrial and Applied Mathematics},
  \bibinfo{address}{USA}, \bibinfo{year}{1998}.
\bibitem[{Hairer et~al.(1993)Hairer, N{\o}rsett, and
  Wanner}]{hairer1993solving}
\bibinfo{author}{E.~Hairer}, \bibinfo{author}{S.~P. N{\o}rsett},
  \bibinfo{author}{G.~Wanner}, \bibinfo{title}{Solving Ordinary Differential
  Equations I}, volume~\bibinfo{volume}{8} of \textit{\bibinfo{series}{Springer
  Series in Computational Mathematics}}, \bibinfo{edition}{2} ed.,
  \bibinfo{publisher}{Springer-Verlag Berlin Heidelberg}, \bibinfo{year}{1993}.
  \DOIprefix\doi{10.1007/978-3-540-78862-1}.
\bibitem[{Hairer and Wanner(1996)}]{hairer1996solving}
\bibinfo{author}{E.~Hairer}, \bibinfo{author}{G.~Wanner},
  \bibinfo{title}{Solving Ordinary Differential Equations {II}},
  volume~\bibinfo{volume}{14} of \textit{\bibinfo{series}{Springer Series in
  Computational Mathematics}}, \bibinfo{edition}{2} ed.,
  \bibinfo{publisher}{Springer-Verlag Berlin Heidelberg}, \bibinfo{year}{1996}.
  \DOIprefix\doi{10.1007/978-3-642-05221-7}.
\bibitem[{Kennedy and Carpenter(2003)}]{kennedy2003additive}
\bibinfo{author}{C.~A. Kennedy}, \bibinfo{author}{M.~H. Carpenter},
\newblock \bibinfo{title}{Additive {R}unge--{K}utta schemes for
  convection--diffusion--reaction equations},
\newblock \bibinfo{journal}{Applied Numerical Mathematics} \bibinfo{volume}{44}
  (\bibinfo{year}{2003}) \bibinfo{pages}{139--181}.
\bibitem[{Julien and Watson(2009)}]{julien2009efficient}
\bibinfo{author}{K.~Julien}, \bibinfo{author}{M.~Watson},
\newblock \bibinfo{title}{Efficient multi-dimensional solution of {PDE}s using
  {C}hebyshev spectral methods},
\newblock \bibinfo{journal}{Journal of Computational Physics}
  \bibinfo{volume}{228} (\bibinfo{year}{2009}) \bibinfo{pages}{1480--1503}.
\bibitem[{Boscarino(2007)}]{boscarino2007error}
\bibinfo{author}{S.~Boscarino},
\newblock \bibinfo{title}{Error analysis of imex runge–kutta methods derived
  from differential-algebraic systems},
\newblock \bibinfo{journal}{SIAM Journal on Numerical Analysis}
  \bibinfo{volume}{45} (\bibinfo{year}{2007}) \bibinfo{pages}{1600–1621}.
\bibitem[{Pareschi and Russo(2005)}]{pareschi2005implicit}
\bibinfo{author}{L.~Pareschi}, \bibinfo{author}{G.~Russo},
\newblock \bibinfo{title}{Implicit–explicit runge–kutta schemes and
  applications to hyperbolic systems with relaxation},
\newblock \bibinfo{journal}{Journal of Scientific Computing}
  \bibinfo{volume}{25} (\bibinfo{year}{2005}) \bibinfo{pages}{129–155}.
\bibitem[{Jameson et~al.(1981)Jameson, Schmidt, and
  Turkel}]{jameson1981numerical}
\bibinfo{author}{A.~Jameson}, \bibinfo{author}{W.~Schmidt},
  \bibinfo{author}{E.~Turkel}, \bibinfo{title}{Numerical solution of the
  {E}uler equations by finite volume methods using {R}unge {K}utta time
  stepping schemes}, \bibinfo{year}{1981}. \URLprefix
  \url{https://arc.aiaa.org/doi/abs/10.2514/6.1981-1259}.
  \DOIprefix\doi{10.2514/6.1981-1259}.
  \href{http://arxiv.org/abs/https://arc.aiaa.org/doi/pdf/10.2514/6.1981-1259}{\tt
  arXiv:https://arc.aiaa.org/doi/pdf/10.2514/6.1981-1259}.
\bibitem[{Boscarino et~al.(2017)Boscarino, Pareschi, and
  Russo}]{boscarino2017unified}
\bibinfo{author}{S.~Boscarino}, \bibinfo{author}{L.~Pareschi},
  \bibinfo{author}{G.~Russo},
\newblock \bibinfo{title}{A unified {IMEX} {R}unge--{K}utta approach for
  hyperbolic systems with multiscale relaxation},
\newblock \bibinfo{journal}{SIAM Journal on Numerical Analysis}
  \bibinfo{volume}{55} (\bibinfo{year}{2017}) \bibinfo{pages}{2085--2109}.
\bibitem[{Boscarino and Russo(2007)}]{boscarino2007uniform}
\bibinfo{author}{S.~Boscarino}, \bibinfo{author}{G.~Russo},
\newblock \bibinfo{title}{{On the uniform accuracy of IMEX Runge-Kutta schemes
  and applications to hyperbolic systems with relaxation}},
\newblock in: \bibinfo{booktitle}{Proceedings of SIMAI2006 VIII Convegno SIMAI
  Ragusa (Italy), May 2006}, Communications to SIMAI Conferences,
  \bibinfo{publisher}{Society for Industrial and Applied Mathematics},
  \bibinfo{address}{Philadelphia, PA}, \bibinfo{year}{2007}.
  \DOIprefix\doi{10.1685/CSC06028}.
\bibitem[{Calvo et~al.(2001)Calvo, de~Frutos, and Novo}]{calvo2001linearly}
\bibinfo{author}{M.~P. Calvo}, \bibinfo{author}{J.~de~Frutos},
  \bibinfo{author}{J.~Novo},
\newblock \bibinfo{title}{Linearly implicit {Runge--Kutta} methods for
  advection--reaction--diffusion equations},
\newblock \bibinfo{journal}{Applied Numerical Mathematics} \bibinfo{volume}{37}
  (\bibinfo{year}{2001}) \bibinfo{pages}{535--549}.
\bibitem[{Kinnmark and Gray(1984)}]{kinnmark1984one}
\bibinfo{author}{I.~P. Kinnmark}, \bibinfo{author}{W.~G. Gray},
\newblock \bibinfo{title}{One step integration methods of third-fourth order
  accuracy with large hyperbolic stability limits},
\newblock \bibinfo{journal}{Mathematics and Computers in Simulation}
  \bibinfo{volume}{26} (\bibinfo{year}{1984}) \bibinfo{pages}{181–188}.
\bibitem[{Liu and Zou(2006)}]{liu2006additive}
\bibinfo{author}{H.~Liu}, \bibinfo{author}{J.~Zou},
\newblock \bibinfo{title}{Some new additive {Runge–Kutta} methods and their
  applications},
\newblock \bibinfo{journal}{Journal of Computational and Applied Mathematics}
  \bibinfo{volume}{190} (\bibinfo{year}{2006}) \bibinfo{pages}{74 -- 98}.
  \bibinfo{note}{Special Issue: International Conference on Mathematics and its
  Application}.
\bibitem[{Kennedy and Carpenter(2019)}]{kennedy2019higher}
\bibinfo{author}{C.~A. Kennedy}, \bibinfo{author}{M.~H. Carpenter},
\newblock \bibinfo{title}{Higher-order additive {R}unge--{K}utta schemes for
  ordinary differential equations},
\newblock \bibinfo{journal}{Applied Numerical Mathematics}
  \bibinfo{volume}{136} (\bibinfo{year}{2019}) \bibinfo{pages}{183--205}.
\bibitem[{Johnston and Doering(2009)}]{johnston2009comparison}
\bibinfo{author}{H.~Johnston}, \bibinfo{author}{C.~R. Doering},
\newblock \bibinfo{title}{Comparison of turbulent thermal convection between
  conditions of constant temperature and constant flux},
\newblock \bibinfo{journal}{Physical review letters} \bibinfo{volume}{102}
  (\bibinfo{year}{2009}) \bibinfo{pages}{064501}.
\bibitem[{Frigo and Johnson(2005)}]{fftw3}
\bibinfo{author}{M.~Frigo}, \bibinfo{author}{S.~G. Johnson},
\newblock \bibinfo{title}{The design and implementation of fftw3},
\newblock \bibinfo{journal}{Proceedings of the IEEE} \bibinfo{volume}{93}
  (\bibinfo{year}{2005}) \bibinfo{pages}{216--231}.
\bibitem[{Anderson et~al.(1999)Anderson, Bai, Bischof, Blackford, Dongarra,
  Croz, Greenbaum, Hammarling, McKenney, and Sorensen}]{anderson1999lapack}
\bibinfo{author}{E.~Anderson}, \bibinfo{author}{Z.~Bai},
  \bibinfo{author}{C.~Bischof}, \bibinfo{author}{S.~Blackford},
  \bibinfo{author}{J.~D.~J. Dongarra}, \bibinfo{author}{J.~D. Croz},
  \bibinfo{author}{A.~Greenbaum}, \bibinfo{author}{S.~Hammarling},
  \bibinfo{author}{A.~McKenney}, \bibinfo{author}{D.~Sorensen},
  \bibinfo{title}{LAPACK Users' Guide}, \bibinfo{edition}{third} ed.,
  \bibinfo{publisher}{SIAM}, \bibinfo{address}{Philadelphia, Pennsylvania,
  USA}, \bibinfo{year}{1999}.
\bibitem[{Alonso et~al.(2000)Alonso, S\'{a}nchez, and
  Net}]{alonso2000transition}
\bibinfo{author}{A.~Alonso}, \bibinfo{author}{J.~S\'{a}nchez},
  \bibinfo{author}{M.~Net},
\newblock \bibinfo{title}{Transition to temporal chaos in an
  $\mathcal{O}(2)$-symmetric convective system for low {Prandtl} numbers},
\newblock \bibinfo{journal}{Progress of Theoretical Physics Supplement}
  \bibinfo{volume}{139} (\bibinfo{year}{2000}) \bibinfo{pages}{315--324}.
\bibitem[{Karniadakis et~al.(1991)Karniadakis, Israeli, and
  Orszag}]{karniadakis1991high}
\bibinfo{author}{G.~E. Karniadakis}, \bibinfo{author}{M.~Israeli},
  \bibinfo{author}{S.~A. Orszag},
\newblock \bibinfo{title}{High-order splitting methods for the incompressible
  {Navier-Stokes} equations},
\newblock \bibinfo{journal}{Journal of Computational Physics}
  \bibinfo{volume}{97} (\bibinfo{year}{1991}) \bibinfo{pages}{414–443}.
\bibitem[{Gaspari and Cohn(1999)}]{gaspari1999construction}
\bibinfo{author}{G.~Gaspari}, \bibinfo{author}{S.~E. Cohn},
\newblock \bibinfo{title}{Construction of correlation functions in two and
  three dimensions},
\newblock \bibinfo{journal}{Quarterly Journal of the Royal Meteorological
  Society} \bibinfo{volume}{125} (\bibinfo{year}{1999})
  \bibinfo{pages}{723--757}.
\bibitem[{Canuto et~al.(2007)Canuto, Hussaini, Quarteroni, and Zang}]{chqz2007}
\bibinfo{author}{C.~Canuto}, \bibinfo{author}{M.~Y. Hussaini},
  \bibinfo{author}{A.~Quarteroni}, \bibinfo{author}{T.~A. Zang},
  \bibinfo{title}{Spectral methods: Evolution to Complex Geometries and
  Applications to Fluid Dynamics}, Scientific Computation,
  \bibinfo{publisher}{Springer}, \bibinfo{address}{Berlin Heidelberg},
  \bibinfo{year}{2007}.
\bibitem[{Virtanen et~al.(2020)Virtanen, Gommers, Oliphant, Haberland, Reddy,
  Cournapeau, Burovski, Peterson, Weckesser, Bright et~al.}]{virtanen2020scipy}
\bibinfo{author}{P.~Virtanen}, \bibinfo{author}{R.~Gommers},
  \bibinfo{author}{T.~E. Oliphant}, \bibinfo{author}{M.~Haberland},
  \bibinfo{author}{T.~Reddy}, \bibinfo{author}{D.~Cournapeau},
  \bibinfo{author}{E.~Burovski}, \bibinfo{author}{P.~Peterson},
  \bibinfo{author}{W.~Weckesser}, \bibinfo{author}{J.~Bright}, et~al.,
\newblock \bibinfo{title}{{SciPy 1.0}: fundamental algorithms for scientific
  computing in python},
\newblock \bibinfo{journal}{Nature methods} \bibinfo{volume}{17}
  (\bibinfo{year}{2020}) \bibinfo{pages}{261--272}.
\bibitem[{Hernandez et~al.(2005)Hernandez, Roman, and
  Vidal}]{hernandez2005slepc}
\bibinfo{author}{V.~Hernandez}, \bibinfo{author}{J.~E. Roman},
  \bibinfo{author}{V.~Vidal},
\newblock \bibinfo{title}{{SLEPc}: A scalable and flexible toolkit for the
  solution of eigenvalue problems},
\newblock \bibinfo{journal}{ACM Trans. Math. Softw.} \bibinfo{volume}{31}
  (\bibinfo{year}{2005}) \bibinfo{pages}{351–362}.
\bibitem[{Dalcin et~al.(2011)Dalcin, Paz, Kler, and
  Cosimo}]{dalcin2011parallel}
\bibinfo{author}{L.~D. Dalcin}, \bibinfo{author}{R.~R. Paz},
  \bibinfo{author}{P.~A. Kler}, \bibinfo{author}{A.~Cosimo},
\newblock \bibinfo{title}{Parallel distributed computing using {P}ython},
\newblock \bibinfo{journal}{Advances in Water Resources} \bibinfo{volume}{34}
  (\bibinfo{year}{2011}) \bibinfo{pages}{1124--1139}. \bibinfo{note}{New
  Computational Methods and Software Tools}.
\bibitem[{Izzo and Jackiewicz(2017)}]{izzo2017highly}
\bibinfo{author}{G.~Izzo}, \bibinfo{author}{Z.~Jackiewicz},
\newblock \bibinfo{title}{Highly stable implicit–explicit {Runge–Kutta}
  methods},
\newblock \bibinfo{journal}{Applied Numerical Mathematics}
  \bibinfo{volume}{113} (\bibinfo{year}{2017}) \bibinfo{pages}{71–92}.
\bibitem[{Kosloff and Tal-Ezer(1993)}]{kosloff1993modified}
\bibinfo{author}{D.~Kosloff}, \bibinfo{author}{H.~Tal-Ezer},
\newblock \bibinfo{title}{A modified {C}hebyshev pseudospectral method with an
  $o(n-1)$ time step restriction},
\newblock \bibinfo{journal}{Journal of Computational Physics}
  \bibinfo{volume}{104} (\bibinfo{year}{1993}) \bibinfo{pages}{457–469}.
\bibitem[{Schwaiger et~al.(2019)Schwaiger, Gastine, and
  Aubert}]{schwaiger2019force}
\bibinfo{author}{T.~Schwaiger}, \bibinfo{author}{T.~Gastine},
  \bibinfo{author}{J.~Aubert},
\newblock \bibinfo{title}{Force balance in numerical geodynamo simulations: a
  systematic study},
\newblock \bibinfo{journal}{Geophysical Journal International}
  \bibinfo{volume}{219} (\bibinfo{year}{2019}) \bibinfo{pages}{S101--S114}.
\bibitem[{Gastine et~al.(2020)Gastine, Aubert, and
  Fournier}]{gastine2020dynamo}
\bibinfo{author}{T.~Gastine}, \bibinfo{author}{J.~Aubert},
  \bibinfo{author}{A.~Fournier},
\newblock \bibinfo{title}{{Dynamo-based limit to the extent of a stable layer
  atop Earth’s core}},
\newblock \bibinfo{journal}{Geophysical Journal International}
  \bibinfo{volume}{222} (\bibinfo{year}{2020}) \bibinfo{pages}{1433--1448}.
\bibitem[{Schaeffer et~al.(2017)Schaeffer, Jault, Nataf, and
  Fournier}]{schaeffer2017turbulent}
\bibinfo{author}{N.~Schaeffer}, \bibinfo{author}{D.~Jault},
  \bibinfo{author}{H.-C. Nataf}, \bibinfo{author}{A.~Fournier},
\newblock \bibinfo{title}{Turbulent geodynamo simulations: a leap towards
  {E}arth's core},
\newblock \bibinfo{journal}{Geophysical Journal International}
  \bibinfo{volume}{211} (\bibinfo{year}{2017}) \bibinfo{pages}{1--29}.
\bibitem[{Sheyko et~al.(2018)Sheyko, Finlay, Favre, and
  Jackson}]{sheyko2018scale}
\bibinfo{author}{A.~Sheyko}, \bibinfo{author}{C.~Finlay},
  \bibinfo{author}{J.~Favre}, \bibinfo{author}{A.~Jackson},
\newblock \bibinfo{title}{Scale separated low viscosity dynamos and dissipation
  within the {E}arth's core},
\newblock \bibinfo{journal}{Scientific Reports} \bibinfo{volume}{8}
  (\bibinfo{year}{2018}).
\bibitem[{Aubert(2019)}]{aubert2019approaching}
\bibinfo{author}{J.~Aubert},
\newblock \bibinfo{title}{Approaching {E}arth's core conditions in
  high-resolution geodynamo simulations},
\newblock \bibinfo{journal}{Geophysical Journal International}
  \bibinfo{volume}{219} (\bibinfo{year}{2019}) \bibinfo{pages}{S137–S151}.
\bibitem[{Zhang et~al.(2016)Zhang, Sandu, and Blaise}]{zhang16high}
\bibinfo{author}{H.~Zhang}, \bibinfo{author}{A.~Sandu},
  \bibinfo{author}{S.~Blaise},
\newblock \bibinfo{title}{High order implicit-explicit general linear methods
  with optimized stability regions},
\newblock \bibinfo{journal}{{SIAM} Journal on Scientific Computing}
  \bibinfo{volume}{38} (\bibinfo{year}{2016}) \bibinfo{pages}{1430--1453}.
\bibitem[{Giraldo and Restelli(2008)}]{giraldo2008study}
\bibinfo{author}{F.~X. Giraldo}, \bibinfo{author}{M.~Restelli},
\newblock \bibinfo{title}{A study of spectral element and discontinuous
  {G}alerkin methods for the {Navier–Stokes} equations in nonhydrostatic
  mesoscale atmospheric modeling: {E}quation sets and test cases},
\newblock \bibinfo{journal}{Journal of Computational Physics}
  \bibinfo{volume}{227} (\bibinfo{year}{2008}) \bibinfo{pages}{3849–3877}.
\bibitem[{Wang and Ruuth(2008)}]{wang2008variable}
\bibinfo{author}{D.~Wang}, \bibinfo{author}{S.~J. Ruuth},
\newblock \bibinfo{title}{Variable step-size implicit-explicit linear multistep
  methods for time-dependent partial differential equations},
\newblock \bibinfo{journal}{Journal of Computational Mathematics}
  \bibinfo{volume}{26} (\bibinfo{year}{2008}) \bibinfo{pages}{838--855}.

\end{thebibliography}

 \newcommand{\noop}[1]{}

\begin{appendix}

\section{Multistep schemes}
\label{app:multistep}
We provide the reader with the 
vectors of coefficients $\mathbf{a}$, $\mathbf{b}$ and $\mathbf{c}$ 
for CNAB2, SBDF2, SBDF3 and SBDF4. These vectors define 
    a linear multistep method with $K$ steps according to
\begin{equation}
\left(1 - \Delta tc_{-1} \linop \right) \statev_{i+1}
  = 
\sum^{K-1}_{j=0}\left[
a_j \statev_{i-j} + 
\Delta t b_j \nonlinop(\statev_{i-j}) 
+ \Delta t c_j \linop\statev_{i-j}\right],
\end{equation}
where $c_{-1} \neq 0$. Table~\ref{tab:multistep} enlists these three 
vectors when a fixed time step size $\Delta t$ is used \citep[see 
e.g.][Table~4.4]{peyret2002spectral}.

\begin{table}
\centering
\caption{Coefficients of the multistep schemes considered in this study 
 when a fixed time step size $\Delta t$ is employed.}
\begin{tabular}{lcrrr} \hline
method & $K$ &  $\mathbf{a}$ & $\mathbf{b}$ & $\mathbf{c}$ \\ \hline 
CNAB2  &  2  &               &  $[3/2,-1/2]$ &   $[1/2,1/2,0]$\\
SBDF2  &  2  &  $[4/3,-1/3]$ &  $[4/3,-2/3]$ &   $[2/3,0,0]$  \\
SBDF3  &  3  &  $[18/11,-9/11,2/11]$ & $[18/11,-18/11,6/11]$ & $[6/11,0,0,0]$ 
\\
SBDF4  &  4  &  $[48/25,-36/25,16/25,-3/25]$ & $[48/25,-72/25,48/25,-12/25]$ 
&$[12/25,0,0,0,0]$  \\
\hline
\end{tabular}
\label{tab:multistep}
\end{table}

Following \cite{wang2008variable}, those vectors can be generalised to 
the case of variable time step sizes. In the following, we define 
\[
 \Delta t_i=t_i-t_{i-1},
\]
and the ratio

\[
 \delta_i =\dfrac{\Delta t_i}{\Delta t_{i-1}}\,.
\]
 For the different IMEX multistep schemes considered here, we then obtain: 

\begin{itemize}
 \item CNAB2
\[
 \mathbf{b}=\left[1+\dfrac{1}{2}\delta_i, 
-\dfrac{1}{2}\delta_i\right], \ 
\mathbf{c}=\left[\dfrac{1}{2},\dfrac{1}{2}\right],
\]
 \item SBDF2
\[ 
\mathbf{a}=\left[\dfrac{(1+\delta_i)^2}{1+2\delta_i},-\dfrac{\delta_i^2}{
1+2\delta_i} \right],\ \mathbf{b}=\left[\dfrac{(1+\delta_i)^2}{1+2\delta_i},
-\dfrac{(1+\delta_i)\delta_i}{1+2\delta_i
} \right], \  
\mathbf{c}=\left[\dfrac{1+\delta_i}{1+2\delta_i}, 
0,0\right ],
\]
 \item SBDF3
 \[
 \begin{aligned}
\mathbf{a}& 
=\dfrac{1}{\alpha}\left[\dfrac{(1+\delta_i)(1+\delta_i+\delta_{i-1})} 
{\delta_i(\delta_i+\delta_{i-1})}, 
-\dfrac{1+\delta_i+\delta_{i-1}}{\delta_i\delta_{i-1}(1+\delta_i)},
\dfrac{1+\delta_i}{\delta_{i-1}(\delta_i+\delta_{i-1})    
(1+\delta_i+\delta_{i-1})}\right ], \\
\mathbf{b} & =\dfrac{1}{\alpha}
\left[\dfrac{(1+\delta_i)(1+\delta_i+\delta_{i-1})}{\delta_i    
(\delta_i+\delta_{i-1})}, 
-\dfrac{1+\delta_i+\delta_{i-1}}{\delta_i\delta_{i-1}},
\dfrac{1+\delta_i}{\delta_{i-1}(\delta_i+\delta_{i-1})}\right], \\
\mathbf{c} & = \left[\dfrac{1}{\alpha},0,0,0\right], \quad \text{with} \quad
\alpha=1+\dfrac{1}{1+\delta_i}+\dfrac{1}{1+\delta_i+\delta_{i-1}},
\end{aligned}
 \]
 \item SBDF4
 \[
 \begin{aligned}
  \mathbf{a}=&\dfrac{1}{\alpha}\left[
  1+\delta_i\left(1+\dfrac{\delta_{i-1}(1+\delta_i)(1+\delta_{i-2}c_2/c_1)}{
1+\delta_{i-1}}\right), -\delta_i\left( 
\dfrac{\delta_i}{1+\delta_i}+\dfrac{\delta_{i-1}\delta_i
(c_3+\delta_{i-2})}{1+\delta_{i-2}}\right), \right. \\
& \left.
\delta_{i-1}^3\delta_i^2\dfrac{1+\delta_i}{1+\delta_{i-1}}\dfrac{c_3}{c_2},
-\dfrac{1+\delta_i}{1+\delta_{i-2}} \dfrac{c_2}{c_1 c_3}\delta_{i-2}^4
            \delta_{i-1}^3\delta_i^2
  \right], \\
  \mathbf{b} = &\dfrac{1}{\alpha}\left[ 
  \delta_{i-1}\dfrac{1+\delta_i}{1+\delta_{i-1}} \dfrac{(1+\delta_i)
        (c_3+\delta_{i-2})+(1+\delta_{i-2})/\delta_{i-1}}{c_1},
        -c_2c_3  \dfrac{\delta_i}{1+\delta_{i-2}},
        c_3\delta_{i-1}^2 \delta_i\dfrac{1+\delta_i}{1+\delta_{i-1}},\right. \\
      &  \left. 
-\delta_{i-2}^3\delta_{i-1}^2\delta_i\dfrac{1+\delta_i}{1+\delta_{i-2}}
            \dfrac{c_2}{c_1}
  \right], \\
  \mathbf{c}= &\left[\dfrac{1}{\alpha},0,0,0,0\right],\quad \text{with} \quad
  \alpha = 1 + 
\dfrac{\delta_i}{1+\delta_i}+\dfrac{\delta_{i-1}\delta_i}{c_2}+\dfrac{
\delta_{i-2}\delta_{i-1}\delta_i}{c_3},
  \end{aligned}
 \]
where the three constants $c_1$, $c_2$ and $c_3$ are expressed by
\[
 c_1 = 1+\delta_{i-2}(1+\delta_{i-1}), \ 
 c_2 = 1+\delta_{i-1}(1+\delta_i), \ 
 c_3 = 1+\delta_{i-2}c_2\,.
\]
\end{itemize}

\section{Butcher tableaux of PC432}
\label{app:pc2}

{%
The PC432 time scheme is assembled using the explicit scheme from 
\cite{jameson1981numerical} for its explicit component and a Crank-Nicolson 
scheme for its implicit part. This is a stiffly accurate second order three 
stage scheme and its Butcher tableaux read}

%
%
\begin{equation}
\begin{array}{c|c}
  \mathbf{c}^E & \mathbf{A}^E \\
  \hline & \mathbf{b}^E
\end{array}
=
\begin{array}{c|cccc}
0 & 0  &  & & \\
1 &  1 & 0  &  & \\
1 &  1/2  &    1/2 & 0 & \\ 
1 &  1/2 & 0 &    1/2 & 0 \\ \hline
      &  1/2 & 0  &  1/2 & 0
\end{array},
\quad 
\begin{array}{c|c}
  \mathbf{c}^I & \mathbf{A}^I \\
  \hline & \mathbf{b}^I
\end{array}
=
\begin{array}{c|cccc}
0 &  0 &  & & \\
1 &  1/2 & 1/2  &  & \\
1 &  1/2 & 0 &   1/2 & \\
1 &  1/2 & 0 & 0 &  1/2 \\ \hline
      &  1/2 & 0  & 0 & 1/2
\end{array}.
\end{equation}

\section{Convergence of explicit Runge--Kutta schemes}
\label{app:rk}
\begin{figure}
\centerline{\includegraphics[width=\linewidth]{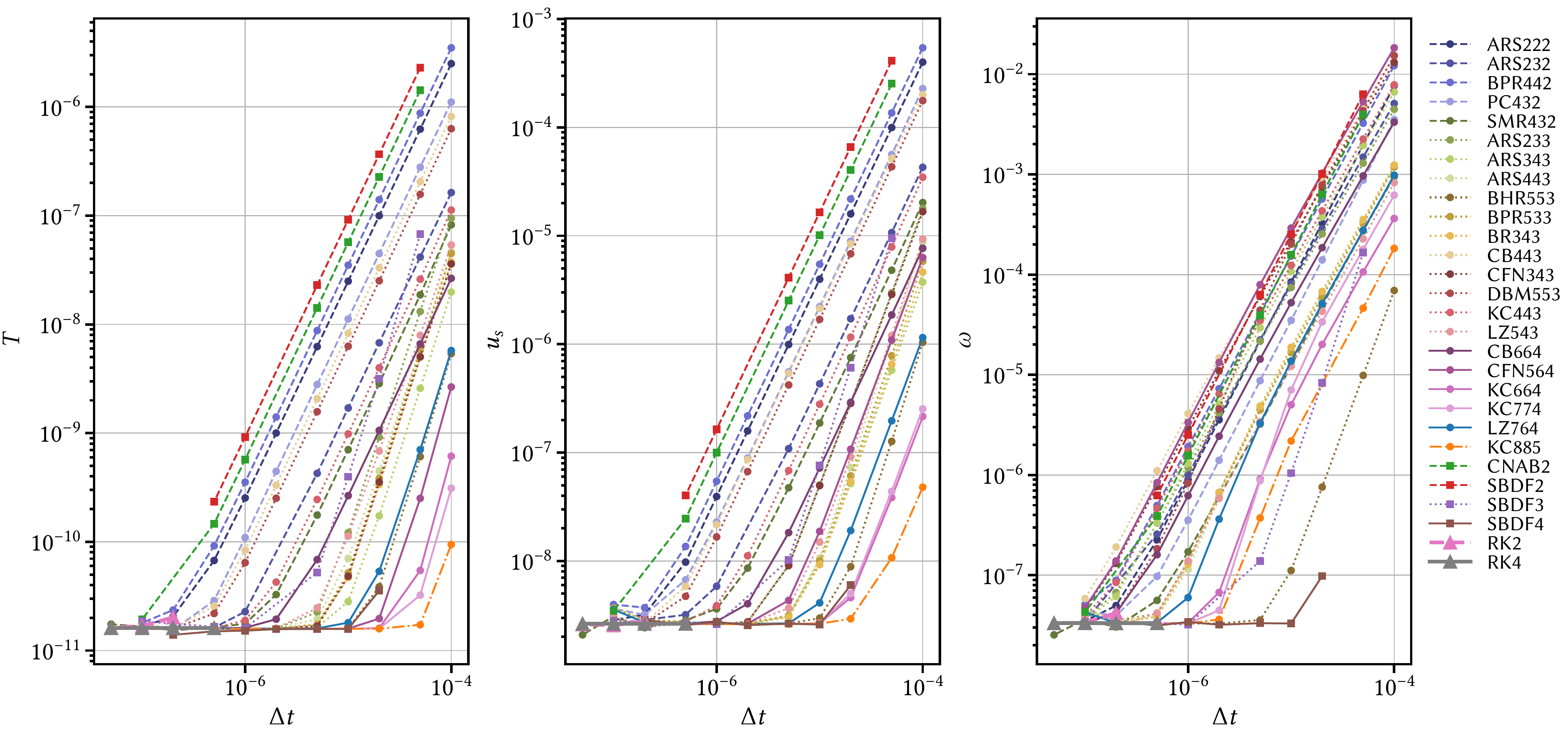}}
\caption{Convergence of the $\mathcal{L}^2$ error for the temperature field
(left panel), the cylindrical radial velocity $u_s$ (middle panel) and the
vorticity $\omega$ (right panel) for Case~3. In addition to the usual 26 
IMEX schemes, this figure also features the fully explicit RK2 and RK4 
methods, whose error levels are marked by large triangles which appear 
in the bottom left corner of each panel.}
\label{fig:error_case3} 
\end{figure}

Our software can also operate in a fully explicit fashion. Convergence results
obtained for case 3 that feature the RK2 and RK4 schemes are shown in 
Figure~\ref{fig:error_case3}, using large triangles located in the bottom left corner of 
each panel.  
For these schemes, the more stringent stability 
requirements due to the explicit treatment of the diffusion terms imply 
that the convergence curves directly land on the roundoff error level plateau. 
It is hence not possible to assess their convergence rates. We 
finally note that RK4 allows larger
values of the time step size $\Delta t$ than RK2.

\section{Stability regions}
\label{app:stab}
\begin{figure}[t!]
\centerline{\includegraphics[width=\linewidth]{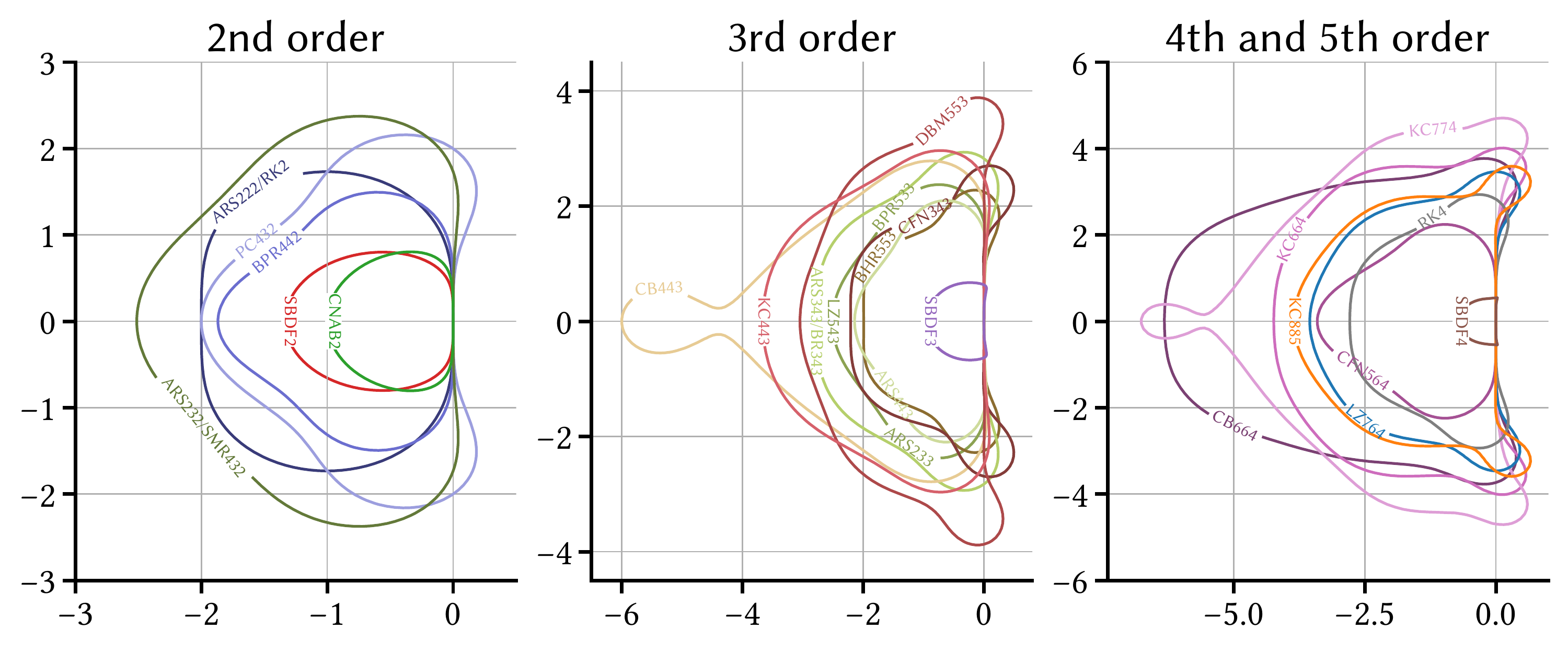}}
\caption{Stability regions of the explicit components of the 22 IMEX-RK, the 4 
multistep and the 2 fully explicit schemes analyzed in this study.
Schemes are stable inside the domains of the complex plane delimited by the curves. Left panel: order~2 schemes.
Middle panel: order~3 schemes. Right panel: schemes of order 4 and 5. Order refers to the expected order
of convergence of the combined IMEX schemes. Note that several schemes 
share the same stability domain for their explicit component: namely ARS222 and 
RK2; ARS232 and SMR432; ARS343, BR343 and RK4; AR233, BPR533 and LZ543.}
\label{fig:stability_regions}
\end{figure}

In this section we give for completeness the stability regions of the explicit components of the
22 implicit explicit Runge--Kutta schemes, of the 4 IMEX multistep and of 
the two fully explicit schemes considered in this study. For an easier 
visual inspection of the stability domains, Fig.~\ref{fig:stability_regions} 
has been split in three panels which gather the different expected orders of 
convergence of the combined IMEX schemes.

\addRone{
\section{Error as a function of time to solution for cases 2 and 10}
\label{sec:std_eff}
We provide in this appendix additional elements
 to assess the efficiency of the 26 schemes of interest in this study, 
 providing metrics that may be  more general than the dissipation-based efficiency
 introduced in Section~\ref{sec:stab_analysis}.  
Figure~\ref{fig:error_runtime} shows error against runtime for
cases 2 and 10, whose convergence is analyzed in Section~\ref{sec:accuracy}. 
This figure complements Figure~\ref{fig:er_2_10} that displays error versus
time step size $\Delta t$. Without getting into too much detail, we can stress 
that ARS343 appears as a good choice for case 2 and case 10. To obtain higher accuracy with
a concomittent moderate increase in the computational cost, SBDF4 (for case 2)
and KC664 (for case 10) should be preferred. 
\begin{figure}
\centerline{\includegraphics[width=.7\linewidth]{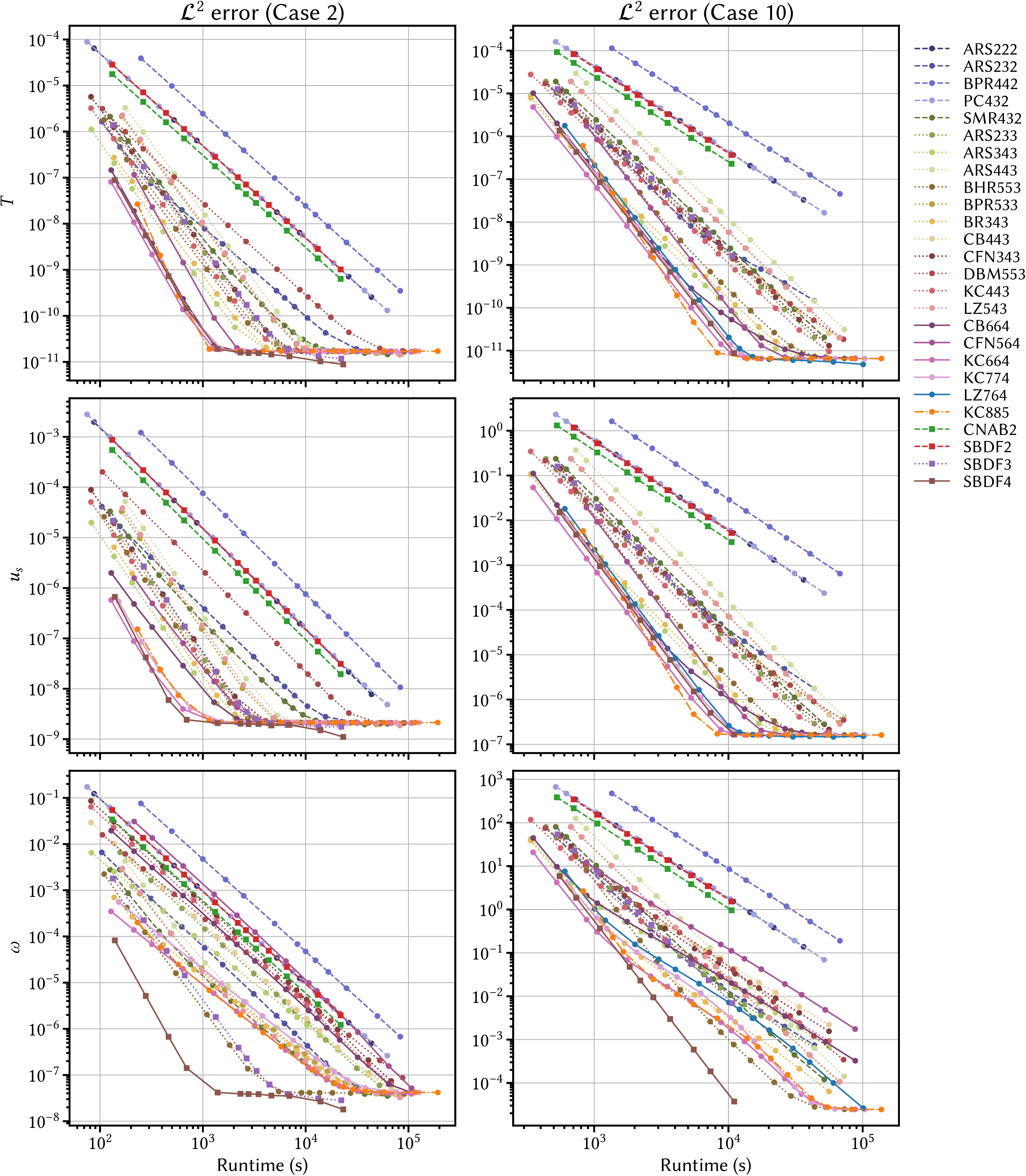}}
\caption{Convergence of the $\mathcal{L}^2$ error for the temperature field
(top panels), the cylindrical radial velocity  $u_s$ (middle panels) and the
vorticity $\omega$ (bottom panels) for Case 2 (left column) and Case 10 (right
column) as a function of total runtime expressed in seconds. 
The markers correspond to
the class of IMEX, with squares denoting IMEX multistep and circles
IMEX-RK multistage schemes. The total runtime is the product of 
the number of iterations times the average walltime, and ignores 
the initial computation and factorization of the requisite matrices.} 
\label{fig:error_runtime}
\end{figure}
}

\clearpage
\section{Expected behaviour of schemes for three dimensional simulations of planetary core dynamics}
\label{sec:pred3d}

\begin{figure}[t!]
 \centerline{\includegraphics[width=\linewidth]{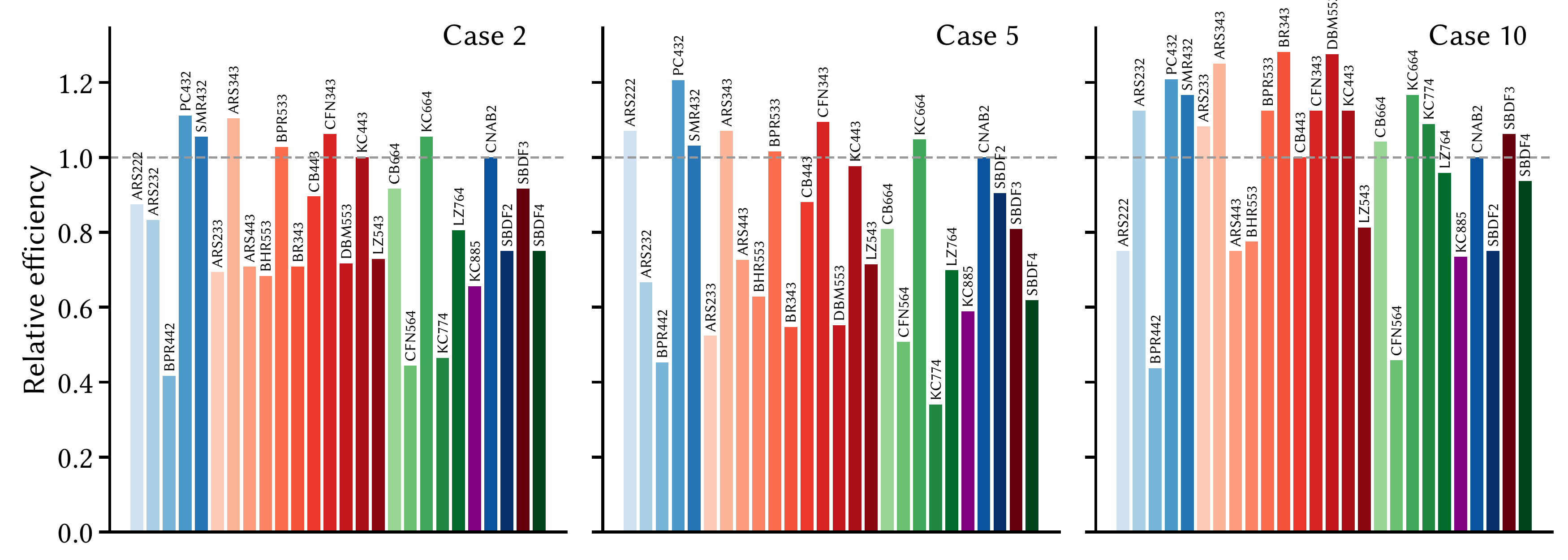}}
 \caption{Conjectured efficiency in three-dimensions  of the time integrators relative to the efficiency of CNAB2 for cases 2, 5 and 10 from left to 
 right. 
 Horizontal dashed lines correspond to a value of unity. 
 See text for details.}
 \label{fig:perf_3D}
 \centerline{\includegraphics[width=\linewidth]{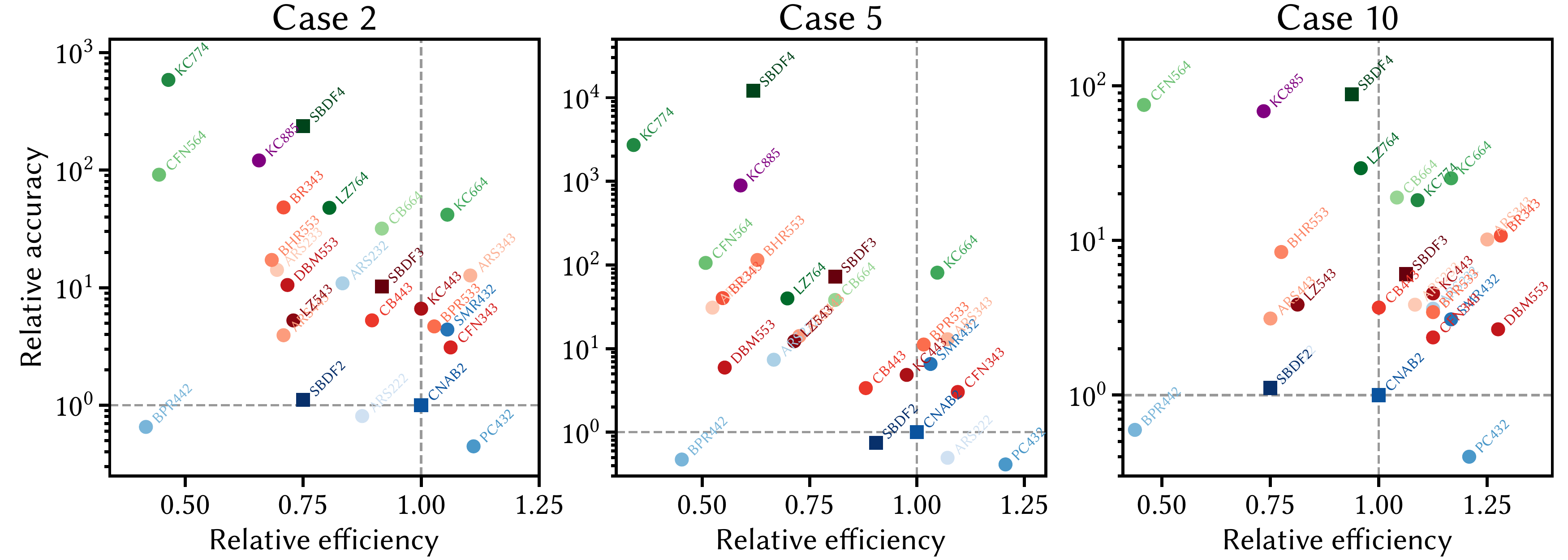}}
 \caption{Conjectured efficiency and accuracy in three-dimensions 
  of the time integrators relative to the efficiency and accuracy 
 of CNAB2 for cases 2, 5 and 10 from left to right. The scale on the $y$-axis is logarithmic. 
 See text for details. 
 Horizontal and vertical dashed lines correspond to a value of unity.}
 \label{fig:perfacc_3D}
 \end{figure}

In this Appendix, we provide the reader with a conjectured efficiency
that the considered time integrators could possibly have in 3-D spherical shell
pseudo-spectral codes. In contrast with the current work where the linear
solves are taking the largest amount of the walltimes, spherical harmonics
transforms involved in the computation of the explicit terms are by a large 
margin the
dominant player of spherical shell code algorithms. As such, under the
asssumption that the CFL coefficients $\alpha^{\mbox{\scriptsize \sc max}}_{\mbox{\scriptsize 
\sc cfl}}$ are
the same in 3-D, the conjectured efficiency of an individual time integrator
for 3-D computations can be estimated using the number of explicit stages
$n^E$ 

\begin{equation}
\eff^{\text{3D}} = \frac{\alpha^{\mbox{\scriptsize \sc max}}_{\mbox{\scriptsize 
\sc cfl}}}{n^E}\,.
\label{eq:eff3D}
\end{equation}
Figure~\ref{fig:perf_3D} shows the conjectured efficiency in 3-D models 
relative to
the efficiency of CNAB2. Compared to Fig.~\ref{fig:perf}, all the schemes which
necessitate an assembly stage have been penalized by the cost of the extra
evaluation of an explicit state, with respect to the stiffly-accurate schemes. 
In
contrast, the schemes with a lower number of explicit stages, such as BPR533, 
present an enhanced efficiency compared to the 2-D computations.
PC432 stands out as the most efficient scheme for cases 2 and 5, while ARS343, 
BR343
and DBM553 become more efficient in the least stiff case 10.
We anticipate that the overall gain in terms of efficiency compared to the
CNAB2 method for 3-D calculations will be smaller than for our 2-D models, 
bounded
to values between 20 and 30\%.

Figure~\ref{fig:perfacc_3D} shows the conjectured efficiency and accuracy in
3-D relative to CNAB2. This figure was obtained making the additional
assumption that the 3-D computations would have similar errors than our 2-D
computations. Focusing our attention on the upper right quadrant, one second
order scheme (SMR432), three third order schemes (ARS343, BPR533 and CFN343)
and one fourth order scheme (KC664) are always more efficient and more accurate
than CNAB2 at the stability limit.

We stress that the reasoning put forward in this Appendix heavily relies on the
assumptions that both the stability coefficients and the convergence curves are
weakly affected by the change from 2-D to 3-D models. While this is a plausible
hypothesis when considering the same physical phenomenon (i.e.
non-rotating convection), the incorporation of additional physical effects such
as rotation or a magnetic field is likely to significantly impact the
convergence and the stability properties of the time integrators.

\end{appendix}

\end{document}